\documentclass[letter,11pt,DIV=12,abstract=true,numbers=noenddot,titlepage=false,twocolumn=false,draft=false]{scrartcl}
\pdfoutput=1 

\makeatletter
\DeclareOldFontCommand{\rm}{\normalfont\rmfamily}{\mathrm}
\DeclareOldFontCommand{\sf}{\normalfont\sffamily}{\mathsf}
\DeclareOldFontCommand{\tt}{\normalfont\ttfamily}{\mathtt}
\DeclareOldFontCommand{\bf}{\normalfont\bfseries}{\mathbf}
\DeclareOldFontCommand{\it}{\normalfont\itshape}{\mathit}
\DeclareOldFontCommand{\sl}{\normalfont\slshape}{\@nomath\sl}

\usepackage[usenames,dvipsnames]{color}
  \definecolor{hgreen}{rgb}{0,.3,0}
  \definecolor{hred}{rgb}{.3,0,0}
  \definecolor{hblue}{rgb}{0,0,.3}
  \definecolor{LightGray}{gray}{0.95}
  \definecolor{gray}{gray}{0.6}
\usepackage[
	    colorlinks=true,
	    linkcolor=hblue,
	    citecolor=hgreen,
	    filecolor=hblue,
	    urlcolor=hred
	    ]{hyperref}
\usepackage{mathrsfs}
\usepackage[intlimits]{amsmath}
\usepackage{amssymb}
\usepackage{slashed}
\usepackage[titletoc,title]{appendix}
\usepackage[affil-it]{authblk}
\usepackage{libertine}
\usepackage[numbers,sort&compress]{natbib}
\usepackage{graphicx}
\usepackage{physics}
\usepackage{bbm}
\usepackage{subcaption}

\newcommand{\jp}{{j_\varphi}}

\setcapindent{1em}
\setkomafont{captionlabel}{\bfseries}
\setkomafont{caption}{\itshape}

\definecolor{Blu}{rgb}{0.,0.,1.}
\definecolor{Red}{rgb}{1.,0.,0.}
\definecolor{Green}{rgb}{0.,1.,0.}
\definecolor{Purple}{rgb}{0.5,0.,0.5}

\allowdisplaybreaks  

\begin{document}
\renewcommand\Authands{, }

\title{\boldmath 
        Two-loop Beta Function for Complex Scalar Electroweak Multiplets
}

\date{\today}
\author[a]{Joachim~Brod%
        \thanks{\texttt{joachim.brod@uc.edu}}}
\author[a]{Zachary~Polonsky%
        \thanks{\texttt{polonsza@mail.uc.edu}}}
	\affil[a]{{\large Department of Physics, University of Cincinnati, Cincinnati, OH 45221, USA}}

\maketitle

\begin{abstract}
We present the general form of the renormalizable four-point
interactions of a complex scalar field furnishing an irreducible
representation of $SU(2)$, and derive a set of algebraic identities
that facilitates the calculation of higher-order radiative
corrections. As an application, we calculate the two-loop beta
function for the SM extended by a scalar multiplet, and provide the
result explicitly in terms of the group invariants. Our results
include the evolution of the Higgs-portal couplings, as well as scalar
``minimal dark matter''. We present numerical results for the two-loop
evolution of the various couplings.
\end{abstract}
\setcounter{page}{1}

\section{Introduction\label{sec:introduction}}

Complex scalar fields furnishing a general representation of the
electroweak gauge group $SU(2) \times U(1)$ of the standard model (SM)
received increased interest in recent years. For instance, they can
provide a viable dark matter candidate in so-called minimal
dark-matter models~\cite{Cirelli:2005uq}.

The renormalization group (RG) evolution of coupling constants is an
invaluable tool in phenomenological analyses~\cite{Collins:1984xc}. It
plays a particularly important role when interpreting and comparing
the results of experiments performed at widely different energy
scales, such as dark matter direct detection and production of dark
matter at particle colliders. A framework for consistent RG analysis
for fermionic dark matter in the context of effective field theories
has been presented in Ref.~\cite{Bishara:2018vix}. The first
consistent and complete basis of effective operators for scalar dark
matter up to mass dimension six has been written down in
Ref.~\cite{Brod:2017bsw}; however, the RG evolution has not yet been
calculated.

For scalar dark matter it is possible to write down self interactions,
as well as interactions with the SM, at the renormalizable level --
the so-called Higgs-portal dark matter~\cite{Silveira:1985rk,
  McDonald:1993ex, Burgess:2000yq}. To our knowledge, the first
classification of the self interactions of scalar fields with
electroweak charges has been given in Ref.~\cite{Chao:2018xwz}. In
this work, we rederive the scalar potential in a slightly different
form that is well suited for the calculation of radiative corrections.

As an application, we calculate the beta functions for all scalar
couplings, as well as the new scalar contributions to all SM
couplings, at the two-loop level. To this end, we prove a set of
algebraic relations that allows to express all two-loop matrix
elements in terms of tree-level matrix elements of the basis operators
in the scalar potential. While these algebraic relations simply rely
on the algebra of Clebsch-Gordan coefficients as well as $SU(2)$ gauge
symmetry, many of them turn out to be quite non-trivial, and have not
been derived before, to the best of our knowledge. Among other
results, we show how to express a product of two $SU(2)$ generators,
contracted over their adjoint indices, in terms of Clebsch-Gordan
coefficients. The resulting relations can be used to manipulate
general representations of the $SU(2)$ algebra in an algorithmic way.

Our results are valid for a scalar field furnishing an arbitrary
irreducible representation of $SU(2)$ and for arbitrary
hypercharge. While these results are known in
principle~\cite{Machacek:1984zw, Jack:1984vj}, we present them in
closed form and explicitly in terms of group invariants for the first
time. We believe that this form of the beta functions makes them more
suitable for practical applications. Auxiliary files with our analytic
results in computer-readable form are available via a \texttt{gitlab}
repository (see Sec.~\ref{sec:conclusions}).

As a cross check of our results, we also calculated the two-loop beta
function for most of the SM Higgs, gauge and Yukawa couplings. We find
a result consistent with the SM beta function extracted from
Ref.~\cite{Machacek:1984zw}, see Ref.~\cite{Arason:1991ic}, if we take
into account the corrections pointed out in Refs.~\cite{Luo:2002ti,
  Luo:2003}. See also Refs.~\cite{Chetyrkin:2012rz, Mihaila:2012,
  Bednyakov:2013cpa, Bednyakov:2014, Davies:2019onf} for recent
results at the three- and four-loop level.

Depending on the representation, the impact of the one- and two-loop
contributions to the running of the scalar as well as the SM couplings
can be sizeable. We discuss a few examples, focusing on a scalar
septuplet (``minimal dark matter'') and the running of the SM quartic
Higgs and $SU(2)$ gauge coupling.

This paper is organized as follows. In Sec.~\ref{sec:op:basis} we
define our setup and construct the scalar potential. In
Sec.~\ref{sec:beta} we present our results for the beta functions. The
required algebraic relations are collected and proven in
Sec.~\ref{grp:thy:rels}. Sec.~\ref{sec:numerics} contains numerical
illustrations of our results. We conclude in
Sec.~\ref{sec:conclusions}. Supplementary material is presented in two
appendices. In App.~\ref{app:checks} we describe the various analytic
checks that we performed on our calculation, and derive explicit
formal expressions for the beta functions. In App.~\ref{app:ren} we
provide all field and mass renormalization constants that are
necessary in intermediate steps of the calculation. For completeness,
we also include all quadratic poles of the coupling renormalization
constants.

\section{Construction of the operator basis}\label{sec:op:basis}

We consider a complex scalar field $\varphi$ with mass $M_\varphi$
which furnishes a $(2\jp+1)$-dimensional irreducible representation of
the Standard Model $SU(2)\times U(1)$ gauge group, where
$\jp=0,1/2,1,\ldots$ is any integer or half integer. The Lagrangian
for this model is given by
\begin{equation}\begin{split}\label{eq:lag}
 \mathcal{L}_\varphi = (D_\mu\varphi)^\dag D^\mu\varphi 
  - M_\varphi^2 \varphi^\dag \varphi
 - \frac{1}{4}W^a_{\mu\nu}W^{a\mu\nu} - 
	\frac{1}{4}B_{\mu\nu}B^{\mu\nu} - V_\varphi \,.
\end{split}\end{equation}
The summation convention over Lorentz and adjoint gauge indices is in
use here and in the following. The covariant derivative acting on the
scalar field is given by
\begin{equation}
 D_\mu \varphi_k = \sum_{l}\left(\delta_{kl}\partial_\mu 
	- ig_2\tilde{\tau}^a_{kl}W^a_\mu
	+ i\frac{Y_\varphi}{2}g_1\delta_{kl}B_\mu\right)\varphi_l \,,
\end{equation}
with the corresponding field strength tensors
\begin{equation}\begin{split}
 W^a_{\mu\nu} = \partial_\mu W^a_\nu-\partial_\nu W^a_\mu +
 g_2\epsilon^{abc}W^b_\mu W^c_\nu \,, \qquad B_{\mu\nu} = \partial_\mu
 B_\nu-\partial_\nu B_\mu \,.
\end{split}\end{equation}
Here, $B_\mu$ and $W_\mu^a$ (with $a=1,2,3$) are the $U(1)$ and
$SU(2)$ gauge fields, respectively. The $\tilde{\tau}^a_{kl}$ are
$SU(2)$ generators in the $(2\jp+1)$-dimensional representation, defined
by
\begin{equation}\begin{split} \label{eq:su2-gen-def}
\big(\tilde \tau^1 \pm i \tilde\tau^2 \big)_{kl} = \delta_{k,l\pm 1}
\sqrt{(\jp \mp l)(\jp \pm l + 1)} \, , \qquad \big( \tilde\tau^3
\big)_{kl} = l \delta_{k,l} \, ,
\end{split}\end{equation}
with $k,l$ running over the values $-\jp, -\jp+1, \ldots, \jp-1, \jp$, while
$Y_\varphi$ is the scalar hypercharge. 

We now derive the general form of the scalar potential
$V_\varphi$. Any Hermitian, renormalizable four-scalar operator has
the general form
\begin{equation}\label{eq:op:gen}
 \mathcal{O}_\varphi = \sum_{irks}\varphi_i^*\varphi_k^*
 \varphi_r\varphi_s v_{irks}^{\jp} \,.
\end{equation}
The form of the real coefficients $v_{irks}^{\jp}$ must be determined
such that the operator $\mathcal{O}_\varphi$ is invariant under the
$SU(2)$ gauge group (the $U(1)$ invariance is immediately
apparent). Ignoring all quantum numbers that do not transform under
$SU(2)$, the operator coefficients can be written as
\begin{equation}
v_{irks}^{\jp} \equiv \bra{\jp,r;\jp,s} V \ket{\jp,i;\jp,k}
\end{equation}
where $V$ are the reduced matrix elements. Inserting two complete sets
of states, we have
\begin{equation}
\begin{split}
v^{\jp}_{irks} & = \sum_{JJ'} \sum_{MM'} C_{\jp\jp}(JM;rs)
C_{\jp\jp}(J'M';ik) \bra{JM}V\ket{J'M'} \\ & \equiv \sum_{JJ'} \sum_{MM'}
C_{\jp\jp}(JM;rs) C_{\jp\jp}(J'M';ik) v^{JJ'}_{MM'} \,,
\end{split}
\end{equation}
where $C_{jj'}(JM;mm')$ are Clebsch-Gordan coefficients (we use the
notation of Ref.~\cite{weinberg_2015}). Defining the composite field
operator
\begin{equation}
\Phi^{(J)}_M\equiv \sum_{mn}\varphi_m\varphi_nC_{\jp\jp}(JM;mn) \,,
\end{equation}
Eq.~\eqref{eq:op:gen} becomes
\begin{equation}
\mathcal{O}_\varphi = \sum_{JJ'}\sum_{MM'} 
\left(\Phi^{(J)}_{M}\right)^*\Phi^{(J')}_{M'} v^{JJ'}_{MM'} \,.
\end{equation}
Writing a general $SU(2)$ transformation as $D^{(J)} = \exp
(i\theta^a\tilde\tau^{(J),a})$, where $\tilde\tau^{(J),a}$ are here the
$SU(2)$ generators in the $2J+1$-dimensional representation, gauge
invariance requires
\begin{equation}
 \mathcal{O}_\varphi\to \sum_{JJ'}\sum_{MM'NN'}
 \left(\Phi^{(J)}_{N}\right)^*\Phi^{(J')}_{N'}
 \left(D^{(J)}_{MN}\right)^*D^{(J')}_{M'N'}v^{JJ'}_{MM'} =
 \mathcal{O}_\varphi \,.
\end{equation}
Using the unitarity of the $D$ matrices, this can be written as the
condition
\begin{equation}
\sum_{M'}v^{JJ'}_{MM'}D^{(J')}_{M'N'} =
\sum_{N}D^{(J)}_{MN}v^{JJ'}_{NN'} \,.
\end{equation}
By Schur's Lemma, $v$ is either zero or has the form
\begin{equation}
 v^{JJ'}_{MM'} = \frac{\lambda^{(J)}_\varphi}{4}\delta^{JJ'}\delta_{MM'} \,,
\end{equation}
where $\lambda^{(J)}_\varphi$ is a constant. We define a set of
``Sigma matrices'' as
\begin{equation}\label{eq:gen:Sigma:def}
	\Sigma^{(J),M}_{mn}\equiv C_{\jp\jp}(JM,mn) \,,
\end{equation}
(note that we regard the isospin $\jp$ of the scalar multiplet to be
fixed in this work). We then write the general potential
as\footnote{We assume the invariance of the Lagrangian under a global
  $U(1)$ symmetry under which only the scalar fields transform
  non-trivially, so that we do not introduce additional ``exotic''
  operators for special values of $Y_\varphi$ (cf.
  Ref.~\cite{Chao:2018xwz}).}
\begin{equation}\label{eq:pot:def}
 V_{\varphi}[\varphi] = \sum_{J} \frac{\lambda^{(J)}_\varphi}{4}
 \sum_{M} \Big| \sum_{mn} \varphi_m \Sigma_{mn}^{(J),M} \varphi_n \Big|^2 \,.
\end{equation}
The symmetry properties of the Clebsch-Gordan coefficients imply the
corresponding properties of the Sigma matrices,
\begin{equation}\label{eq:symm:gen:Pauli}
\Sigma_{mm'}^{(J),a} = (-1)^{J-2j_\varphi} \Sigma_{m'm}^{(J),a} \,.
\end{equation}
This restricts the number of independent operators in the
basis. Obviously, the coefficients $v_{irks}^\jp$ in
Eq.~\eqref{eq:op:gen} can be chosen symmetric under exchange of $i
\leftrightarrow k$ and $r \leftrightarrow s$. Hence, the only non-zero
operators in our basis are those involving Sigma matrices that are
symmetric in their lower indices,
\begin{equation}\label{eq:symmetric}
 \Sigma_{mm'}^{(J)a}=\Sigma_{m'm}^{(J)a} \,.
\end{equation}
This immediately tells us that there are $N_\varphi \equiv
\text{floor}(\jp+1)$ operators in our basis. As a related consequence,
the sum over $J$ in Eq.~\eqref{eq:pot:def} effectively runs only over
even values for integer $\jp$, while for half-integer $\jp$ only terms
with odd $J$ contribute.

We illustrate this construction by the example of an electroweak
doublet. The Sigma matrices for $\jp=1/2$ are
\begin{equation}
\Sigma^{(0),0} =
\frac{1}{\sqrt{2}}
\begin{pmatrix} 
0&1\\-1&0
\end{pmatrix};
\quad 
\Sigma^{(1),1} =
\begin{pmatrix} 
1&0\\0&0
\end{pmatrix},
\quad 
\Sigma^{(1),0} =
\frac{1}{\sqrt{2}}
\begin{pmatrix} 
0&1\\1&0
\end{pmatrix},
\quad 
\Sigma^{(1),-1} =
\begin{pmatrix} 
0&0\\0&1
\end{pmatrix}.
\end{equation}
The potential operator for $J=0$ vanishes identically:
\begin{equation}
  {\cal O}^{(0)} = \sum_{ikrs=-1/2}^{1/2} \varphi^*_i \varphi_r
  \varphi^*_k \varphi_s \Sigma_{ik}^{(0),0} \Sigma_{rs}^{(0),0} =
  \frac{1}{2} \big| \varphi_{1/2}\varphi_{-1/2} -
  \varphi_{1/2}\varphi_{-1/2} \big|^2 \equiv 0\,,
\end{equation}
and only the operator for $J=1$ remains:
\begin{equation}
\begin{split}
  {\cal O}^{(1)} & = \sum_{ikrs=-1/2}^{1/2} \sum_{a=-1}^1 \varphi^*_i
  \varphi_r \varphi^*_k \varphi_s \Sigma_{ik}^{(1),a}
  \Sigma_{rs}^{(1),a} \\ & = \big| \varphi_{1/2} \big|^4 + \big|
  \varphi_{-1/2} \big|^4 + 2 \big| \varphi_{1/2} \big|^2 \big|
  \varphi_{-1/2} \big|^2 \equiv (\varphi^\dag \varphi)^2 \,.
\end{split}
\end{equation}
This is equivalent to the fact that we can, employing the more
standard definition of operators, express $(\varphi^\dag \sigma^a
\varphi)^2$ in terms of $(\varphi^\dag \varphi)^2$, using the Fierz
relation $\sigma_{ij}^a \sigma_{kl}^a = 2 \delta_{il} \delta_{kj} -
\delta_{ij} \delta_{kl}$. Here, $\sigma_{ij}^a$ are the usual Pauli
matrices.

\section{Beta function for a scalar multiplet}\label{sec:beta}

\begin{figure}
	\centering
	\begin{subfigure}{0.45\textwidth}
	\centering
	\includegraphics[width=0.6\linewidth]{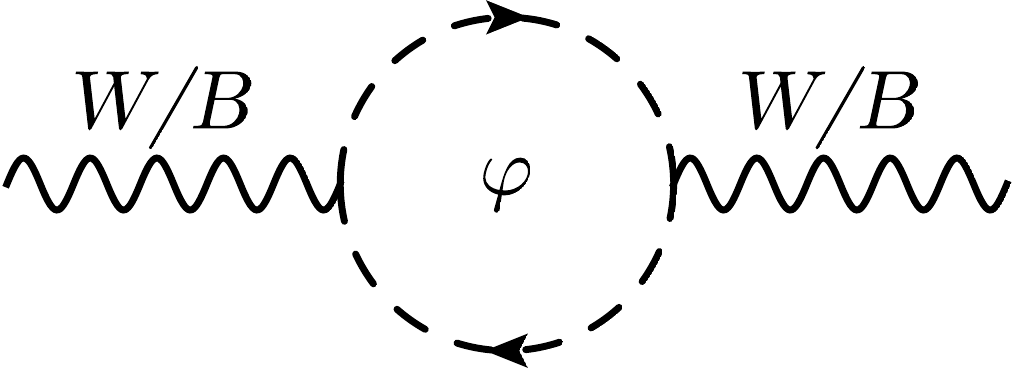}
	\caption{}
	\label{fig:1loop_gauge}
	\end{subfigure}
	\begin{subfigure}{0.45\textwidth}
	\centering
	\includegraphics[width=0.6\linewidth]{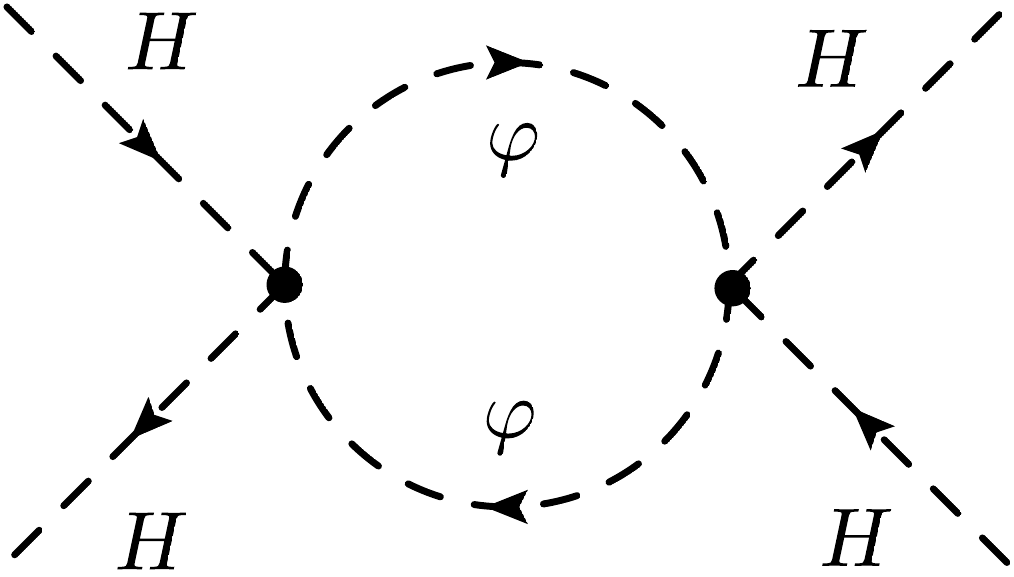}
	\caption{}
	\label{fig:1loop_Higgs}
	\end{subfigure}
	\caption{Feynman diagrams corresponding to the contributions
          from the scalar field $\varphi$ to the one-loop standard
          model beta function. Fig. (a) shows the contribution to the
          gauge boson field counterterms which must be subtracted when
          gauge bosons appear in external states in Green's
          functions. Fig. (b) shows the $\varphi$ loop contributing to
          the one-loop Higgs quartic coupling beta function.}
	 \label{fig:SM_1loop}
\end{figure}

In this section, we present the beta function of the full SM extended
by a scalar $\varphi$ furnishing a representation $(0,\jp,Y_\varphi)$
under the SM $SU(3)_c\times SU(2)_L\times U(1)_Y$ gauge group. The
Lagrangian we consider is given by
\begin{equation}\label{eq:fulllag}
	\mathcal{L}=\mathcal{L}_\varphi + \mathcal{L}_\psi 
	+\mathcal{L}_H + \mathcal{L}_Y + \mathcal{L}_\text{QCD}
	+ \mathcal{L}_{\text{portal}}
\end{equation}
where $\mathcal{L}_\varphi$ is given in Eq.~\eqref{eq:lag},
\begin{equation}
\mathcal{L}_\text{QCD} = - \frac{1}{4}G^A_{\mu\nu}G^{A\,\mu\nu}
\end{equation}
is the gluonic QCD Lagrangian, and
\begin{equation}
	\mathcal{L}_\psi = \sum_{k}\overline{Q}_{L,k}i
	\slashed{D}Q_{L,k}
	+ \sum_{k}\overline{u}_{R,k}i
	\slashed{D}u_{R,k}
	+ \sum_{k}\overline{d}_{R,k}i
	\slashed{D}d_{R,k}
	+\sum_{k}\overline{L}_{L,k}i
	\slashed{D}L_{L,k}+
	\sum_{k}\overline{\ell}_{R,i}i
	\slashed{D}\ell_{R,k}
\end{equation}
are the kinetic terms for the SM fermions, where $Q_L$ and $L_L$
denote the left-handed quark and lepton doublets, and $u_R$, $d_R$,
and $\ell_R$ the right-handed up-quark, down-quark, and lepton
fields. The sums run over the three fermion generations,
$k=1,2,3$. Furthermore,
\begin{equation}
	\mathcal{L}_H=(D_\mu H)^\dag D^\mu H + \mu^2 H^\dag H
	- \frac{\lambda_H}{4}\big(H^\dag H\big)^2
\end{equation}
is the Higgs doublet Lagrangian, and the Yukawa Lagrangian is given by
\begin{equation}
	\mathcal{L}_Y = -\sum_{kl}\overline{Q}_{L,k}
	Y_u^{kl} H^c u_{R,l} - \sum_{kl}
	\overline{Q}_{L,k}Y_d^{kl} H d_{R,l} - \sum_{kl}
	\overline{L}_{L,k}Y_\ell^{kl} H \ell_{R,l} +\text{h.c.}\,,
\end{equation}
where $H^c=i\sigma_2H^*$ is the charge-conjugated Higgs field. In this
work, we neglect the Yukawa couplings of all light fermions, keeping
only the top, bottom, charm, and $\tau$ Yukawas $y_t$, $y_b$, $y_c$,
and $y_\tau$ non-zero. This implies that we can assume the Yukawa
matrices to be diagonal and neglect CKM mixing. Finally, the
Higgs-portal Lagrangian is given by
\begin{equation}\label{eq:higgs:portal}
 {\mathcal L}_{\text{portal}} = - \frac{\lambda_{\varphi H}}{4}
	\big(\varphi^\dag\varphi\big)\big(H^\dag H\big)
	- \frac{\lambda_{\varphi H}'}{4}
	\big(\varphi^\dag\tilde{\tau}^a\varphi\big)
	\big(H^\dag\tau^a H\big)\,.
\end{equation}
Here, $\tau^a \equiv \sigma^a/2$ in terms of the usual Pauli
matrices. Note that the second term in Eq.~\eqref{eq:higgs:portal} is
absent in the case $\jp=0$.

The Lagrangian~\eqref{eq:fulllag} is renormalized in the usual way by
introducing field and coupling renormalization constants. For
instance, we express the unrenormalized scalar couplings (denoted by
the superscript ``0'') in terms of renormalized couplings as
\begin{equation}
  \lambda_\varphi^{(J),0} = Z_{\lambda_\varphi^{(J)}} \lambda_\varphi^{(J)} 
  = (1 + \delta Z_{\lambda_\varphi^{(J)}}^{(1)} 
       + \delta Z_{\lambda_\varphi^{(J)}}^{(2)} 
       + \ldots) \lambda_\varphi^{(J)} \,,
\end{equation}
and similarly for all other couplings and fields. The superscripts
$(1)$ and $(2)$ denote the one- and two-loop contributions,
respectively. The ellipsis stands for higher-order terms. We extract
the beta function in the $\overline{\mathrm{MS}}$ scheme from the
$1/\epsilon$ poles of the coupling counterterms, as explained in
App.~\ref{app:checks}. We employ dimensional regularization in $d = 4
- 2\epsilon$ space-time dimensions, and we can treat all particles as
massless in our calculation.

\begin{figure}
	\centering
	\includegraphics[width=0.87\linewidth]{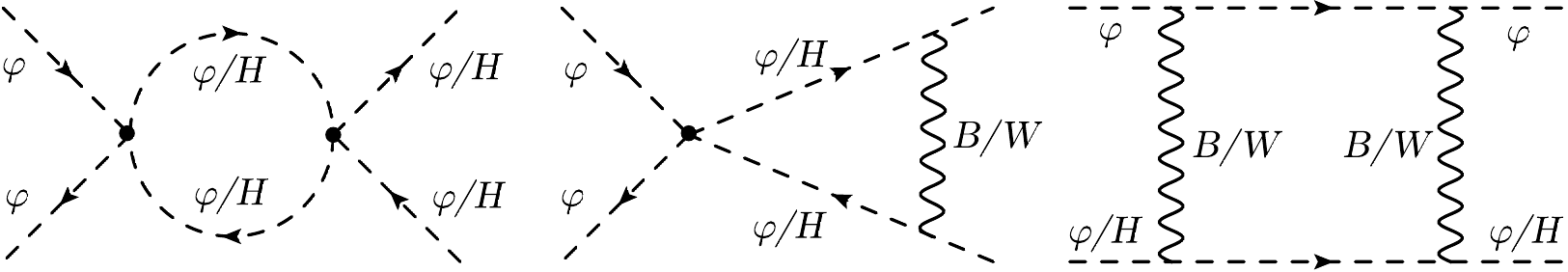}
	\caption{Sample one-loop Feynman diagrams for the calculation
          of the quartic scalar coupling and the Higgs-portal
          couplings.}
	\label{fig:QSC_1l}
\end{figure}

\begin{figure}
	\centering
	\begin{subfigure}{0.66\textwidth}
	\centering
	\includegraphics[width=0.9\linewidth]{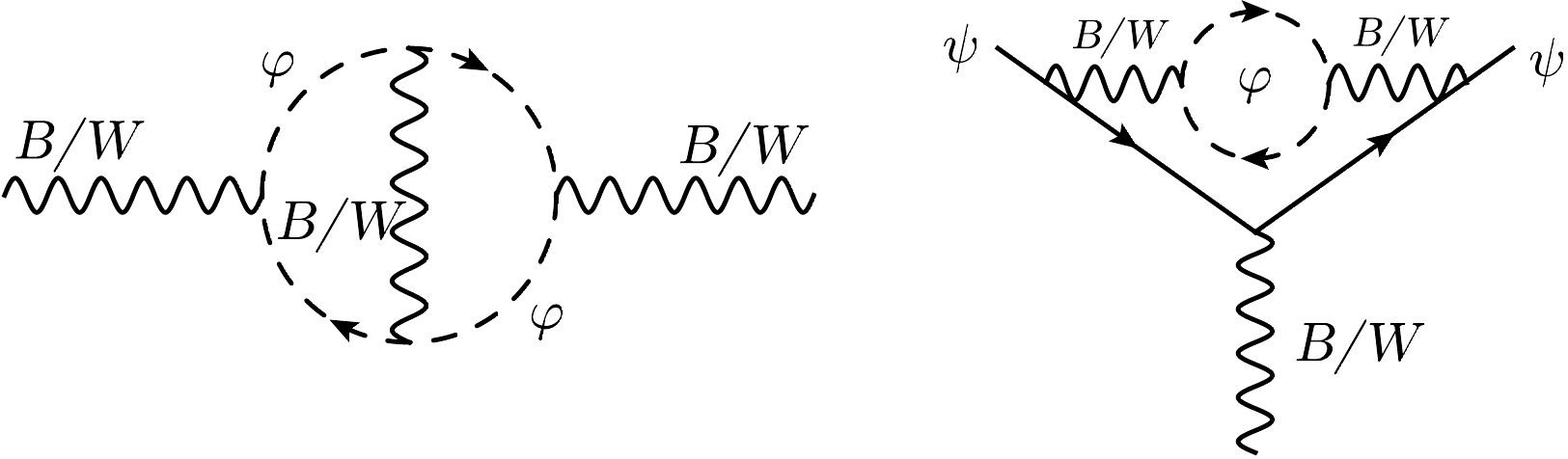}
	\caption{}
	\label{fig:2loop_gauge}
	\end{subfigure}
	\begin{subfigure}{0.33\textwidth}
	\centering
	\includegraphics[width=\linewidth]{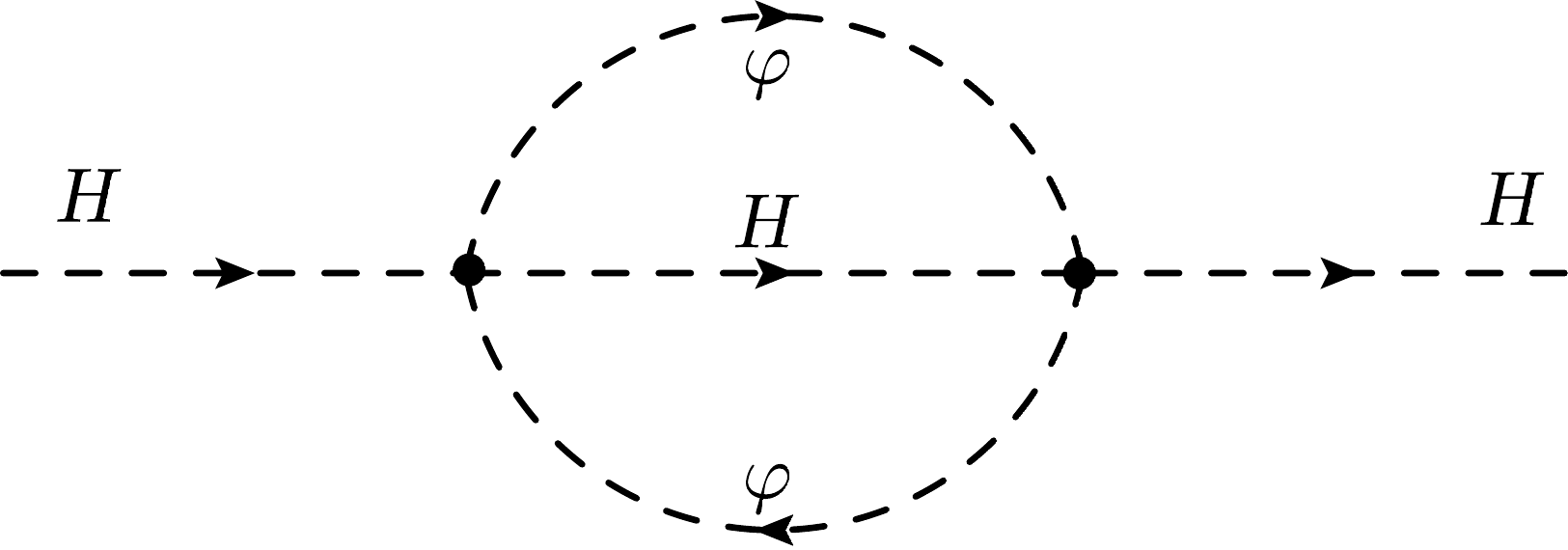}
	\caption{}
	\label{fig:2loop_Higgs}
	\end{subfigure}
	\caption{Sample Feynman diagrams showing two-loop
          contributions from the scalar field $\varphi$ to gauge and
          Yukawa coupling beta functions. Fig.  (a) shows
          contributions to gauge coupling beta functions from gauge
          boson self-energies as well as $\varphi$ insertions to the
          three-point Green's function with a single external gauge
          boson. Here, $\psi$ represents any fermion which couples to
          the gauge fields. Fig. (b) shows a diagram from the two-loop
          Higgs self-energy which contributes to the Yukawa coupling
          beta functions.}
\end{figure}

\begin{figure}
	\centering
	\includegraphics[width=0.75\linewidth]{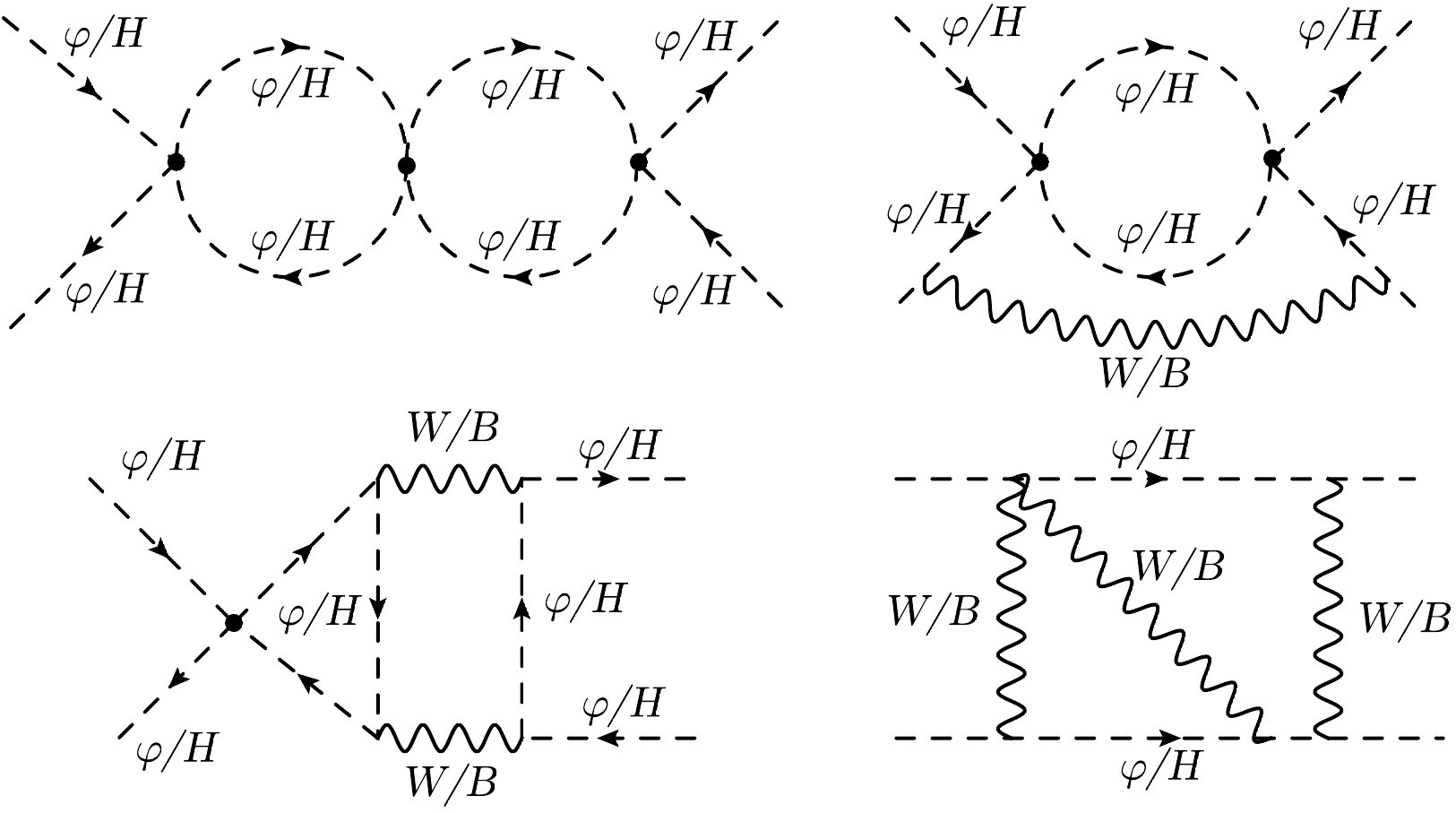}
	\caption{Sample Feynman diagrams which give contributions to
	quartic scalar couplings $\lambda_H$, $\lambda_{\varphi H}$,
	$\lambda_{\varphi H}'$, and $\lambda_\varphi^{(J)}$ from
	$\varphi$. These diagrams divide into four classes: zero 
	(top left), one (top right), two (bottom left), and three
	(bottom right) gauge boson insertions.}
	\label{fig:QSC_2loop}
\end{figure}

We determine all renormalization constants by calculating the
divergent parts of Green's functions with suitably chosen external
states (sample Feynman diagrams are shown in
Figs.~\ref{fig:SM_1loop}-\ref{fig:QSC_2loop}). In the
calculation of the coupling counterterms, it is necessary to subtract
field counterterms corresponding to the external fields. For this
reason, all field renormalization constants are calculated in addition
to the coupling renormalization constants (the results are
collected in App.~\ref{app:ren}).

In order to isolate the ultraviolet poles, we employ the infrared (IR)
rearrangement described in Ref.~\cite{Chetyrkin:1997fm}, to which we
refer for more details. In short, the method amounts to an exact
decomposition of all propagators in terms of propagators with a common
IR regulator mass, which we call $M_\text{IRA}$. Effectively, we
introduce a common mass $M_\text{IRA}$ for the scalar, the
gauge-boson, and the ghost fields,
\begin{equation}
 \mathcal{L}_\text{IRA} = 
   \frac{1}{2} M_\text{IRA}^2 W^a_\mu W^{a\,\mu} 
 + \frac{1}{2} M_\text{IRA}^2 B_\mu B^{\mu} 
 - M_\text{IRA}^2 \varphi^\dag_i\varphi_i
 - M_\text{IRA}^2 \bar{u}_W^a u_W^a\,.
\end{equation}
These masses get renormalized at higher orders, and we introduce
corresponding mass counterterms $Z_{M_\text{IRA},i}$, $i=W,B,\varphi$,
in the usual way ($M^2_\text{bare} = Z_{M^2} M^2$). The explicit
results needed for our work are collected in App.~\ref{app:ren}. We
explicitly verified that all our results are independent of the
regulator mass $M_\text{IRA}$, as it should be.

All ${\mathcal O}(10\,000)$ Feynman diagrams were calculated
using self-written \texttt{FORM}~\cite{Vermaseren:2000nd} routines,
encoding the algorithm presented in Ref.~\cite{Davydychev:1992mt}. The
Feynman diagrams were generated using
\texttt{qgraf}~\cite{Nogueira:1991ex}. The $SU(2)$ group algebra and
renormalization was performed independently by the two authors; the
results are in complete agreement. We describe further analytic checks
of our calculation in App.~\ref{app:checks}.

The beta functions are defined as the logarithmic derivatives of the
couplings with respect to the renormalization scale,
\begin{equation}
 \mu \frac{d}{d\mu} g_i = \beta_{g_i} \,.
\end{equation}
They are given in terms of the coupling counterterms by
\begin{equation}
 \beta_{g_i} = 
	g_i\sum_k a_k g_k \frac{\partial Z_{g_i,1}}{\partial g_k}
\end{equation}
for all couplings, denoted here collectively by $g_i = g_1, g_2, g_s,
\lambda_\varphi^{(J)}, \lambda_{\varphi H}, \lambda_{\varphi H}',
\lambda_{H}, y_t, y_b, y_c, y_\tau$. Here, $Z_{g_i,1}$ is the residue of 
the $1/\epsilon$ pole of the counterterm and $a_k=1$ when $g_k$ is a 
gauge or Yukawa coupling while $a_k=2$ when $g_k$ is a quartic scalar 
coupling. Expanding the beta function by loop order as 
$\beta_{g_i} = \beta_{g_i}^{(1)} + \beta_{g_i}^{(2)} + \ldots$, we find 
for the one-loop contributions
\begin{align}\label{eq:beta:1loop}
 \beta_{g_1}^{(1)} & =
   \frac{g_1^3}{16\pi^2}
	\left(\frac{Y_\varphi^2}{12}\mathcal{D}(\jp)
	+ \frac{1}{6} + \frac{20}{9} n_g \right)\,, \\
 \beta_{g_2}^{(1)} & = 
   \frac{g_2^3}{16\pi^2}
 \left(\frac{1}{9}\mathcal{J}(\jp)\mathcal{D}(\jp)
	-\frac{43}{6} + \frac{4}{3} n_g \right) \,, \\
 \beta_{g_s}^{(1)} & =
  \frac{g_s^3}{16\pi^2}\left(\frac{4}{3} n_g - 11\right) \,, \\
 \beta_{y_t}^{(1)} & =
  \frac{y_t}{16\pi^2}
	\left(-\frac{17 g_1^2}{12}-\frac{9g_2^2}{4}-8g_s^2
	+ \frac{9y_t^2}{2} + 3y_c^2 + \frac{3y_b^2}{2} + y_\tau^2\right) \,, \\
 \beta_{y_b}^{(1)} & =
  \frac{y_b}{16\pi^2}
	\left(-\frac{5 g_1^2}{12}-\frac{9g_2^2}{4}-8g_s^2
	+ \frac{3y_t^2}{2} + 3y_c^2 + \frac{9y_b^2}{2} + y_\tau^2\right) \,, \\
 \beta_{y_c}^{(1)} & =
  \frac{y_c}{16\pi^2}
	\left(-\frac{17 g_1^2}{12}-\frac{9g_2^2}{4}-8g_s^2
	+ 3y_t^2 + \frac{9y_c^2}{2} + 3y_b^2 + y_\tau^2\right) \,, \\
 \beta_{y_\tau}^{(1)} & =
  \frac{y_\tau}{16\pi^2}
	\left(-\frac{15 g_1^2}{4}-\frac{9g_2^2}{4}
	+ 3y_t^2 + 3y_c^2 + 3y_b^2 + \frac{5y_\tau^2}{2}\right) \,, \\
\begin{split}
 \beta_{\lambda_H}^{(1)} & = 
  \frac{g_1^2}{16\pi^2}\left(\frac{3g_1^2}{2}-3\lambda_H\right)
  +\frac{g_2^2}{16\pi^2}\left(\frac{9g_2^2}{2}-9\lambda_H\right)
  +\frac{3g_1^2g_2^2}{16\pi^2} \\
 & \quad - \frac{1}{2\pi^2}\left(3y_t^4+3y_c^4+3y_b^4+y_\tau^4\right)
	+\frac{\lambda_H}{4\pi^2}\left(3y_t^2+3y_c^2+3y_b^2+y_\tau^2\right) \\
 & \quad + \frac{3\lambda_H^2}{8\pi^2}
  +\frac{\lambda_{\varphi H}^2}{64\pi^2}\mathcal{D}(\jp)
  +\frac{\lambda_{\varphi H}'^2}{768\pi^2}\mathcal{D}(\jp)\mathcal{J}(\jp)\,,
\end{split} \\
\begin{split}
 \beta_{\lambda_{\varphi H}}^{(1)} & =
  \frac{g_1^2}{16\pi^2}\left[3g_1^2Y_\varphi^2
	-\frac{3\lambda_{\varphi H}}{2}\left(1+Y_\varphi^2\right)\right]
  + \frac{g_2^2}{16\pi^2}\left[12g_2^2\mathcal{J}(\jp)
	-\lambda_{\varphi H}\left(\frac{9}{2}+6\mathcal{J}(\jp)\right)\right] \\
 & \quad + \frac{\lambda_{\varphi H}}{8\pi^2}\left(3y_t^2+3y_c^2+3y_b^2+y_\tau^2\right) 
  + \frac{3\lambda_H\lambda_{\varphi H}}{16\pi^2}
  + \frac{\lambda_{\varphi H}^2}{16\pi^2} \\
 & \quad + \frac{\lambda_{\varphi H}'^2}{64\pi^2}\mathcal{J}(\jp)
  + \frac{\lambda_{\varphi H}}{8\pi^2}{\sum_J}'\lambda_{\varphi}^{(J)}
	\frac{\mathcal{D}(J)}{\mathcal{D}(\jp)}\,,
\end{split} \\
\begin{split}
 \beta_{\lambda_{\varphi H}'}^{(1)} & =
  -\frac{3g_1^2\lambda_{\varphi H}'}{32\pi^2}\left(1+Y_\varphi^2\right)
  -\frac{g_2^2\lambda_{\varphi H}'}{16\pi^2}\left(\frac{9}{2}+6\mathcal{J}(\jp)\right)
  +\frac{3g_1^2g_2^2}{2\pi^2}Y_\varphi \\
 & \quad + \frac{\lambda_{\varphi H}'}{8\pi^2}\left(3y_t^2+3y_c^2+3y_b^2+y_\tau^2\right)
  +\frac{\lambda_{\varphi H}'\lambda_H}{16\pi^2}
  +\frac{\lambda_{\varphi H}'\lambda_{\varphi H}}{8\pi^2} \\
 & \quad +\frac{\lambda_{\varphi H}'}{16\pi^2}{\sum_J}'\lambda_{\varphi}^{(J)}
	\frac{\mathcal{D}(J)\big(\mathcal{J}(J) 
       - 2 \mathcal{J}(\jp)\big)}{\mathcal{J}(\jp)\mathcal{D}(\jp)} \,,
\end{split} \\
\begin{split}
\beta_{\lambda_\varphi^{(J)}}^{(1)} & = 
   \left(\frac{\lambda_{\varphi}^{(J)}}{4\pi}\right)^2
 + \frac{3g_2^2}{8\pi^2}
   \left(g_2^2 \big[\big(\mathcal{J}(J) - 2 \mathcal{J}(\jp)\big)^2
                   +\big(\mathcal{J}(J) - 2 \mathcal{J}(\jp)\big)\big] 
              -2\lambda_\varphi^{(J)}\mathcal{J}(\jp)\right) \\
 & \quad + \frac{1}{4\pi^2}
   {\sum_{J_1,J_2}}^\prime
   \lambda_\varphi^{(J_1)}\lambda_\varphi^{(J_2)}K\left(J_1,J_2,J\right)
 + \frac{\lambda_{\varphi H}^2}{32\pi^2}
 + \frac{\lambda_{\varphi H}'^2}{256\pi^2}\big(\mathcal{J}(J) - 2 \mathcal{J}(\jp)\big) \\
 & \quad + \frac{3g_1^4}{32\pi^2} Y_\varphi^4
	 - \frac{3g_1^2\lambda_\varphi^{(J)}}{16\pi^2} Y_\varphi^2
         + \frac{3g_1^2 g_2^2}{8\pi^2}
           \big(\mathcal{J}(J) - 2 \mathcal{J}(\jp)\big) Y_\varphi^2 \,.
\end{split}
\end{align}
Here and in the following, a prime on the summation sign indicates a
restricted sum over indices, defined by
\begin{equation}
{\sum_J}^\prime \ldots \equiv \sum_{J=0}^{2\jp}
\frac{1+(-1)^{2\jp-J}}{2} \ldots \,.
\end{equation}
The sum effectively runs over even or odd values of $J$ only, if the
weak isospin $\jp$ of the scalar multiplet is integer or half-integer,
respectively. (For instance, for a $SU(2)$ septuplet with $\jp=3$ we
have $J=0,2,4,6$.) The group-theory functions are defined as
$\mathcal{J}(\jp) \equiv \jp(\jp+1)$, $\mathcal{D}(\jp) \equiv
2\jp+1$, and
\begin{equation}
        K(J_1,J_2,J_3) \equiv \mathcal{D}(J_1)\mathcal{D}(J_2)\left\{
        \begin{array}{ccc}
        J_1 & \jp & \jp \\
        \jp & J_2 & \jp \\
        \jp & \jp & J_3
        \end{array}
        \right\}\,,
\end{equation}
in terms of the Wigner $9j$ symbol~\cite{Messiah:1962} -- see
Sec.~\ref{grp:thy:rels} for more details. Moreover, $n_g = 3$ denotes
the number of SM fermion generations. Our one-loop results for the
pure SM contributions agree with those in
Ref.~\cite{Arason:1991ic}. The scalar contribution to
$\beta_{g_2}^{(1)}$ agrees with the expression given in
Ref.~\cite{Chao:2018xwz}. The remaining results are new.

We note here that, at one-loop, the only beta functions which receive
contributions from the complex scalar are the gauge and quartic scalar
couplings. The contributions to the gauge coupling beta functions
arise in our calculation from the gauge boson field counterterms
(Fig.~\ref{fig:1loop_gauge}). In addition to SM terms, the Higgs
quartic coupling beta function gains two terms from diagrams with
scalar loops, shown in Fig.~\ref{fig:1loop_Higgs}.

The beta functions for the Higgs-portal couplings and quartic scalar
couplings are subdivided into three classes: scalar only terms, mixed
scalar-gauge terms, and gauge-only terms. Sample diagrams of each of
these classes are shown in Fig.~\ref{fig:QSC_1l}. The Higgs-portal
coupling beta functions also receive contributions from Yukawa
couplings, coming from the field counterms for the external Higgs
fields in the four-point Green's functions.

In order to express these contributions in terms of the operators in
the scalar potential~\eqref{eq:pot:def}, we rewrite all $SU(2)$
generators appearing in the $W$-boson vertices in terms of the Sigma
matrices defined in Eq.~\eqref{eq:gen:Sigma:def}, and use completeness
relations for the Clebsch-Gordan coefficients to simplify the terms. A
similar strategy is applied for the ``mixed'' contributions involving
both gauge and scalar interactions. The detailed relations that we use
are discussed in Sec.~\ref{grp:thy:rels}. In several cases, particular
care has to be taken, as the sum over indices in the completeness
relations runs over all possible values of the $J$ spin quantum
number, while the local scalar interactions can only involve the
restricted sums over odd or even values. Gauge invariance ensures that
the final result can be expressed in terms of restricted sums only.

A comment on our treatment of $\gamma_5$ is in order. Diagrams
containing fermion triangles can contribute terms with an odd number
of $\gamma_5$ matrices to the gauge-boson field counterterm and gauge
coupling counterterms. We took the corresponding contributions to the
gauge coupling beta functions from the
literature~\cite{Arason:1991ic,Machacek:1984zw}, and calculated only
the additional scalar contributions at one- and two-loop (sample
Feynman diagrams showing these contributions are given in Fig.
\ref{fig:2loop_gauge}). For all other (scalar and Yukawa) beta
functions, we performed the two-loop calculation including also the
full set of SM particles. We verified explicitly that, in our
calculation, only traces with an even number of $\gamma_5$ matrices in
closed fermion loops appeared. According to common
lore~\cite{Trueman:1995ca}, we evaluated these traces using naive
anticommuting $\gamma_5$. We find the following two-loop results:
\begin{align}
  \begin{split}
  \beta_{g_1}^{(2)} &= 
    \quad \frac{g_1^5}{(16\pi^2)^2}
        \bigg(   \frac{Y_\varphi^4}{4} \mathcal{D}(\jp)
               + \frac{95}{27} n_g
               + \frac{1}{2} \bigg)\\
    & \quad + \frac{g_1^3 g_2^2}{(16\pi^2)^2}
        \bigg(   Y_\varphi^2 \mathcal{D}(\jp) \mathcal{J}(\jp)
               + n_g
               + \frac{3}{2} \bigg)
            + \frac{g_1^3 g_s^2}{(16\pi^2)^2} \frac{44}{9} n_g\\
    & \quad - \frac{g_1^3}{(16\pi^2)^2}
        \bigg(   \frac{17}{6} \big( y_t^2 + y_c^2 \big)
               + \frac{5}{6} y_b^2
               + \frac{5}{2} y_\tau^2 \bigg)\,,
  \end{split}\\
  \begin{split}
  \beta_{g_2}^{(2)} &= 
    \quad \frac{g_2^5}{(16\pi^2)^2}
        \bigg(   \frac{4}{3} \mathcal{D}(\jp) \mathcal{J}(\jp)^2
               + \frac{4}{9} \mathcal{D}(\jp) \mathcal{J}(\jp)
               - \frac{136}{3} 
               + \frac{49}{3} n_g 
               + \frac{13}{6} \bigg)\\
    & \quad + \frac{g_1^2 g_2^3}{(16\pi^2)^2}
        \bigg(   \frac{Y_\varphi^2}{3} \mathcal{D}(\jp) \mathcal{J}(\jp)
               + \frac{1}{3} n_g
               + \frac{1}{2} \bigg)
            + \frac{g_s^2 g_2^3}{(16\pi^2)^2} 4 n_g\\
    & \quad - \frac{g_2^3}{(16\pi^2)^2}
        \bigg(   \frac{3}{2} \big( y_t^2 + y_c^2 + y_b^2 \big)
               + \frac{1}{2} y_\tau^2 \bigg) \,,
  \end{split}\\
  \begin{split}
  \beta_{g_s}^{(2)} &= 
    \quad \frac{g_s^5}{(16\pi^2)^2}
        \bigg( \frac{76}{3} n_g - 102 \bigg)
            + \frac{g_1^2 g_s^3}{(16\pi^2)^2} \frac{11}{18} n_g
            + \frac{g_2^2 g_s^3}{(16\pi^2)^2} \frac{3}{2} n_g\\
    & \quad - 2 \frac{g_s^3}{(16\pi^2)^2}
        \big( y_t^2 + y_c^2 + y_b^2 \big) \,,
  \end{split}\\
  \begin{split}
  \beta_{y_t}^{(2)} &= 
      \quad \frac{y_t g_1^4}{(16\pi^2)^2}
        \bigg(   \frac{1}{8}
	       + \frac{145}{81} n_g
               + \frac{5}{27} \mathcal{D}(\jp) Y_\varphi^2 \bigg)\\
    & \quad + \frac{y_t g_2^4}{(16\pi^2)^2}
        \bigg(   n_g
               + \frac{1}{3} \mathcal{J}(\jp) \mathcal{D}(\jp)
               - \frac{35}{4} \bigg)
            - \frac{3}{4} \frac{y_t g_1^2 g_2^2}{(16\pi^2)^2}\\
    & \quad + \frac{y_t g_s^4}{(16\pi^2)^2}
        \bigg(   \frac{80}{9} n_g
               - \frac{404}{3} \bigg)\\
    & \quad + \frac{y_t g_1^2}{(16\pi^2)^2}
        \bigg(   \frac{131}{16} y_t^2
               + \frac{7}{48} y_b^2
               + \frac{85}{24} y_c^2
               + \frac{25}{8} y_\tau^2 \bigg)
            + \frac{19}{9} \frac{y_t g_1^2 g_s^2}{(16\pi^2)^2}\\
    & \quad + \frac{y_t g_2^2}{(16\pi^2)^2}
        \bigg(   \frac{225}{16} y_t^2
               + \frac{99}{16} y_b^2
               + \frac{45}{8} y_c^2
               + \frac{15}{8} y_\tau^2 \bigg)
            + 9 \frac{y_t g_2^2 g_s^2}{(16\pi^2)^2}\\
    & \quad + \frac{y_t g_s^2}{(16\pi^2)^2}
        \bigg(   36 y_t^2
               + 4 y_b^2
               + 20 y_c^2 \bigg)
            - \frac{y_t^3}{(16\pi^2)^2}
        \bigg(   12 y_t^2
               + \frac{11}{4} y_b^2
               + \frac{27}{4} y_c^2
               + \frac{9}{4} y_\tau^2
               + 3 \lambda_{H} \bigg)\\
    & \quad + \frac{y_t}{(16\pi^2)^2}
        \bigg(   \frac{1}{32} \mathcal{D}(\jp) \lambda_{\varphi H}^2
               + \frac{1}{128} \mathcal{J}(\jp) \mathcal{D}(\jp) \big(\lambda_{\varphi H}'\big)^2
               + \frac{3}{8} \lambda_{H}^2  \bigg)\\
    & \quad - \frac{y_t}{(16\pi^2)^2}
        \bigg(   \frac{1}{4} y_b^4
               + \frac{27}{4} y_c^4
               + \frac{9}{4} y_\tau^4
               - \frac{15}{4} y_b^2 y_c^2
               - \frac{5}{4} y_b^2 y_\tau^2 \bigg) \,,
  \end{split}\\
  \begin{split}
  \beta_{y_b}^{(2)} &= 
      \quad - \frac{y_b g_1^4}{(16\pi^2)^2}
        \bigg(   \frac{29}{72}
	       + \frac{5}{81} n_g
               - \frac{7}{216} \mathcal{D}(\jp) Y_\varphi^2 \bigg)\\
    & \quad + \frac{y_b g_2^4}{(16\pi^2)^2}
        \bigg(   n_g
               + \frac{1}{3} \mathcal{J}(\jp) \mathcal{D}(\jp)
               - \frac{35}{4} \bigg)
            - \frac{9}{4} \frac{y_b g_1^2 g_2^2}{(16\pi^2)^2}\\
    & \quad + \frac{y_b g_s^4}{(16\pi^2)^2}
        \bigg(   \frac{80}{9} n_g
               - \frac{404}{3} \bigg)\\
    & \quad + \frac{y_b g_1^2}{(16\pi^2)^2}
        \bigg(   \frac{91}{48} y_t^2
               + \frac{79}{16} y_b^2
               + \frac{85}{24} y_c^2
               + \frac{25}{8} y_\tau^2 \bigg)
            + \frac{31}{9} \frac{y_b g_1^2 g_s^2}{(16\pi^2)^2}\\
    & \quad + \frac{y_b g_2^2}{(16\pi^2)^2}
        \bigg(   \frac{99}{16} y_t^2
               + \frac{225}{16} y_b^2
               + \frac{45}{8} y_c^2
               + \frac{15}{8} y_\tau^2 \bigg)
            + 9 \frac{y_b g_2^2 g_s^2}{(16\pi^2)^2}\\
    & \quad + \frac{y_b g_s^2}{(16\pi^2)^2}
        \bigg(   4 y_t^2
               + 36 y_b^2
               + 20 y_c^2 \bigg)
            - \frac{y_b^3}{(16\pi^2)^2}
	  \bigg(   \frac{11}{4} y_t^2
               + 12 y_b^2
               + \frac{27}{4} y_c^2
               + \frac{9}{4} y_\tau^2
               + 3 \lambda_{H} \bigg)\\
    & \quad + \frac{y_b}{(16\pi^2)^2}
        \bigg(   \frac{1}{32} \mathcal{D}(\jp) \lambda_{\varphi H}^2
               + \frac{1}{128} \mathcal{J}(\jp) \mathcal{D}(\jp) \big(\lambda_{\varphi H}'\big)^2
               + \frac{3}{8} \lambda_{H}^2  \bigg)\\
    & \quad - \frac{y_b}{(16\pi^2)^2}
        \bigg(   \frac{1}{4} y_t^4
               + \frac{27}{4} y_c^4
               + \frac{9}{4} y_\tau^4
               - \frac{15}{4} y_t^2 y_c^2
               - \frac{5}{4} y_t^2 y_\tau^2 \bigg) \,,
  \end{split}\\
  \begin{split}
  \beta_{y_c}^{(2)} &= 
      \quad  \frac{y_c g_1^4}{(16\pi^2)^2}
        \bigg(   \frac{1}{8}
	       + \frac{145}{81} n_g
               + \frac{5}{27} \mathcal{D}(\jp) Y_\varphi^2 \bigg)\\
    & \quad + \frac{y_c g_2^4}{(16\pi^2)^2}
        \bigg(   n_g
               + \frac{1}{3} \mathcal{J}(\jp) \mathcal{D}(\jp)
               - \frac{35}{4} \bigg)
            - \frac{3}{4} \frac{y_c g_1^2 g_2^2}{(16\pi^2)^2}\\
    & \quad + \frac{y_c g_s^4}{(16\pi^2)^2}
        \bigg(   \frac{80}{9} n_g
               - \frac{404}{3} \bigg)\\
    & \quad + \frac{y_c g_1^2}{(16\pi^2)^2}
        \bigg(   \frac{85}{24} y_t^2
               + \frac{25}{24} y_b^2
               + \frac{131}{16} y_c^2
               + \frac{25}{8} y_\tau^2 \bigg)
            + \frac{19}{9} \frac{y_c g_1^2 g_s^2}{(16\pi^2)^2}\\
    & \quad + \frac{y_c g_2^2}{(16\pi^2)^2}
        \bigg(   \frac{45}{8} y_t^2
               + \frac{45}{8} y_b^2
               + \frac{225}{16} y_c^2
               + \frac{15}{8} y_\tau^2 \bigg)
            + 9 \frac{y_c g_2^2 g_s^2}{(16\pi^2)^2}\\
    & \quad + \frac{y_c g_s^2}{(16\pi^2)^2}
        \bigg(   20 y_t^2
               + 20 y_b^2
               + 36 y_c^2 \bigg)
            - \frac{y_c^3}{(16\pi^2)^2}
	  \bigg( \frac{27}{4} y_t^2
	       + \frac{27}{4} y_b^2
               + 12 y_c^2
               + \frac{9}{4} y_\tau^2
               + 3 \lambda_{H} \bigg)\\
    & \quad + \frac{y_c}{(16\pi^2)^2}
        \bigg(   \frac{1}{32} \mathcal{D}(\jp) \lambda_{\varphi H}^2
               + \frac{1}{128} \mathcal{J}(\jp) \mathcal{D}(\jp) \big(\lambda_{\varphi H}'\big)^2
               + \frac{3}{8} \lambda_{H}^2  \bigg)\\
    & \quad - \frac{y_c}{(16\pi^2)^2}
        \bigg(   \frac{27}{4} y_t^4
               + \frac{27}{4} y_b^4
               + \frac{9}{4} y_\tau^4
               - \frac{3}{2} y_t^2 y_b^2 \bigg) \,,
  \end{split}\\
  \begin{split}
  \beta_{y_\tau}^{(2)} &= 
      \quad  \frac{y_\tau g_1^4}{(16\pi^2)^2}
        \bigg(   \frac{17}{24}
	       + \frac{55}{9} n_g
               + \frac{13}{24} \mathcal{D}(\jp) Y_\varphi^2 \bigg)\\
    & \quad + \frac{y_\tau g_2^4}{(16\pi^2)^2}
        \bigg(   n_g
               + \frac{1}{3} \mathcal{J}(\jp) \mathcal{D}(\jp)
               - \frac{35}{4} \bigg)
            + \frac{9}{4} \frac{y_\tau g_1^2 g_2^2}{(16\pi^2)^2}\\
    & \quad + \frac{y_\tau g_1^2}{(16\pi^2)^2}
        \bigg(   \frac{85}{24} y_t^2
               + \frac{25}{24} y_b^2
               + \frac{85}{24} y_c^2
               + \frac{179}{16} y_\tau^2 \bigg)\\
    & \quad + \frac{y_\tau g_2^2}{(16\pi^2)^2}
        \bigg(   \frac{45}{8} y_t^2
               + \frac{45}{8} y_b^2
	       + \frac{45}{8} y_c^2
               + \frac{165}{16} y_\tau^2 \bigg)\\
    & \quad + \frac{y_\tau g_s^2}{(16\pi^2)^2}
        \bigg(   20 y_t^2
               + 20 y_b^2
               + 20 y_c^2 \bigg)
            - \frac{y_\tau^3}{(16\pi^2)^2}
	  \bigg( \frac{27}{4} y_t^2
	       + \frac{27}{4} y_b^2
	       + \frac{27}{4} y_c^2
               + 3 y_\tau^2
               + 3 \lambda_{H} \bigg)\\
    & \quad + \frac{y_\tau}{(16\pi^2)^2}
        \bigg(   \frac{1}{32} \mathcal{D}(\jp) \lambda_{\varphi H}^2
               + \frac{1}{128} \mathcal{J}(\jp) \mathcal{D}(\jp) \big(\lambda_{\varphi H}'\big)^2
               + \frac{3}{8} \lambda_{H}^2  \bigg)\\
    & \quad - \frac{y_\tau}{(16\pi^2)^2}
        \bigg(   \frac{27}{4} y_t^4
               + \frac{27}{4} y_b^4
               + \frac{27}{4} y_c^4
               - \frac{3}{2} y_t^2 y_b^2 \bigg) \,,
  \end{split}\\
  \begin{split}
  \beta_{k}^{(2)} &= 
       \frac{1}{(16\pi^2)^2}\bigg(
        g_1^6 B_{k,60} + g_1^4g_2^2 B_{k,42} + g_2^2g_2^4 B_{k,24} + g_2^6 B_{k,06} + B_{k,00}\\
	&\qquad\qquad\quad + 
        g_1^4 B_{k,40} + g_1^2g_2^2 B_{k,22} + g_2^4B_{k,04} + g_1^2 B_{k,20} + g_2^2 B_{k,02} \bigg) \,,
  \end{split}
\end{align}
where the coefficients are given, for $k = \lambda_{\varphi}^{(J)},
\lambda_{\varphi H}, \lambda_{\varphi H}', \lambda_{H}$, by
\begin{align}
  B^{(J)}_{\varphi,60} &= 
        - \frac{Y_\varphi^6}{4} \left(\frac{7}{3}\mathcal{D}(\jp) + 15 \right)
        - Y_\varphi^4 \bigg(   \frac{7}{6} + \frac{80}{9} n_g \bigg) \,,\\
  \begin{split}
  B^{(J)}_{\varphi,42} &= 
        Y_\varphi^4 \left[  \mathcal{J}(\jp) \left(\frac{7}{3} \mathcal{D}(\jp)
                                               + 15 \right)
                        - \mathcal{J}(J) \left(\frac{7}{6} \mathcal{D}(\jp)
                                               + 15 \right) \right] \\
       & \quad - Y_\varphi^2 \bigg(   \frac{7}{3} + \frac{160}{9} n_g \bigg) 
                         \Big( \mathcal{J}(J) - 2 \mathcal{J}(\jp)\Big) \,,
  \end{split}\\
  \begin{split}
  B^{(J)}_{\varphi,24} &=
        Y_\varphi^2 \bigg[  \mathcal{J}(\jp) \left(  60 \mathcal{J}(\jp)
                                        + \frac{28}{9} \mathcal{D}(\jp) \mathcal{J}(\jp)
                                        - \frac{218}{3} + \frac{64}{3} n_g \right) \\
  & \qquad \quad  - \mathcal{J}(J) \left( 15 \mathcal{J}(J)
                                        + \frac{14}{9} \mathcal{D}(\jp) \mathcal{J}(\jp)
                                        - \frac{109}{3} + \frac{32}{3} n_g \right) \bigg] \,,
  \end{split}\\
  \begin{split}
  B^{(J)}_{\varphi,06} &=       \left( \frac{\mathcal{J}(J)}{2} - \mathcal{J}(\jp) \right)
                              \left( \frac{\mathcal{J}(J)}{2} - \mathcal{J}(\jp) + \frac{1}{2} \right)\\
                     & \quad  \times \left( \frac{584}{3} - \frac{112}{9} \mathcal{J}(\jp) \mathcal{D}(\jp)
                                     - 240 \mathcal{J}(\jp) - \frac{256}{3} n_g \right)
                              + 216 \mathcal{J}(\jp)^2 \,,
  \end{split}\\
  \begin{split}
  B^{(J)}_{\varphi,20} &=
        Y_\varphi^2 \bigg( 8 {\sum_{J_1,J_2}}^\prime K(J_1, J_2, J)
                                    \lambda_\varphi^{(J_1)}\lambda_\varphi^{(J_2)} 
                                  - \big(\lambda_\varphi^{(J)}\big)^2\bigg)\\
  & \quad + \frac{1}{8} \big( \lambda_{\varphi H}'\big)^2 
                         \Big( \mathcal{J}(J) - 2 \mathcal{J}(\jp) \Big)
		       + \lambda_{\varphi H}^2\,,
  \end{split}\\
  \begin{split}
  B^{(J)}_{\varphi,02} &=
               \big(\lambda_\varphi^{(J)}\big)^2
               \bigg( 8\mathcal{J}(\jp) - 3\mathcal{J}(J) \bigg)
               - 16 \mathcal{J}(\jp) {\sum_{J_1,J_2}}^\prime K(J_1, J_2, J)
               \lambda_\varphi^{(J_1)}\lambda_\varphi^{(J_2)} 
               \\
  & \qquad \quad + 12 {\sum_{J_1,J_2}}^\prime \sum_{J_3,J_4} (-1)^{J_4-2j}
                         \lambda_\varphi^{(J_1)} \lambda_\varphi^{(J_2)}
                         \mathcal{J}(J_4)
			  K(J_1, J_2, J_3) K(J_3, J_4, J) 
                 + 3\lambda_{\varphi H}^2\,,
  \end{split}\\
  \begin{split}
  B^{(J)}_{\varphi,40} &=
        Y_\varphi^4 \left( \frac{9}{8} \lambda_\varphi^{(J)}
                        + \frac{11\mathcal{D}(\jp)}{24} \lambda_\varphi^{(J)} 
                        + 5 {\sum_{J_1}}^\prime \lambda_\varphi^{(J_1)}
                          \frac{\mathcal{D}(J_1)}{\mathcal{D}(\jp)} \right) \\
	  & \quad + Y_\varphi^2\lambda_\varphi^{(J)} 
            \bigg( \frac{11}{12} + \frac{50}{9} n_g \bigg) 
                  + \frac{5\lambda_{\varphi H}}{2}Y_\varphi^2\,,
  \end{split}\\
  \begin{split}
  B^{(J)}_{\varphi,22} &= Y_{\varphi}^2 {\sum_{J_1}}^\prime \lambda_\varphi^{(J_1)}
                        \frac{\mathcal{D}(J_1)}{\mathcal{D}(\jp)}
        \bigg[ 10 \frac{\mathcal{J}(J_1)\mathcal{J}(J)}
                                {\mathcal{J}(\jp)}
               -20 \big( \mathcal{J}(J_1) + \mathcal{J}(J) \big)
               +40 \mathcal{J}(\jp) \bigg]  \\
                     & \quad + Y_{\varphi}^2 \lambda_\varphi^{(J)}
		       \bigg( 2 \mathcal{J}(J) + \mathcal{J}(\jp)\bigg)
                             + \frac{5}{2} Y_{\varphi} \lambda_{\varphi H}'
                       \big( \mathcal{J}(J) - 2 \mathcal{J}(\jp) \big) \,,
  \end{split}\\
  \begin{split}
  B^{(J)}_{\varphi,04} &= \lambda_\varphi^{(J)}
        \bigg[     \mathcal{J}(\jp) \left( 18 \mathcal{J}(\jp) 
                                       + \frac{22}{9} \mathcal{D}(\jp) \mathcal{J}(\jp)
                                       - \frac{275}{3}
                                       + \frac{40}{3} n_g \right) \\
  & \hspace{4em} + \mathcal{J}(J) \left( \mathcal{J}(J)
				          - 4 \mathcal{J}(\jp) + 2 \right) \bigg]
                 + 10 \lambda_{\varphi H} \mathcal{J}(\jp) \\
  & \quad + 4 {\sum_{J_1,J_2}}^\prime \lambda_\varphi^{(J_2)} K(J_1, J_2, J) 
                \left( \mathcal{J}(J_1)^2 
                       - 4 \mathcal{J}(J_1) \mathcal{J}(\jp) \right) \\
  & \quad + \sum_{J_1} {\sum_{J_2}}^\prime \lambda_\varphi^{(J_2)} K(J_1, J_2, J) 
                \left( 18 \mathcal{J}(J_1)^2 
                       - 72 \mathcal{J}(J_1) \mathcal{J}(\jp) \right) \\
  & \quad + {\sum_{J_1}}^\prime \frac{\lambda_\varphi^{(J_1)}}{\mathcal{D}(\jp)} \bigg[
                80 \mathcal{D}(J_1)\mathcal{J}(\jp)^2
              + 40 \mathcal{D}(J_1)\mathcal{J}(\jp) \\
  & \hspace{6.5em}
              - 20  \mathcal{D}(J_1)\mathcal{J}(J_1)
              - 20  \mathcal{D}(J_1)\mathcal{J}(J)
              + 10 \frac{\mathcal{J}(J)\mathcal{J}(J_1)\mathcal{D}(J_1)}
                                 {\mathcal{J}(\jp)}
            \bigg] \,,
  \end{split}\\
  \begin{split}
  B^{(J)}_{\varphi,00} &= 
        \lambda_\varphi^{(J)} {\sum_{J_1}}^\prime \big(\lambda_\varphi^{(J_1)}\big)^2
        \frac{\mathcal{D}(J_1)}{\mathcal{D}(\jp)}
        - 4{\sum_{J_1,J_2}}^\prime K(J_1, J_2, J)
        \lambda_\varphi^{(J_1)} \lambda_\varphi^{(J_2)}
        \big( \lambda_\varphi^{(J_1)} + \lambda_\varphi^{(J)} \big)  \\
  & \quad - 8 {\sum_{J_1,J_2,J_3}}^\prime \sum_{J_4} (-1)^{J_4-2j}
            K(J_1, J_2, J_4) K(J_4, J_3, J)
            \lambda_\varphi^{(J_1)} \lambda_\varphi^{(J_2)} \lambda_\varphi^{(J_3)} \\
  & \quad - \frac{\lambda_\varphi^{(J)} \big(\lambda_{\varphi H}' \big)^2}{16}
            \big( 2 \mathcal{J}(J) - 3 \mathcal{J}(\jp) \big)
          + \frac{\lambda_{\varphi H}\big(\lambda_{\varphi H}'\big)^2}{8}
	  \left(\mathcal{J}(\jp) - \mathcal{J}(J)\right)\\
  & \quad - \frac{\big(\lambda_{\varphi H}' \big)^2}{8}
            \big( 3 y_t^2 + 3 y_t^2 + 3 y_t^2 + y_\tau^2 \big)
            \big( \mathcal{J}(J) - 2 \mathcal{J}(\jp)\big)\\
  & \quad - \lambda_{\varphi H}^2\left(
	  3y_t^2+3y_b^2+3y_c^2+y_\tau^2+\frac{5}{4}\lambda_{\varphi}^{(J)}\right)
	  -\frac{\lambda_{\varphi H}^3}{2}\,.
  \end{split}
\end{align}
For the Higgs-portal couplings we find
\begin{align}
  B_{\varphi H,60} &= 
        - Y_\varphi^4 \bigg(\frac{7}{6}\mathcal{D}(\jp) + \frac{15}{4} \bigg)
        - Y_\varphi^2 \bigg(   \frac{73}{12} + \frac{160}{9} n_g \bigg) \,,\\
  B_{\varphi H,42} &= 
        - Y_\varphi^2 \bigg(15 \mathcal{J}(\jp) + \frac{45}{4} \bigg) \,,\\
  B_{\varphi H,24} &= 
        - 15 \mathcal{J}(\jp) \big( 1 + Y_\varphi^2 \big) \,,\\
  B_{\varphi H,06} &= 
          \mathcal{J}(\jp) \bigg(   \frac{1129}{3} - \frac{128}{3} n_g
                                - 60 \mathcal{J}(\jp)
                                - \frac{56}{9} \mathcal{J}(\jp) \mathcal{D}(\jp) \bigg) \,,\\
  \begin{split}
  B_{\varphi H,40} &= 
          Y_\varphi^2 \bigg(   \frac{15}{2} \lambda_H
                           - \big( 19 y_t^2 - 5 y_b^2 + 19 y_c^2 + 25 y_\tau^2 \big)
                           + 5 {\sum_{J_1}}^\prime \lambda_\varphi^{(J_1)}
                             \frac{\mathcal{D}(J_1)}{\mathcal{D}(\jp)} \bigg)\\
  & \quad + Y_\varphi^2 \lambda_{\varphi H} 
            \bigg( \frac{23}{24} + \frac{11}{48} \mathcal{D}(\jp) \bigg)
          + Y_\varphi^4 \lambda_{\varphi H}
            \bigg( \frac{5}{16} + \frac{71}{48} \mathcal{D}(\jp) \bigg) \\
  & \quad + \lambda_{\varphi H} \bigg[
                   \frac{157}{48}
                 + \frac{25}{9} n_g \big( 1 + Y_\varphi^2 \big) \bigg] \,,
  \end{split}\\
  B_{\varphi H,22} &= 
          Y_\varphi \mathcal{J}(\jp) \lambda_{\varphi H}' 
	  + \lambda_{\varphi H}\left(\frac{15}{8}+\frac{5}{2}\mathcal{J}(\jp)Y_\varphi^2\right)\,,\\
  \begin{split}
  B_{\varphi H,04} &= 
          \mathcal{J}(\jp) \bigg(   30 \lambda_H
                                - 4 \big( 3 y_t^2 + 3 y_b^2 + 3 y_c^2 + y_\tau^2 \big)
				+ 20 {\sum_{J_1}}^\prime \lambda_{\varphi}^{(J_1)}
				    \frac{\mathcal{D}(J_1)}{\mathcal{D}(\jp)} \bigg)\\
  & \quad - \lambda_{\varphi H}\bigg[\frac{385}{16} - 5 n_g
	  + \mathcal{J}(\jp)\left(\frac{263}{6}
	  - \frac{20}{3} n_g - \frac{11}{12}\mathcal{D}(\jp)\right)
          - \mathcal{J}(\jp)^2\bigg(5 + \frac{71}{9}\mathcal{D}(\jp)\bigg)\bigg]\,,
  \end{split}\\
  \begin{split}
  B_{\varphi H,20} &= 
            \frac{\big( \lambda_{\varphi H}' \big)^2}{16}
            \mathcal{J}(\jp) \big( 1 + Y_\varphi^2 \big)
          + \frac{\lambda_{\varphi H}^2}{4} \big( 1 + Y_\varphi^2 \big)
          + \frac{\lambda_{\varphi H}}{12} 
            \big( 85 y_t^2 + 25 y_b^2 + 85 y_c^2 + 75 y_\tau^2 \big)\\
  & \quad + 6 \lambda_{\varphi H} \lambda_{H}
          + 4 Y_\varphi^2 \lambda_{\varphi H} 
            {\sum_{J_1}}^\prime  \lambda_\varphi^{(J_1)}
            \frac{\mathcal{D}(J_1)}{\mathcal{D}(\jp)} \,,
  \end{split}\\
  \begin{split}
  B_{\varphi H,02} &= 
            \frac{\big( \lambda_{\varphi H}' \big)^2}{16}
            \mathcal{J}(\jp) \big( 15 + 4 \mathcal{J}(\jp) \big)
          + \frac{\lambda_{\varphi H}^2}{4} \big( 3 + 4 \mathcal{J}(\jp) \big)\\
  & \quad + \frac{15\lambda_{\varphi H}}{4} 
            \big( 3 y_t^2 + 3 y_b^2 + 3 y_c^2 + y_\tau^2 \big)\\
  & \quad + 18 \lambda_{\varphi H} \lambda_{H}
          + 16 \lambda_{\varphi H} 
            {\sum_{J_1}}^\prime  \lambda_\varphi^{(J_1)}
            \frac{\mathcal{D}(J_1)\mathcal{J}(\jp)}{\mathcal{D}(\jp)} \,,
  \end{split}\\
  \begin{split}
  B_{\varphi H,00} &= 
          - \bigg(   \frac{\big(\lambda_{\varphi H}'\big)^2}{4} \mathcal{J}(\jp) 
                   + 6 \lambda_{\varphi H} \lambda_{H}
                   + \lambda_{\varphi H}^2
            \bigg) \big( 3 y_t^2 + 3 y_b^2 + 3 y_c^2 + y_\tau^2 \big)\\
  & \quad + 40 \lambda_{\varphi H} g_s^2 \big( y_t^2 + y_b^2 + y_c^2 \big)
          - \frac{\lambda_{\varphi H}}{2}
            \big( 27 y_t^4 + 27 y_b^4 + 27 y_c^4 + 9 y_\tau^4 + 42 y_t^2 y_b^2 \big)\\
  & \quad - \frac{5\mathcal{J}(\jp)}{8} \big(\lambda_{\varphi H}'\big)^2 \lambda_{H}
          - \frac{15}{4} \lambda_{\varphi H} \lambda_{H}^2
          - \lambda_{\varphi H} \big(\lambda_{\varphi H}'\big)^2
            \bigg( \frac{13}{32} + \frac{\mathcal{D}(\jp)}{64} \bigg) \mathcal{J}(\jp)\\
  & \quad - \frac{9}{2} \lambda_{\varphi H}^2 \lambda_{H}
          - \lambda_{\varphi H}^3
            \bigg( \frac{5}{8} + \frac{\mathcal{D}(\jp)}{16} \bigg) \\
  & \quad - \frac{5\lambda_{\varphi H}}{2}
                    {\sum_{J_1}}^\prime \big(\lambda_\varphi^{(J_1)}\big)^2
                                     \frac{\mathcal{D}(J_1)}{\mathcal{D}(\jp)}
          - \frac{\big(\lambda_{\varphi H}'\big)^2}{4}
                    {\sum_{J_1}}^\prime \lambda_\varphi^{(J_1)}
                    \frac{\mathcal{D}(J_1)\big(\mathcal{J}(J_1)-\mathcal{J}(\jp)\big)}
                         {\mathcal{D}(\jp)}\\
  & \quad - 3 \lambda_{\varphi H}^2
            {\sum_{J_1}}^\prime \lambda_\varphi^{(J_1)}
            \frac{\mathcal{D}(J_1)}{\mathcal{D}(\jp)}\,;
  \end{split}
\end{align}
\begin{align}
  B^{\prime}_{\varphi H,60} &= 0 \,,\\
  B^{\prime}_{\varphi H,42} &=
        - Y_\varphi^3 \bigg(\frac{14}{3} \mathcal{D}(\jp) + 30 \bigg)
        - Y_\varphi   \bigg(   \frac{118}{3} + \frac{640}{9} n_g \bigg) \,,\\
  B^{\prime}_{\varphi H,24} &=
          Y_\varphi   \bigg(   \frac{346}{3} - \frac{128}{3} n_g
                           - 120 \mathcal{J}(\jp)
                           - \frac{56}{9} \mathcal{D}(\jp) \mathcal{J}(\jp) \bigg) \,,\\
  B^{\prime}_{\varphi H,06} &= 0 \,,\\
  B^{\prime}_{\varphi H,40} &=
            \frac{11}{48} Y_\varphi^2 \big( 1 + Y_\varphi^2 \big) \lambda_{\varphi H}' \mathcal{D}(\jp) 
	  + \frac{5}{16} Y_\varphi^4\lambda_{\varphi H}'
          + \lambda_{\varphi H}' \bigg[ 
                   \frac{37}{48} + \frac{23Y_\varphi^2}{24}
                 + \frac{25}{9} n_g \big( 1 + Y_\varphi^2 \big) \bigg] \,,\\
  \begin{split}
  B^{\prime}_{\varphi H,22} &=
            \lambda_{\varphi H}'
            \bigg[   \frac{47}{8}
                   + Y_\varphi^2 \bigg(   \frac{10}{3} \mathcal{J}(\jp) \mathcal{D}(\jp) 
				      + \frac{5}{2}\mathcal{J}(\jp) - 1 \bigg) \bigg] 
		   + 4 Y_\varphi \left(\lambda_{\varphi H} + 5\lambda_H\right)\\
  & \quad + 4 Y_\varphi \big( 42 y_t^2 + 18 y_b^2 + 42 y_c^2 + 22 y_\tau^2 \big) 
          + 20 Y_\varphi {\sum_{J_1}}^\prime \lambda_\varphi^{(J_1)} 
                      \frac{\big(\mathcal{J}(J_1)-2\mathcal{J}(\jp)\big)\mathcal{D}(J_1)}
                           {\mathcal{J}(\jp)\mathcal{D}(\jp)} \,,
  \end{split}\\
  \begin{split}
  B^{\prime}_{\varphi H,04} &= \lambda_{\varphi H}' \bigg[
            5 n_g - \frac{457}{16}
          + \mathcal{J}(\jp) 
             \bigg(   \frac{20}{3} n_g
                    - \frac{287}{6} + \frac{11}{12} \mathcal{D}(\jp)
                    + \mathcal{J}(\jp)
                      \bigg( 5 + \frac{11}{9} \mathcal{D}(\jp) \bigg) \bigg) \bigg]\,,
  \end{split}\\
  \begin{split}
  B^{\prime}_{\varphi H,20} &= 
            \frac{3 Y_\varphi}{4} \big( \lambda_{\varphi H}' \big)^2
          + \frac{1}{2} \lambda_{\varphi H} \lambda_{\varphi H}' 
            \big( 1 + Y_\varphi^2 \big)
          + \frac{\lambda_{\varphi H}'}{12} 
            \big( 85 y_t^2 + 25 y_b^2 + 85 y_c^2 + 75 y_\tau^2 \big)\\
  & \quad + 2 \lambda_{\varphi H}' \lambda_{H}
          + 2 Y_\varphi^2 \lambda_{\varphi H}'
            {\sum_{J_1}}^\prime  \lambda_\varphi^{(J_1)}
            \frac{\big(\mathcal{J}(J_1) - 2 \mathcal{J}(\jp)\big)
                  \mathcal{D}(J_1)}{\mathcal{J}(\jp)\mathcal{D}(\jp)} \,,
  \end{split}\\
  \begin{split}
  B^{\prime}_{\varphi H,02} &= 
            \frac{15\lambda_{\varphi H}'}{4}
            \big( 3 y_t^2 + 3 y_b^2 + 3 y_c^2 + y_\tau^2 \big)
          + \lambda_{\varphi H}' \lambda_{\varphi H} 
            \bigg( \frac{15}{2} + 2 \mathcal{J}(\jp) \bigg)\\
  & \quad + \bigg( 8 \mathcal{J}(\jp) - 6 \bigg) \lambda_{\varphi H}'
            {\sum_{J_1}}^\prime  \lambda_\varphi^{(J_1)}
            \frac{\big(\mathcal{J}(J_1) - 2 \mathcal{J}(\jp)\big)
                  \mathcal{D}(J_1)}{\mathcal{J}(\jp)\mathcal{D}(\jp)} \,,
  \end{split}\\
  \begin{split}
  B^{\prime}_{\varphi H,00} &= 
          - \big(   \lambda_{\varphi H}' \lambda_{H}
                  + \lambda_{\varphi H} \lambda_{\varphi H}'
            \big)
            \big( 6 y_t^2 + 6 y_b^2 + 6 y_c^2 + 2 y_\tau^2 \big)\\
  & \quad + 40 \lambda_{\varphi H}' g_s^2 \big( y_t^2 + y_b^2 + y_c^2 \big)
          - \frac{\lambda_{\varphi H}'}{2}
            \big( 27 y_t^4 + 27 y_b^4 + 27 y_c^4 + 9 y_\tau^4 - 54 y_t^2 y_b^2 \big)\\
  & \quad - 5 \lambda_{\varphi H} \lambda_{\varphi H}' \lambda_{H}
          - \frac{7}{4} \lambda_{\varphi H}' \lambda_{H}^2
          - \lambda_{\varphi H}' \lambda_{\varphi H}^2
            \bigg( \frac{13}{8} + \frac{\mathcal{D}(\jp)}{16} \bigg)\\
  & \quad + \big( \lambda_{\varphi H}' \big)^3
            \bigg(   \frac{5\mathcal{J}(\jp)\mathcal{D}(\jp)}{192}
                   - \frac{5\mathcal{J}(\jp)}{32}
                   + \frac{3}{16} \bigg)\\
  & \quad - \frac{\lambda_{\varphi H}'}{2}
            {\sum_{J_1}}^\prime \big(\lambda_\varphi^{(J_1)}\big)^2
            \bigg(  2\frac{\mathcal{J}(J_1)\mathcal{D}(J_1)}
                          {\mathcal{J}(\jp)\mathcal{D}(\jp)}
                   -3\frac{\mathcal{D}(J_1)}
                          {\mathcal{D}(\jp)}
            \bigg)\\
  & \quad - 2 \lambda_{\varphi H} \lambda_{\varphi H}'
            {\sum_{J_1}}^\prime \lambda_\varphi^{(J_1)}
            \bigg(   \frac{\mathcal{J}(J_1)\mathcal{D}(J_1)}
                          {\mathcal{J}(\jp)\mathcal{D}(\jp)}
                   - \frac{\mathcal{D}(J_1)}
                          {\mathcal{D}(\jp)}
            \bigg)\,.
  \end{split}
\end{align}
Our results for the quartic Higgs self coupling are
\begin{align}
  B_{H,60} &= 
        - \frac{7}{12} Y_\varphi^2 \mathcal{D}(\jp)
        - \frac{59}{12}
        - \frac{80}{9} n_g \,,\\
  B_{H,42} &= 
        - \frac{7}{12} Y_\varphi^2 \mathcal{D}(\jp)
        - \frac{239}{12} - \frac{80}{9} n_g \,,\\
  B_{H,24} &= 
        - \frac{7}{9} \mathcal{J}(\jp) \mathcal{D}(\jp)
        - \frac{97}{12} - \frac{16}{3} n_g \,,\\
  B_{H,06} &= 
        - \frac{7}{3} \mathcal{J}(\jp) \mathcal{D}(\jp)
        + \frac{497}{4} - 16 n_g \,,\\
  \begin{split}
  B_{H,40} &= 
            \frac{1}{24} \lambda_H 
            \big( 11 Y_\varphi^2 \mathcal{D}(\jp) + 229 \big)
          + \frac{5}{4} \lambda_{\varphi H} Y_\varphi^2 \mathcal{D}(\jp)\\
  & \quad + \frac{50}{9} n_g \lambda_H
          - \big( 19 y_t^2 - 5 y_b^2 + 19 y_c^2 + 25 y_\tau^2 \big)\,,
  \end{split}\\
  B_{H,22} &= 
            \frac{39}{4} \lambda_{H}
          + \frac{5}{6} \lambda_{\varphi H}' Y_\varphi \mathcal{J}(\jp) \mathcal{D}(\jp)
	  + 2 \big( 21 y_t^2 + 9 y_b^2 + 21 y_c^2 + 11 y_\tau^2 \big)\,,\\
  \begin{split}
  B_{H,04} &= 
            5 \lambda_{\varphi H} \mathcal{J}(\jp) \mathcal{D}(\jp)
          - 3 \big( 3 y_t^2 + 3 y_b^2 + 3 y_c^2 + y_\tau^2 \big)\\
  & \quad + \lambda_H \bigg(   10 n_g
                             - \frac{313}{8} 
                             + \frac{11}{6} \mathcal{J}(\jp) \mathcal{D}(\jp)\bigg)\,,
  \end{split}\\
  \begin{split}
  B_{H,20} &= 
            9 \lambda_{H}^2
          + \frac{1}{2} \lambda_{\varphi H}^2 Y_\varphi^2 \mathcal{D}(\jp)
          + \frac{1}{24} \big(\lambda_{\varphi H}'\big)^2 Y_\varphi^2 \mathcal{J}(\jp) \mathcal{D}(\jp)\\
  & \quad + \frac{\lambda_{H}}{6} \big( 85 y_t^2 + 25 y_b^2 + 85 y_c^2 + 75 y_\tau^2 \big)
          - \frac{4}{3} \big( 8 y_t^4 - 4 y_b^4 + 8 y_c^4 + 12 y_\tau^4 \big)\,,
  \end{split}\\
  \begin{split}
  B_{H,02} &= 
            27 \lambda_{H}^2
          + 2 \lambda_{\varphi H}^2 \mathcal{J}(\jp) \mathcal{D}(\jp)
          + \big(\lambda_{\varphi H}'\big)^2 \mathcal{J}(\jp) \mathcal{D}(\jp)
            \bigg( \frac{\mathcal{J}(\jp)}{6} - \frac{1}{8} \bigg) \\
  & \quad + \frac{15}{2} \lambda_{H} \big( 3 y_t^2 + 3 y_b^2 + 3 y_c^2 +  y_\tau^2 \big)\,,
  \end{split}\\
  \begin{split}
  B_{H,00} &= 
          - \frac{39}{2} \lambda_{H}^3
          - \frac{\mathcal{D}(\jp)}{4} \lambda_{\varphi H}^3
          - \frac{5}{8} \lambda_{\varphi H}^2 \lambda_{H} \mathcal{D}(\jp)
          - \frac{7}{96} \big(\lambda_{\varphi H}'\big)^2 \lambda_{H} 
            \mathcal{J}(\jp) \mathcal{D}(\jp)\\
  & \quad - \frac{5}{48} \big(\lambda_{\varphi H}'\big)^2 \lambda_{\varphi H} 
            \mathcal{J}(\jp) \mathcal{D}(\jp)\\
  & \quad - 12 \lambda_{H}^2 \big( 3 y_t^2 + 3 y_b^2 + 3 y_c^2 + y_\tau^2 \big)
          + 80 \lambda_{H} g_s^2 \big( y_t^2 + y_b^2 + y_c^2 \big)\\
  & \quad - \lambda_{H}
            \big( 3 y_t^4 + 3 y_b^4 + 3 y_c^4 + y_\tau^4 + 42 y_t^2 y_b^2 \big)\\
  & \quad + 120 \big( y_t^6 + y_b^6 + y_c^6 \big) + 40 y_\tau^6 
          - 24 \big( y_t^2 y_b^4 + y_t^4 y_b^2 \big) 
          - 128 g_s^2 \big( y_t^4 + y_b^4 + y_c^4 \big)\,;
  \end{split}
\end{align}
Our pure SM results agree with those in Refs.~\cite{Arason:1991ic,
  Machacek:1984zw}, apart from three terms which are consistent with
the corrections made in Ref.~\cite{Luo:2003}. All other analytic
results are presented here in closed form for the first
time.\footnote{As a cross check, we compared all our one- and two-loop
  beta functions with the results obtained using the code
  \texttt{PyR@TE3}~\cite{Sartore:2020gou} for the two cases $\jp=1$,
  $Y_\varphi=0$ and $\jp=1$, $Y_\varphi=1$, and find complete
  agreement.}

For the two-loop calculation, we use the same strategy to express all
$SU(2)$ generators in terms of Sigma matrices and to simplify the
expressions using the relations given in Sec.~\ref{grp:thy:rels}. It
is again possible to express all results in terms of the operators in
the scalar potential~\eqref{eq:pot:def}, as required by gauge invariance.

At two-loops, all beta functions receive contributions from scalar
fields. In the Yukawa beta functions, the only additional terms from
scalar fields arise from the external Higgs field counterterm (Fig.
\ref{fig:2loop_Higgs}). The Feynman diagrams required to extract the
quartic scalar coupling beta functions again split into different
classes: those including zero, one, two, or three internal gauge
bosons. In Fig.  \ref{fig:QSC_2loop}, we give sample diagrams from
each class which give contributions to the quartic scalar coupling
beta functions.

\section{Group Theory Relations}\label{grp:thy:rels}

To express all results in terms of matrix elements of our basis
operators, and to check the gauge-parameter independence and locality
of our two-loop counterterms explicitly, we had to use a number of
algebraic relations. These relations arise from the gauge invariance
of the underlying theory as well as the properties of the
Clebsch-Gordan coefficients, and are collected and proven below. For
clarity, the summation convention is {\em suspended} in this
section. All summations are indicated explicitly.

To begin, we collect some orthogonality properties of the Sigma
matrices that follow directly from the corresponding standard
properties of the Clebsch-Gordan coefficients:
\begin{align}
 \sum_{km} \Sigma^{(J),M}_{km} \Sigma^{(J'),M'}_{km} 
  &=\delta^{JJ'}\delta^{MM'} \,, \label{eq:ortho:km}\\
 \sum_{Mm} \Sigma^{(J),M}_{km} \Sigma^{(J),M}_{k'm} 
  &=\frac{2J+1}{2\jp+1}\delta_{kk'} \,, \label{eq:ortho:Mm}\\
 \sum_{JM} \Sigma^{(J),M}_{km} \Sigma^{(J),M}_{k'm'} 
  &=\delta_{kk'}\delta_{mm'} \,. \label{eq:ortho:JM}
\end{align}

The exchange of $W$ gauge bosons introduces explicit $SU(2)$
generators that need to be rewritten in terms of Sigma matrices. Since
the Clebsch-Gordan coefficients describe a transformation between two
complete sets of orthonormal state vectors, they are used to rewrite
the product of two $SU(2)$ generators:
\begin{equation}\begin{split}\label{eq:Ceqn}
\sum_a \tilde{\tau}^a_{ir}\tilde{\tau}^a_{kl}
 =\sum_{JM} C(J) \Sigma^{(J),M}_{ik}\Sigma^{(J),M}_{rl}\,,
\end{split}\end{equation}
where the coefficient $C(J)$ is a function of $J$. The
relation~\eqref{eq:Ceqn} is convenient since it can be applied
recursively. Consider, for instance, the product of four generators:
\begin{equation}
\sum_{ab} \sum_{lk} 
 \tilde{\tau}^a_{il}\tilde{\tau}^b_{lr}
 \tilde{\tau}^a_{mk}\tilde{\tau}^b_{kn}
= \sum_{JM}\sum_{J'M'} \sum_{lk} C(J)C(J') 
 \Sigma^{(J),M}_{im} \Sigma^{(J),M}_{lk} 
 \Sigma^{(J'),M'}_{lk} \Sigma^{(J'),M'}_{rn} \,.
\end{equation}
After applying the orthogonality relations, this becomes a linear
combination of the basis operators,
\begin{equation}
\sum_{lk}
\tilde{\tau}^a_{il}\tilde{\tau}^b_{lr}\tilde{\tau}^a_{mk}\tilde{\tau}^b_{kn}
= \sum_{JM} C(J)^2\,\Sigma^{(J),M}_{im}\Sigma^{(J),M}_{rn} \,.
\end{equation}
In fact, for a product of $2n$ generators Eq.~\eqref{eq:Ceqn} implies
\begin{equation}
\sum_{a_1 a_2 \ldots a_n} 
\left(\tilde{\tau}^{a_1}\tilde{\tau}^{a_2}\dots\tilde{\tau}^{a_n}\right)_{ir}
\left(\tilde{\tau}^{a_1}\tilde{\tau}^{a_2}\dots\tilde{\tau}^{a_n}\right)_{mq}
=\sum_{JM} C(J)^n\,\Sigma^{(J),M}_{im}\Sigma^{(J),M}_{rq} \,.
\end{equation}

Diagrams with multiple scalar couplings likewise need to be expressed
in terms of the basis operators. This is facilitated by the following
``sum rule'' for Sigma matrices:
\begin{equation}\begin{split}\label{eq:K1eqn}
 \sum_{M_1M_2} \sum_{mn} \Sigma^{(J_1),M_1}_{im}\Sigma^{(J_1),M_1}_{rn} 
                         \Sigma^{(J_2),M_2}_{kn}\Sigma^{(J_2),M_2}_{lm}
 =\sum_{J_3,M_3} K(J_1,J_2,J_3) \Sigma^{(J_3),M_3}_{ik}\Sigma^{(J_3),M_3}_{rl} \,.
\end{split}\end{equation}

In the following, we give explicit expressions for $C(J)$ and
$K(J_1,J_2,J_3)$. We then derive further relations between these
quantities that can be used to simplify the results of our calculation. 
Our general strategy is to express all results in terms of our 
operator basis and the group theory invariants $\mathcal{J}(\jp) \equiv 
\jp(\jp+1)$, the eigenvalue of the $SU(2)$ Casimir operator,
\begin{equation}
 \sum_l \tilde{\tau}^a_{il}\tilde{\tau}^a_{lk}=\mathcal{J}(\jp)\delta_{ik} \,,
\end{equation}
and $\mathcal{D}(\jp) \equiv 2\jp+1$, the dimension of the $SU(2)$
multiplet representation with isospin $\jp$.

We begin by showing
\begin{equation}\label{eq:K}
        \boxed{K(J_1,J_2,J_3) = \mathcal{D}(J_1)\mathcal{D}(J_2)\left\{
        \begin{array}{ccc}
        J_1 & \jp & \jp \\
        \jp & J_2 & \jp \\
        \jp & \jp & J_3
        \end{array}
        \right\}}
\end{equation}
in terms of the Wigner $9j$ symbol~\cite{Messiah:1962}. Starting with
Eq.~\eqref{eq:K1eqn}, we multiply both sides by
$\Sigma^{(J),M}_{rl}$ and sum over $r,l$, to obtain
\begin{equation}\label{eq:K1}
 \sum_{M_1,M_2} \sum_{mnrl} \Sigma^{(J_1),M_1}_{im}\Sigma^{(J_1),M_1}_{rn} 
                     \Sigma^{(J_2),M_2}_{kn}\Sigma^{(J_2),M_2}_{lm} \Sigma^{(J),M}_{rl}
  = K(J_1,J_2,J)\Sigma^{(J),M}_{ik} \,.
\end{equation}
The Sigma matrices can be written in terms of the Wigner $3j$ symbols
as~\cite{Messiah:1962}
\begin{equation}
 \Sigma^{(J),M}_{mm'} = (-1)^{M}\sqrt{\mathcal{D}(J)}\left(
 \begin{array}{ccc}
  \jp & \jp & J \\
  m & m' & -M
 \end{array}
 \right) \,.
\end{equation}
In this way, Eq.~\eqref{eq:K1} becomes
\begin{equation}\begin{split}
 K(J_1,J_2,J)&\left(
 \begin{array}{ccc}
  \jp & \jp & J \\
  i & k & -M
 \end{array}
 \right) = \sum_{M_1,M_2} \sum_{mnrl} (-1)^{-2M_1-2M_2}\mathcal{D}(J_1)\mathcal{D}(J_2)\\
 &\times\left(
 \begin{array}{ccc}
  \jp & \jp & J_1 \\
  i & m & -M_1
 \end{array}
 \right)\left(
 \begin{array}{ccc}
  \jp & \jp & J_1 \\
  r & n & -M_1
 \end{array}
 \right)\\
 &\times\left(
 \begin{array}{ccc}
  \jp & \jp & J_2 \\
  k & n & -M_2
 \end{array}
 \right)\left(
 \begin{array}{ccc}
  \jp & \jp & J_2 \\
  l & m & -M_2
 \end{array}
 \right)\left(
 \begin{array}{ccc}
  \jp & \jp & J \\
  r & l & -M
 \end{array}
 \right)\,.
\end{split}\end{equation}
Since $M_1, M_2 \in \mathbb{Z}$, the factor of $-1$ disappears. We can
also freely change $-M_1, -M_2\to M_1, M_2$ since these indices are
summed over. We also take $M \to -M$ on both sides. Now, we use
the symmetry properties of the $3j$ symbols
\begin{equation}
 \left(
 \begin{array}{ccc}
  j_1 & j_2 & j_3 \\
  m_1 & m_2 & m_3
 \end{array}
 \right) = \left(
 \begin{array}{ccc}
   j_2 & j_3 & j_1 \\
   m_2 & m_3 & m_1
 \end{array}
 \right) = (-1)^{j_1+j_2+j_3}\left(
 \begin{array}{ccc}
  j_1 & j_3 & j_2 \\
  m_1 & m_3 & m_2
 \end{array}
 \right)
\end{equation}
to rewrite
\begin{equation}\begin{split}
 K(J_1,J_2,J)&\left(
 \begin{array}{ccc}
  \jp & \jp & J \\
  i & k & M
 \end{array}
 \right) = \sum_{M_1,M_2} \sum_{mnrl} (-1)^{2J_1+2J_2+8\jp}\mathcal{D}(J_1)\mathcal{D}(J_2)\\
 &\times\left(
 \begin{array}{ccc}
  J_1 & \jp & \jp \\
  M_1 & n & r
 \end{array}
 \right)\left(
 \begin{array}{ccc}
  \jp & J_2 & \jp \\
  m & M_2 & l
 \end{array}
 \right)\\
 &\times\left(
 \begin{array}{ccc}
  J_1 & \jp & \jp \\
  M_1 & m & i
 \end{array}
 \right)\left(
 \begin{array}{ccc}
  \jp & J_2 & \jp \\
  n & M_2 & k
 \end{array}
 \right)\left(
 \begin{array}{ccc}
  \jp & \jp & J \\
  r & l & M
 \end{array}
 \right)\,.
\end{split}\end{equation}
The Wigner $9j$ symbols are written in terms of the $3j$ symbols as
\cite{Messiah:1962}
\begin{equation}\begin{split}
 &\left(
 \begin{array}{ccc}
  J_{13} & J_{24} & J \\
  M_{13} & M_{24} & M
 \end{array}
 \right)\left\{
 \begin{array}{ccc}
  j_1 & j_2 & J_{12} \\
  j_3 & j_4 & J_{34} \\
  J_{13} & J_{24} & J
 \end{array}
 \right\} \\ & = \sum_{\substack{m_1\,m_2\,m_3\,m_4 \\ M_{12}\,M_{34}}}\left(
 \begin{array}{ccc}
  j_1 & j_2 & J_{12} \\
  m_1 & m_2 & M_{12} 
 \end{array}
 \right)\left(
 \begin{array}{ccc}
  j_3 & j_4 & J_{34} \\
  m_3 & m_4 & M_{34}
 \end{array}
 \right) \\
 &\times\left(
 \begin{array}{ccc}
  j_1 & j_3 & J_{13} \\
  m_1 & m_3 & M_{13} 
 \end{array}
 \right)\left(
 \begin{array}{ccc}
  j_2 & j_4 & J_{24} \\
  m_2 & m_4 & M_{24}
 \end{array}
 \right)\left(
 \begin{array}{ccc}
  J_{12} & J_{34} & J \\
  M_{12} & M_{34} & M
 \end{array}
 \right) \,.
\end{split}\end{equation}
Comparison of the last two equation yields Eq.~\eqref{eq:K}. Note
that, as expected, $K(J_1, J_2, J_3)$ is symmetric in its first two
indices.

Next, we show
\begin{equation}\label{eq:C}
\boxed{C(J)=\frac{1}{2}\mathcal{J}(J)-\mathcal{J}(\jp)}
\end{equation}

First, we contract Eq.~\eqref{eq:Ceqn} with two Sigma matrices and use
the orthogonality relation~\eqref{eq:ortho:km} to arrive at
\begin{equation}\label{eq:Ccondef}
 \mathcal{D}(J)C(J) = \sum_a \sum_{M} \sum_{irkl} 
\tilde{\tau}^a_{ir} \tilde{\tau}^a_{kl}
\Sigma^{(J),M}_{ik} \Sigma^{(J),M}_{rl} \,.
\end{equation}
We find $C(J)$ by writing Eq.~\eqref{eq:Ccondef} explicitly in
terms of Clebsch-Gordan coefficients as
\begin{equation}\label{eq:braketexpand1}
\begin{split}
 \mathcal{D}(J)C(J) = 
 \sum_a\sum_M\sum_{mnm'n'}
 & C_{\jp\jp}(JM; mn) C_{\jp\jp}(JM; m'n') \\ & \times
 \bra{\jp m} \tilde\tau^{(\jp),a} \ket{\jp m'}
 \bra{\jp n} \tilde\tau^{(\jp),a} \ket{\jp n'} \,.
\end{split}
\end{equation}
Noting that, by definition, $\tilde\tau^{(\jp),a}$ are the spin-$\jp$
generators, we introduce the notation $\tilde\tau_{mn}^{a} \equiv
\bra{\jp m} \tilde\tau^{(\jp),a} \ket{\jp n}$, i.e. we label the
generators with $\jp$ as well as $a=1,2,3$, and the states by $\jp$ and
their ``magnetic'' quantum numbers $m,n,m,'n'=-\jp,\ldots,\jp$. Using the
symmetry relation of the Clebsch-Gordan coefficients
\begin{equation}\label{eq:CG:sym:rel}
 C_{jj}(JM; mn) = (-1)^{j+m}
 \sqrt{\frac{\mathcal{D}(J)}{\mathcal{D}(j)}} C_{jJ}(j,-n;m,-M)\,,
\end{equation}
Eq.~\eqref{eq:braketexpand1} becomes
\begin{equation}\label{eq:braketexpand2}
\begin{split}
 \mathcal{D}(J)C(J) = 
 \frac{\mathcal{D}(J)}{\mathcal{D}(\jp)}
 \sum_a\sum_M\sum_{mm'nn'} & (-1)^{2\jp-n-n'} \\ & \times
 C_{\jp J}(\jp,-n; m M) C_{\jp J}(\jp,-n'; m' M) \\ & \times
 \bra{\jp m} \tilde\tau^{(\jp),a} \ket{\jp m'}
 \bra{\jp n} \tilde\tau^{(\jp),a} \ket{\jp n'} \,,
\end{split}
\end{equation}
where we use $m=M-n$ and $m'=M-n'$, as well as the fact that $M$ is
always integer, to rewrite the phase factor. We also take $-M\to M$
using the symmetry of the sum over $M$. Now, we artificially
regard each spin-$\jp$ state as belonging to the spin-$\jp$ subspace
of the Clebsch-Gordan decomposition of the tensor product of a
spin-$\jp$ and a spin-$J$ state. For instance, $\bra{\jp m} =
\sum_{nN} C_{\jp J}(\jp m; nN) \bra{\jp n; JN}$ and analogous
relations lead to
\begin{equation}
\begin{split}
 & \bra{\jp m} \tilde\tau^{(\jp),a} \ket{\jp m'} \\
 & =\sum_{nN} \sum_{n'N'}
 C_{\jp J}(\jp m; nN) C_{\jp J}(\jp m'; n'N')
 \bra{\jp n, JN} 
    \big(\tilde\tau^{(\jp),a} \otimes\mathbbm{1}_J 
  + \mathbbm{1}_\jp \otimes \tilde\tau^{(J),a}\big)
 \ket{\jp n', JN'} \,.
\end{split}
\end{equation}
Hence, we find the explicit tensor decomposition for the generators
\begin{equation}\begin{split}\label{eq:tensor:decomp}
 \tilde{\tau}^{(\jp),a}_{mn} & = 
 \sum_{\tilde M} \sum_{\tilde m \tilde n}
  C_{\jp J}(\jp m; \tilde m \tilde M) C_{\jp J}(\jp n; \tilde n \tilde M)
  \tilde{\tau}^{(\jp),a}_{\tilde m \tilde n} \\
 & \quad + \sum_{\tilde M \tilde N} \sum_{\tilde m}
  C_{\jp J}(\jp m; \tilde m \tilde M) C_{\jp J}(\jp n; \tilde m \tilde N)
  \tilde{\tau}^{(J),a}_{\tilde M \tilde N} \,.
\end{split}\end{equation}
Inserting this identity into Eq.~\eqref{eq:braketexpand2} gives
\begin{equation}\begin{split}
 \mathcal{D}(J) & C(J) 
 = \frac{\mathcal{D}(J)}{\mathcal{D}(\jp)}
 \sum_a\sum_{nn'} (-1)^{2\jp-n-n'}
 \tilde{\tau}^{(\jp),a}_{-n,-n'}\tilde{\tau}^{(\jp),a}_{nn'}\\
 &-\frac{\mathcal{D}(J)}{\mathcal{D}(\jp)}
 \sum_{a}\sum_{\tilde m}\sum_{MM'nn'}(-1)^{2\jp-n-n'}
 C_{\jp J}(\jp,-n;\tilde m M) C_{\jp J}(\jp,-n';\tilde m M')
 \tilde{\tau}^{(J),a}_{M M'}\tilde{\tau}^{(\jp),a}_{nn'}\\
 =&\frac{\mathcal{D}(J)}{\mathcal{D}(\jp)}
 \sum_a\sum_{nn'} (-1)^{2\jp+n+n'}
 \tilde{\tau}^{(\jp),a}_{-n,-n'}\tilde{\tau}^{(\jp),a}_{nn'}\\
 &-\sum_{a}\sum_{\tilde m}\sum_{MM'nn'}(-1)^{2J+M+M'}
 C_{\jp\jp}(J,-M;\tilde m n) C_{\jp\jp}(J,-M';\tilde m n') 
 \tilde{\tau}^{(J),a}_{MM'}\tilde{\tau}^{(\jp),a}_{nn'}
\end{split}\end{equation}
where we again use Eq.~\eqref{eq:CG:sym:rel} and rewrite the phase in
the second term noting that $\tilde m = -n-M = -n'-M'$ and $(-1)^{2J}
= 1$. We then use an identity analogous to~\eqref{eq:tensor:decomp},
namely, 
\begin{equation}\begin{split}
 \tilde{\tau}^{(J),a}_{MN} & = 
 \sum_{\tilde m} \sum_{m' n'}
  C_{\jp\jp}(JM; m' \tilde m) C_{\jp\jp}(JN; n' \tilde m)
  \tilde{\tau}^{(\jp),a}_{m' n'} \\
 & \quad + \sum_{\tilde m} \sum_{m' n'}
  C_{\jp\jp}(JM; \tilde m m') C_{\jp\jp}(JN; \tilde m n')
  \tilde{\tau}^{(\jp),a}_{m' n'} \,,
\end{split}\end{equation}
to obtain
\begin{equation}
 \sum_{\tilde m} \sum_{m' n'}
  C_{\jp\jp}(JM; \tilde m m') C_{\jp\jp}(JN; \tilde m n')
  \tilde{\tau}^{(\jp),a}_{m' n'}
 = \frac{1}{2} \tilde{\tau}^{(J),a}_{MN}\,,
\end{equation}
and so
\begin{equation}\label{eq:preDC}
\begin{split}
 \mathcal{D}(J)C(J)
 &=\frac{\mathcal{D}(J)}{\mathcal{D}(\jp)}
 \sum_a\sum_{nn'}(-1)^{2\jp+n+n'}
  \tilde{\tau}^{(\jp),a}_{-n,-n'}\tilde{\tau}^{(\jp),a}_{nn'} \\
 & \quad -\frac{1}{2}\sum_a\sum_{MM'}(-1)^{2J+M+M'}
 \tilde{\tau}^{(J),a}_{-M,-M'}\tilde{\tau}^{(J),a}_{MM'} \,.
\end{split}
\end{equation}
A straightforward calculation using the explicit expressions for the
spin-$\jp$ generators, Eq.~\eqref{eq:su2-gen-def}, gives
\begin{equation}\begin{split}
 & (-1)^{2\jp+n+n'}\tilde{\tau}^{(\jp),1}_{-n,-n'}\tilde{\tau}^{(\jp),1}_{nn'}
 = (-1)^{2\jp+n+n'}\tilde{\tau}^{(\jp),2}_{-n,-n'}\tilde{\tau}^{(\jp),2}_{nn'} \\
& \qquad =-\frac{1}{4}\left[\delta_{n,n'+1}^2(\jp+n'+1)(\jp-n')+\delta_{n,n'-1}^2(\jp-n'+1)(\jp+n')\right]\,,\\
 &(-1)^{2\jp+n+n'}\tilde{\tau}^{(\jp),3}_{-n,-n'}\tilde{\tau}^{(\jp),3}_{nn'}=-n^2\delta_{nn'}^2\,.
\end{split}\end{equation}
Comparing these to the analogous expressions
\begin{equation}\begin{split}
 & \tilde{\tau}^{(\jp),1}_{n'n}\tilde{\tau}^{(\jp),1}_{nn'}
 = \tilde{\tau}^{(\jp),2}_{n'n}\tilde{\tau}^{(\jp),2}_{nn'} \\
& \quad =\frac{1}{4}\left[\delta_{n,n'+1}^2(\jp+n'+1)(\jp-n')+\delta_{n,n'-1}^2(\jp-n'+1)(\jp+n')\right]\,\\
 &\tilde{\tau}^{(\jp),3}_{n'n}\tilde{\tau}^{(\jp),3}_{nn'}=n^2\delta_{nn'}^2\,,
\end{split}\end{equation}
we see that
\begin{equation}
 \sum_a\sum_{nn'} (-1)^{2\jp+k+l}
 \tilde{\tau}^{(\jp),a}_{-n,-n'} \tilde{\tau}^{(\jp),a}_{nn'}
 = -\sum_a\sum_{nn'}
 \tilde{\tau}^{(\jp),a}_{n'n} \tilde{\tau}^{(\jp),a}_{nn'} 
 = -\jp(\jp+1)(2\jp+1) \,.
\end{equation}
This yields the desired result~\eqref{eq:C}.

In the following, we collect several sum rules involving the
coefficient $C(J)$. The first is
\begin{equation}\label{eq:CD}
\boxed{\sum_J C(J)\mathcal{D}(J) = 0}
\end{equation}
It is derived by summing Eq.~\eqref{eq:Ccondef} over $J$ and using
the orthogonality relations, to obtain
\begin{equation}
 \sum_{J}C(J)\mathcal{D}(J) = 
 \sum_{a} \sum_{irkl}
 \tilde{\tau}^a_{ir}\tilde{\tau}^a_{kl}\delta_{ir}\delta_{kl} = 0 \,,
\end{equation}
where we use the fact that the generators are traceless. In practice,
the interchange of indices in the Sigma matrices gives rise to
additional phase factors, hence we also need the relation
\begin{equation}\label{eq:mCD}
\boxed{\sum_J (-1)^{J-2\jp}C(J)\mathcal{D}(J) = \mathcal{D}(\jp)\mathcal{J}(\jp)}
\end{equation}
To prove this relation, we again sum Eq.~\eqref{eq:Ccondef} over $J$,
now taking into account the symmetry properties of the Clebsch-Gordan
coefficients
\begin{equation}
\sum_{J}(-1)^{J-2\jp}C(J)\mathcal{D}(J) = 
 \sum_{a} \sum_{irkl}
\tilde{\tau}^a_{ir}\tilde{\tau}^a_{kl}\delta_{il}\delta_{kr} = 
 \sum_{a} \sum_{ir}
\tilde{\tau}^a_{ir}\tilde{\tau}^a_{ri} = 
\mathcal{D}(\jp)\mathcal{J}(\jp) \,.
\end{equation}
We note in passing that this relation can be used to calculate the
``restricted'' sum over $J$, as
\begin{equation}
{\sum_{J}}^\prime C(J)\mathcal{D}(J) \equiv
\sum_{J} \frac{1+(-1)^{J-2\jp}}{2} C(J)\mathcal{D}(J) =
\frac{1}{2} \mathcal{D}(\jp)\mathcal{J}(\jp) \,.
\end{equation}

A similar sum rule, quadratic in $C(J)$, reads
\begin{equation}\label{eq:CCD}
\boxed{\sum_J \mathcal{D}(J)C(J)^2 =
  \frac{\mathcal{J}(\jp)^2\mathcal{D}(\jp)^2}{3}}
\end{equation}
To see this, we rewrite the left-hand side as
\begin{equation}
 \sum_J\mathcal{D}(J)C(J)^2 
 = \sum_{ab} \sum_{JM} \sum_{irklmn}
 \tilde{\tau}^a_{im}\tilde{\tau}^b_{mr}\tilde{\tau}^a_{kn}\tilde{\tau}^b_{nl}
 \Sigma^{(J),M}_{ik}\Sigma^{(J),M}_{rl} 
 = \sum_{ab} \sum_{imkn}
 \tilde{\tau}^a_{im}\tilde{\tau}^b_{mi}\tilde{\tau}^a_{kn}\tilde{\tau}^b_{nk} \,.
\end{equation}
Now, we use the $SU(2)$ relation
\begin{equation}
 \sum_{im} \tilde{\tau}^a_{im}\tilde{\tau}^b_{mi}
 = \frac{\mathcal{J}(\jp)\mathcal{D}(\jp)}{3}\delta^{ab}
\end{equation}
to find Eq.~\eqref{eq:CCD}.
The analogous sum rule with phase factor reads
\begin{equation}\label{eq:mCCD}
\boxed{\sum_J(-1)^{J-2\jp}\mathcal{D}(J)C(J)^2 = \mathcal{J}(\jp)\mathcal{D}(\jp)\left(\mathcal{J}(\jp)-1\right)}
\end{equation}
The proof proceeds similar to the above, except we must use the
(anti-)symmetry of the Clebsch-Gordan coefficients:
\begin{equation}
 \sum_J(-1)^{J-2\jp} \mathcal{D}(J)C(J)^2 = 
 \sum_{ab} \sum_{imkn}
 \tilde{\tau}^a_{im} \tilde{\tau}^b_{mk} 
 \tilde{\tau}^a_{kn}\tilde{\tau}^b_{ni} \,.
\end{equation}
We then use the $SU(2)$ algebra to re-write
\begin{equation}
\begin{split}
 \sum_{ab} \sum_{imkn}
 \tilde{\tau}^a_{im}\tilde{\tau}^b_{mk}
 \tilde{\tau}^a_{kn}\tilde{\tau}^b_{ni} & = 
 \sum_{ab} \sum_{imkn}
 \tilde{\tau}^a_{im}\tilde{\tau}^a_{mk}
 \tilde{\tau}^b_{kn}\tilde{\tau}^b_{ni} 
 + \sum_{abc} \sum_{imn}
 i\epsilon^{bac}\tilde{\tau}^a_{im}\tilde{\tau}^c_{mn}\tilde{\tau}^b_{ni} \\
 & = \mathcal{J}(\jp)^2\mathcal{D}(\jp)-\mathcal{J}(\jp)\mathcal{D}(\jp) \,,
\end{split}
\end{equation}
which gives the relation~\eqref{eq:mCCD}.
Again, we use this to calculate the restricted sum over $J$, as
\begin{equation}
\begin{split}
 {\sum_J}^\prime \mathcal{D}(J)C(J)^2 & = 
 \sum_J \frac{1+(-1)^{J-2\jp}}{2} \mathcal{D}(J)C(J)^2 \\ & = 
 \frac{1}{6} \mathcal{J}(\jp)\mathcal{D}(\jp)
 \left(\mathcal{J}(\jp)\mathcal{D}(\jp) + 3\mathcal{J}(\jp) -3\right) \,.
\end{split}
\end{equation}

Another pair of rules cubic in $C(J)$ is necessary to reduce the
algebra in diagrams involving the SM Higgs. The first is given by
\begin{equation}\label{eq:Ccube}
\boxed{\sum_J\mathcal{D}(J)C(J)^3 = -\frac{\mathcal{J}(\jp)^2
       \mathcal{D}(\jp)^2}{6}}
\end{equation}
Rewriting the left-hand side and using the appopriate orthogonality
relations for the Clebsch-Gordan coefficients gives
\begin{equation}
       \sum_J\mathcal{D}(J)C(J)^3 = 
       \sum_{abc}\Tr\left[\tilde{\tau}^a\tilde{\tau}^b\tilde{\tau}^c\right]
       \Tr\left[\tilde{\tau}^a\tilde{\tau}^b\tilde{\tau}^c\right]\,.
\end{equation}
The trace of three generators is expressed as
\begin{equation}
	\Tr\left[\tilde{\tau}^a\tilde{\tau}^b\tilde{\tau}^c\right] =
	\frac{1}{2}\left(\Tr\left[\left\{\tilde{\tau}^a,\tilde{\tau}^b\right\}\tilde{\tau}^c\right]
	+\Tr\left[\left[\tilde{\tau}^a,\tilde{\tau}^b\right]\tilde{\tau}^c\right]\right) 
	= \frac{1}{2}\Tr\left[\left[\tilde{\tau}^a,\tilde{\tau}^b\right]\tilde{\tau}^c\right]\,,
\end{equation}
where we make use of the definition of the totally symmetric tensor,
\begin{equation}
	d^{abc}\propto\Tr\left[\left\{\tilde{\tau}^a,\tilde{\tau}^b\right\}\tilde{\tau}^c\right]\,,
\end{equation}
which vanishes for $SU(2)$. Next, we simply use the group algebra to find
\begin{equation}
	\Tr\left[\tilde{\tau}^a\tilde{\tau}^b\tilde{\tau}^c\right] =
	\frac{i\mathcal{J}(\jp)\mathcal{D}(\jp)}{6}\epsilon^{abc} \,.
\end{equation}
Squaring this expression and using
\begin{equation}
	\sum_{abc}\epsilon^{abc}\epsilon^{abc} = 6
\end{equation}
gives the result in Eq.~\eqref{eq:Ccube}. The second relation cubic in
$C(J)$ is
\begin{equation}\label{eq:CcubeNeg}
	\boxed{\sum_J (-1)^{J-2\jp}\mathcal{D}(J)C(J)^3 = 
	\mathcal{J}(\jp)\mathcal{D}(\jp)\left(\mathcal{J}(\jp)-1\right)
	\left(\mathcal{J}(\jp)-2\right)}
\end{equation}
As before, we re-write the left-hand side of this expression and use
orthogonality relations
\begin{equation}
	\sum_J (-1)^{J-2\jp}\mathcal{D}(J)C(J)^3 =
	\sum_{abc}\Tr\left[\tilde{\tau}^a\tilde{\tau}^b\tilde{\tau}^c
	\tilde{\tau}^a\tilde{\tau}^b\tilde{\tau}^c\right]\,.
\end{equation}
The product in the trace simplifies using the $SU(2)$ algebra
\begin{equation}
	\sum_a\tilde{\tau}^a\tilde{\tau}^b\tilde{\tau}^c\tilde{\tau}^a = 
	\sum_a\left(\tilde{\tau}^a\tilde{\tau}^b\tilde{\tau}^a\tilde{\tau}^c
	+i\epsilon^{cad}\tilde{\tau}^a\tilde{\tau}^b\tilde{\tau}^d\right)\,.
\end{equation}
Then, using
\begin{equation}
	\sum_a\tilde{\tau}^a\tilde{\tau}^b\tilde{\tau}^a = 
	\left(\mathcal{J}(\jp)-1\right)\tilde{\tau}^b
\end{equation}
the trace is reduced to
\begin{equation}
	\sum_{abc}\Tr\left[\tilde{\tau}^a\tilde{\tau}^b\tilde{\tau}^c
	\tilde{\tau}^a\tilde{\tau}^b\tilde{\tau}^c\right] =
	\Tr\left[\mathcal{J}(\jp)\left(\mathcal{J}(\jp)-1\right)^2\mathbbm{1}
	- \mathcal{J}(\jp)\left(\mathcal{J}(\jp)-1\right)\mathbbm{1}\right]
\end{equation}
which, when the trace is performed over the identity, gives 
Eq.~\eqref{eq:CcubeNeg}.

In order to derive the necessary algebraic relations involving the
factor $K$, we first prove the following useful relation:
\begin{equation}\label{eq:SigSigtt}
\boxed{\sum_M \sum_{kl} \Sigma^{(J),M}_{ik}\Sigma^{(J),M}_{rl}\tilde{\tau}^a_{lk} = 
\frac{\mathcal{D}(J)C(J)}{\mathcal{J}(\jp)\mathcal{D}(\jp)}\tilde{\tau}^a_{ir}}
\end{equation}
To derive this, we note that the only object in our basis with a
single adjoint representation index and two isospin-$\jp$
representation indices is the generator $\tilde{\tau}^a_{ir}$ (any
other objects with only these free indices can be reduced to this
generator). Therefore, we make the ansatz
\begin{equation}
 \sum_{M} \sum_{kl}
 \Sigma^{(J),M}_{ik}\Sigma^{(J),M}_{rl}
 \tilde{\tau}^a_{lk} 
 = G(J) \tilde{\tau}^a_{ir} \,.
\end{equation}
Multiplying both sides by $\tilde{\tau}^a_{ri}$, summing over $a,i,r$,
and using Eq.~\eqref{eq:Ccondef} gives the
relation~\eqref{eq:SigSigtt}. Using this result, we now prove
\begin{equation}\label{eq:CK}
  \boxed{\sum_{J_1}C(J_1)K(J_1,J_2,J_3) = 
  \frac{\mathcal{D}(J_2)C(J_2)C(J_3)}{\mathcal{J}(\jp)\mathcal{D}(\jp)}}
\end{equation}
To this end, we consider the product
\begin{equation}
\begin{split}
 \sum_{a,mn} & \sum_{M}
 \tilde{\tau}^a_{ir}\tilde{\tau}^a_{mn}
 \Sigma^{(J),M}_{kn}\Sigma^{(J),M}_{lm} = 
 \frac{\mathcal{D}(J)C(J)}{\mathcal{J}(\jp)\mathcal{D}(\jp)}
 \sum_a \tilde{\tau}^a_{ir}\tilde{\tau}^a_{kl} \\ & =
 \frac{\mathcal{D}(J)C(J)}{\mathcal{J}(\jp)\mathcal{D}(\jp)}
 \sum_{J_1 M_1} C(J_1)
 \Sigma^{(J_1),M_1}_{ik}\Sigma^{(J_1),M_1}_{rl} \,,
\end{split}
\end{equation}
making use of Eq.s~\eqref{eq:SigSigtt} and ~\eqref{eq:Ceqn}. However,
this is alternatively written as
\begin{equation}
\begin{split}
 \sum_{a,mn} \sum_M
 \tilde{\tau}^a_{ir} & \tilde{\tau}^a_{mn}
 \Sigma^{(J),M}_{kn}\Sigma^{(J),M}_{lm} = 
 \sum_{mn} \sum_{J_1,M_1,M} C(J_1) 
 \Sigma^{(J),M}_{kn} \Sigma^{(J),M}_{lm}
 \Sigma^{(J_1),M_1}_{im} \Sigma^{(J_1),M_1}_{rn} \\ & =
 \sum_{J_1,J_2,M_2}C(J_1)K(J_1,J,J_2)
 \Sigma^{(J_2),M_2}_{ik} \Sigma^{(J_2),M_2}_{rl} \,.
\end{split}
\end{equation}
Equating these expressions and using the orthogonality relations, the
result~\eqref{eq:CK} follows.

A further important relation incorporates the condition of gauge
invariance:
\begin{equation}\label{eq:mCK}
\boxed{\sum_{J_1}(-1)^{J_1-2\jp}\left(1+(-1)^{J-2\jp}(-1)^{J_2-2\jp}\right)C(J_1)K(J,J_1,J_2) = 
\left(\mathcal{J}(j)+C(J)\right)\delta^{J,J_2}}
\end{equation}
All interaction vertices must be gauge-invariant. A scalar $SU(2)$
spin-$\jp$ field multiplet transforms as $\varphi_i \to \varphi_i^\prime
= \varphi_i + \delta \varphi_i$ under an infinitesimal gauge
transformation, where
\begin{equation}
  \delta \varphi_i = \sum_{a,k} i \epsilon^a \tilde \tau_{ik}^a \varphi_k
        - \frac{1}{2} \sum_{ab,lk} \epsilon^a \epsilon^b \tilde\tau_{il}^a\tilde{\tau}_{lk}^b\varphi_k
        + {\mathcal O}(\epsilon^3)\,,
\end{equation}
while
\begin{equation}
  \delta \varphi_i^* = - \sum_{a,k} i \epsilon^a \tilde \tau_{ki}^a \varphi_k^* 
        - \frac{1}{2} \sum_{ab,lk} \epsilon^a \epsilon^b \tilde\tau_{kl}^a\tilde{\tau}_{li}^b\varphi_k^*
        + {\mathcal O}(\epsilon^3)\,.
\end{equation}
Hence, we have the relation
\begin{equation}\begin{split}
  0 = \sum_{mM} \Big(
      & \quad \tilde\tau_{mi}^a \Sigma_{mk}^{(J),M} \Sigma_{rl}^{(J),M}
      + \tilde\tau_{mk}^a \Sigma_{im}^{(J),M} \Sigma_{rl}^{(J),M} \\
      & - \tilde\tau_{rm}^a \Sigma_{ik}^{(J),M} \Sigma_{ml}^{(J),M}
      - \tilde\tau_{lm}^a \Sigma_{ik}^{(J),M} \Sigma_{rm}^{(J),M} \Big) \,,
\end{split}\end{equation}
which leads, upon contraction with a $SU(2)$ generator, to
\begin{equation}\begin{split}\label{eq:gaugeinv1}
 \mathcal{J}(\jp) & \sum_{M}
        \Sigma_{ik}^{(J),M}\Sigma_{rn}^{(J),M} \\ & = 
 \sum_{lm,a,M} \Big(
        \tilde{\tau}^a_{nl}\tilde{\tau}^a_{mi}\Sigma_{mk}^{(J),M}\Sigma_{rl}^{(J),M}
      + \tilde{\tau}^a_{nl}\tilde{\tau}^a_{mk}\Sigma_{im}^{(J),M}\Sigma_{rl}^{(J),M}
      - \tilde{\tau}^a_{nl}\tilde{\tau}^a_{rm}\Sigma_{ik}^{(J),M}\Sigma_{ml}^{(J),M} \Big) \,.
\end{split}\end{equation}
We use Eq.s~\eqref{eq:Ceqn} and~\eqref{eq:K1eqn} to rewrite this condition as
\begin{equation}\begin{split}
 & \mathcal{J}(\jp) \sum_{M}
 \Sigma^{(J),M}_{ik} \Sigma^{(J),M}_{rn}
  = \sum_{J_1} \sum_{M,M_1} \sum_{ml} C(J_1) 
 \Big( \Sigma^{(J),M}_{mk}\Sigma^{(J),M}_{rl}\Sigma^{(J_1),M_1}_{nm}\Sigma^{(J_1),M_1}_{li} \\
       & \hspace{18em} +\Sigma^{(J),M}_{im}\Sigma^{(J),M}_{rl}\Sigma^{(J_1),M_1}_{nm}\Sigma^{(J_1),M_1}_{lk} \\
       & \hspace{18em} -\Sigma^{(J),M}_{ik}\Sigma^{(J),M}_{ml}\Sigma^{(J_1),M_1}_{nr}\Sigma^{(J_1),M_1}_{lm}\Big)\\
 = & \sum_{J_1,J_2,M_2}(-1)^{J_1-2\jp} C(J_1)K(J,J_1,J_2)
   \Big[(-1)^{J-2\jp} \Sigma^{(J_2),M_2}_{ik} \Sigma^{(J_2),M_2}_{nr}
        +\Sigma^{(J_2),M_2}_{ik}\Sigma^{(J_2),M_2}_{rn}\Big] \\ 
  & - C(J) \sum_{M} \Sigma^{(J),M}_{ik}\Sigma^{(J),M}_{rn} \,.
\end{split}\end{equation}
Multiplying both sides by $\sum_{M_3}
\Sigma^{(J_3),M_3}_{ik}\Sigma^{(J_3),M_3}_{rn}$ and summing over
$i,k,r,n$ yields Eq.~\eqref{eq:mCK}.

We now derive a few relations involving $K$ and two powers of $C$. The
first is
\begin{equation}\label{eq:CCK}
\boxed{\sum_{J_1,J_2}(-1)^{J_1-2\jp}(-1)^{J_2-2\jp}C(J_1)C(J_2)K(J_1,J_2,J')
 = (-1)^{J'-2\jp}C(J')(C(J')+1)}
\end{equation}
For a proof, consider the product of generators
\begin{equation}
 \left\{\tilde{\tau}^a,\tilde{\tau}^b\right\}_{ir}
 \left\{\tilde{\tau}^a,\tilde{\tau}^b\right\}_{kl}
 = 2 \sum_{mn}
  \left(\tilde{\tau}^a_{im}\tilde{\tau}^b_{mr}
        \tilde{\tau}^a_{kn}\tilde{\tau}^b_{nl} + 
        \tilde{\tau}^a_{im}\tilde{\tau}^b_{mr}
        \tilde{\tau}^b_{kn}\tilde{\tau}^a_{nl}\right) \,.
\end{equation}
Using $SU(2)$ commutation relations and Eq.~\eqref{eq:Ceqn}, it is
easy to see that this becomes
\begin{equation}
 \sum_{ab}
 \left\{\tilde{\tau}^a,\tilde{\tau}^b\right\}_{ir}
 \left\{\tilde{\tau}^a,\tilde{\tau}^b\right\}_{kl} = 
 2\sum_{JM}C(J)\left(2C(J)+1\right)
  \Sigma^{(J),M}_{ik}\Sigma^{(J),M}_{rl} \,,
\end{equation}
but it is also expressed as
\begin{equation}\begin{split}
 2 & \sum_{J_1,J_2} \sum_{M_1,M_2} \sum_{m,n} C(J_1) C(J_2) \\
   & \times
     \left(\Sigma^{(J_1),M_1}_{ik}\Sigma^{(J_1),M_1}_{mn}
           \Sigma^{(J_2),M_2}_{mn}\Sigma^{(J_2),M_2}_{rl} + 
           \Sigma^{(J_1),M_1}_{in}\Sigma^{(J_1),M_1}_{ml}
           \Sigma^{(J_2),M_2}_{mk}\Sigma^{(J_2),M_2}_{rn}\right) \\
 = 2 & \sum_{J_1, M_1}C(J_1)^2 \Sigma^{(J_1),M_1}_{ik}\Sigma^{(J_1),M_1}_{rl}\\
   & + 2 \sum_{J_1,J_2,J_3} \sum_{M_3} (-1)^{J_1+J_2+J_3-6\jp}
    C(J_1) C(J_2) K(J_1,J_2,J_3) 
    \Sigma^{(J_3),M_3}_{ik}\Sigma^{(J_3),M_3}_{rl} \,.
\end{split}\end{equation}
Equating these two expressions, multiplying both sides by $\sum_{M'}
\Sigma^{(J'),M'}_{ik}\Sigma^{(J'),M'}_{rl}$ and summing over $i,k,r,n$
yields Eq.~\eqref{eq:CCK}. A variant of this relation involving only
one phase factor reads
\begin{equation}\label{eq:mCCK}
\boxed{\sum_{J_1,J_2}(-1)^{J_1-2\jp}C(J_1)C(J_2)K(J_1,J_2,J_3)
 = C(J_3)(\mathcal{J}(\jp)-1)}
\end{equation}
To prove it, we perform the sum over $J_2$ using Eq.~\eqref{eq:CK} and
the fact that $K$ is symmetric in its first two indices to find
\begin{equation}
\begin{split}
 & \sum_{J_1,J_2} (-1)^{J_1-2\jp} 
 C(J_1) C(J_2) K(J_1,J_2,J_3) \\ = &  
 \sum_{J_1} (-1)^{J_1-2\jp}
 \frac{\mathcal{D}(J_1)C(J_1)^2C(J_3)}{\mathcal{J}(\jp)\mathcal{D}(\jp)} = 
 C(J_3) (\mathcal{J}(\jp)-1) \,,
\end{split}
\end{equation}
where in the last equality we used Eq.~\eqref{eq:mCCD}.
The relation without phase factors
\begin{equation}
\boxed{\sum_{J_1,J_2}C(J_1)C(J_2)K(J_1,J_2,J_3) = \frac{1}{3}\mathcal{J}(\jp)\mathcal{D}(\jp)C(J_3)}
\end{equation}
is shown similar to the above, performing the sum over $J_2$ and using
Eq.~\eqref{eq:CCD}.

Finally, we prove the following symmetry relation for a contraction of
two $K$ factors:
\begin{equation}\label{eq:KK}
\boxed{\sum_{J_4}K(J_1,J_2,J_4)K(J_4,J_3,J_5) = \sum_{J_4}K(J_1,J_3,J_4)K(J_4,J_2,J_5)}
\end{equation}
First, consider the sum
\begin{equation}\begin{split}\label{eq:KsqStart}
	\sum_{imsl}\sum_{LMN}\Sigma^{(J_1),L}_{im}\Sigma^{(J_1),L}_{sl}
		\Sigma^{(J_2),M}_{qs}\Sigma^{(J_2),M}_{ri}
		\Sigma^{(J_3),N}_{nm}\Sigma^{(J_3),N}_{kl} \,.
\end{split}\end{equation}
This simplifies to
\begin{equation}\begin{split}\label{eq:KsqFin1}
	\sum_{J_4}\sum_{lm}\sum_{MN}&K(J_1,J_2,J_4)\Sigma^{(J_3),N}_{nm}\Sigma^{(J_3),N}_{kl}
		\Sigma^{(J_4),M}_{mq}\Sigma^{(J_4),M}_{lr} \\
		& = \sum_{J_4,J_5}\sum_M K(J_1,J_2,J_4)K(J_4,J_3,J_5)\Sigma^{(J_5),M}_{nr}\Sigma^{(J_5),M}_{kq} \,.
\end{split}\end{equation}
However, Eq.~\eqref{eq:KsqStart} also reduces in a different way
\begin{equation}\begin{split}\label{eq:KsqFin2}
	\sum_{J_4}\sum_{is}\sum_{MN}&K(J_1,J_3,J_4)\Sigma^{(J_2),N}_{qs}\Sigma^{(J_2),N}_{ri}
		\Sigma^{(J_4),M}_{ik}\Sigma^{(J_4),M}_{sn} \\
		& = \sum_{J_4,J_5}\sum_M K(J_1,J_3,J_4)K(J_4,J_2,J_5)\Sigma^{(J_5),M}_{nr}\Sigma^{(J_5),M}_{kq} \,.
\end{split}\end{equation}
Equating these two expressions gives the final
result~\eqref{eq:KK}.

\section{Numerics}
\label{sec:numerics}

\begin{figure}
\centering
\includegraphics[width=0.49\textwidth]{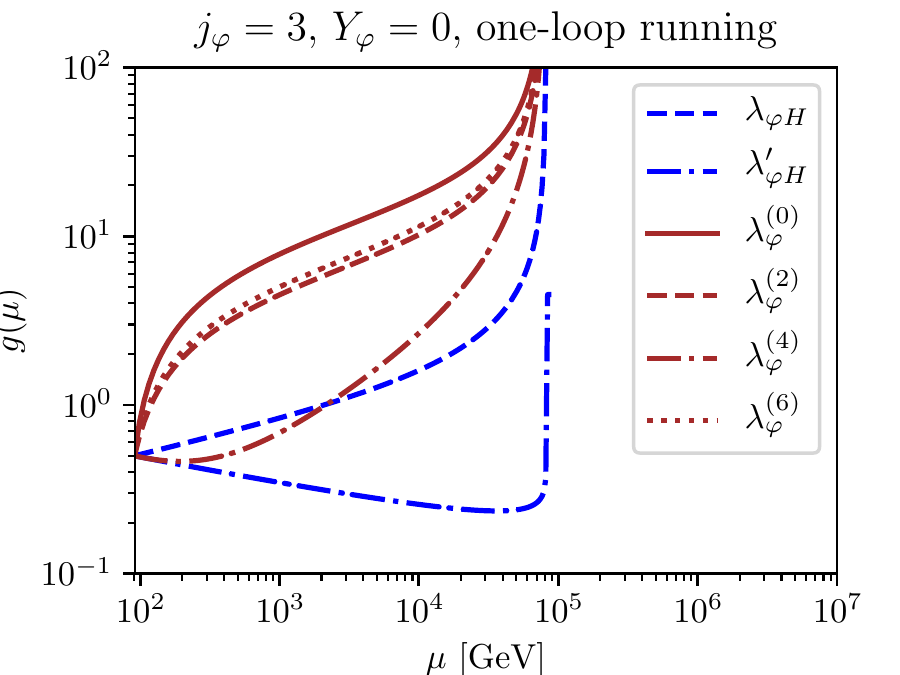}
\includegraphics[width=0.49\textwidth]{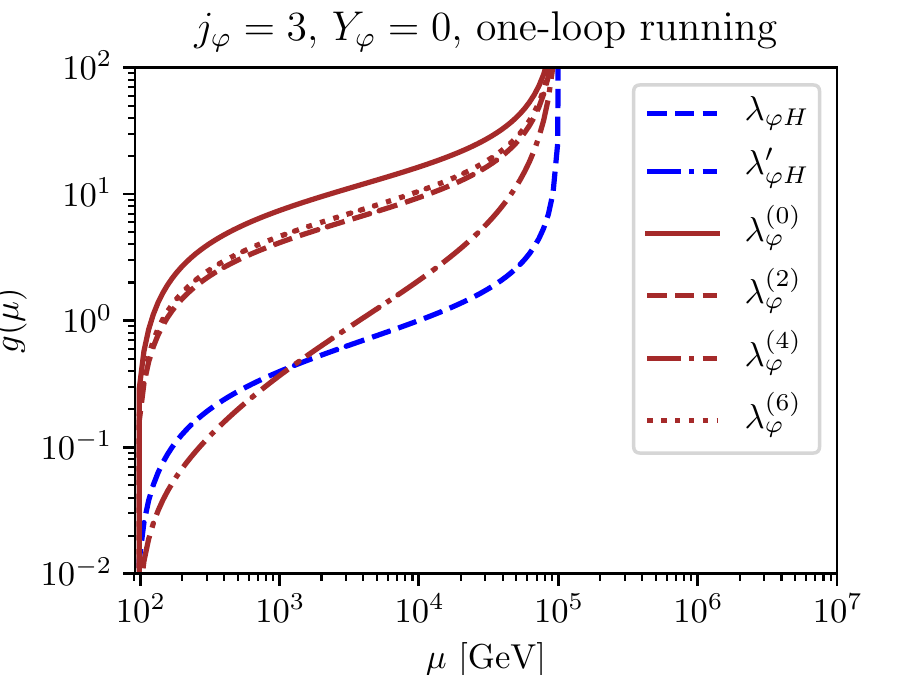}
\caption{One-loop running of scalar quartic couplings for $\jp=3$ with
  $Y_\varphi=0$ (``Minimal Scalar Dark Matter''). The dashed and
  dash-dotted lines denote the Higgs-portal couplings
  $\lambda_{\varphi H}$ and $\lambda_{\varphi H}'$, respectively,
  while the solid lines denote the four scalar couplings
  $\lambda_\varphi^{(J)}$, for $J=0,2,4,6$. The running at one-loop
  exhibits a Landau pole around $\mu=10^5\,$GeV. Left panel: all
  initial conditions are set to 0.5 at $\mu=M_Z$. Right panel:
  vanishing initial conditions at $\mu=M_Z$.}
\label{fig:1loop:scalar}
\end{figure}

\begin{table}
\begin{center}
\begin{tabular}{|l|l|l|}
\hline
$m_t(\text{pole}) = 172.4(7)\,$GeV & $m_b(m_b) = 4.18^{+0.03}_{-0.02}\,$GeV & $m_c(m_c) = 1.27(2)\,$GeV\\
\hline
$m_\tau = 1.77686(12)\,$GeV & $M_h = 125.10(14)\,$GeV & $M_Z = 91.1876(21)\,$GeV\\
\hline
$\alpha^{(5)}(M_Z)^{-1} = 127.952(9)$ & $\sin^2\theta (M_Z) = 0.23121(4)$ & $\alpha_s(M_Z) = 0.1179(10)$\\
\hline
$G_F = 1.11663787(6) \times 10^{-5}\,\text{GeV}^{-2}$ & & \\
\hline
\end{tabular}
\end{center}
\caption{Numerical input used to determine the initial conditions of
  the coupling constants. All values are taken from
  Ref.~\cite{PDG2020}.}
\label{tab:num}
\end{table}

In this section, we present numerical results for the running of the
scalar and gauge couplings. All the numerical inputs are taken from
Ref.~\cite{PDG2020}, see Tab.~\ref{tab:num}. We employ the expressions
given in Ref.~\cite{Degrassi:2012ry} to determine the initial
conditions for the strong coupling $g_s(M_Z) = 1.1626$, the top Yukawa
coupling $y_t(M_Z) = 0.9320$, and the quartic Higgs coupling
$\lambda_H(M_Z) = 0.5040$. We determine $g_1(M_Z)$ and $g_2(M_Z)$
directly via the relation
\begin{equation}
  \sin^2\theta_w (\mu) \equiv \frac{g_1^2(\mu)}{g_1^2(\mu) +
    g_2^2(\mu)} \,
\end{equation}
to find $g_1 (M_Z) = 0.3574$, $g_2 (M_Z) = 0.6517$. To determine
$y_\tau(M_Z) = 0.0102$ we used $m_\tau = 1.77686(12)\,$GeV, and the
relations
\begin{equation}\label{eq:convert_ytau}
  y_\tau = \frac{\sqrt{2} m_\tau}{v_{\rm EW}} \,, \qquad G_F = \frac{1}{\sqrt{2}v_{\rm EW}^2} \,.
\end{equation}
Note that $G_F$ is RG invariant, and we neglect the QED running of
$m_\tau$. We obtain $y_c(M_Z) = 0.0036$ and $y_b(M_Z) = 0.0164$ in the
six-flavor theory by four-loop QCD running and decoupling of the
corresponding quark masses and subsequent conversion using an
expression analogous to Eq.~\eqref{eq:convert_ytau}. As we are only
interested in the qualitative behaviour of our results, we neglect
uncertainties throughout. We solve the coupled system of RG equations
numerically, using the python package
\texttt{pywigxjpf}~\cite{Johansson:2015cca} and the Mathematica code
found in Ref. \cite{Wodtke:1999} for the numerical
evaluation of the Wigner $9j$ symbols.

In Fig.~\ref{fig:1loop:scalar} we show the one-loop running of all
scalar couplings for $\jp=3$, with scalar hypercharge
$Y_\varphi=0$. This case corresponds to the ``minimal scalar dark
matter'' (MSDM) scenario in Ref.~\cite{Cirelli:2009uv}, amended by the
two Higgs-portal couplings $\lambda_{\varphi H}$ and $\lambda_{\varphi
  H}'$. In the left panel, we assumed an initial condition of
$\lambda_i (M_Z) = 0.5$ for all four scalar couplings and the two
Higgs-portal couplings. The high-energy behaviour is largely
independent of these assumptions; in fact, even if the couplings are
all zero at the weak scale, large values get generated via weak
gauge-boson exchange (with the exception of $\lambda_{\varphi
  H}'$). The couplings quickly enter a non-perturbative regime and run
into a Landau pole around $10^{5}\,$GeV.

\begin{figure}
\centering
\includegraphics[width=0.49\textwidth]{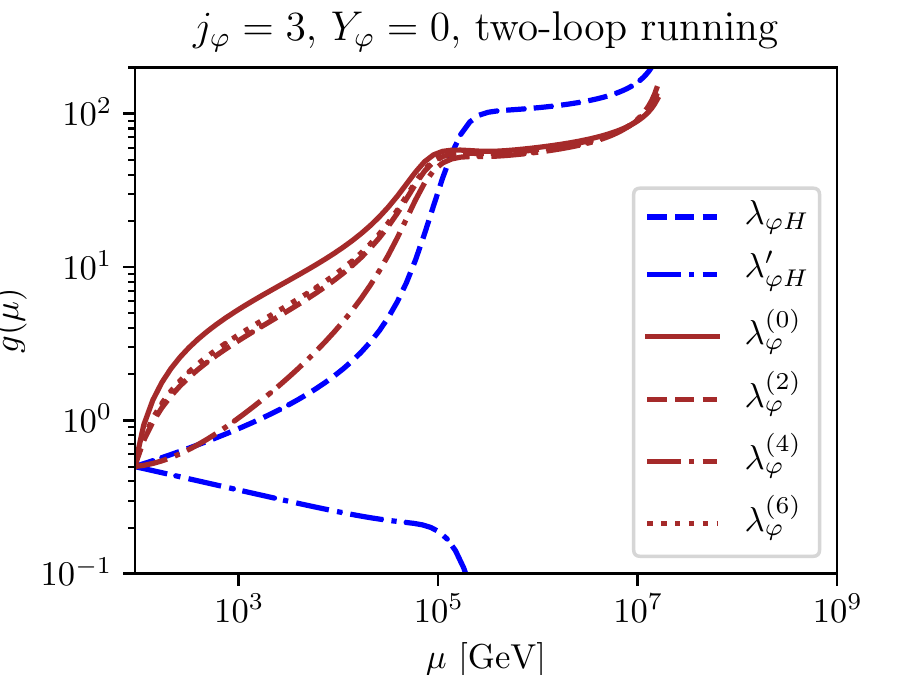}
\includegraphics[width=0.49\textwidth]{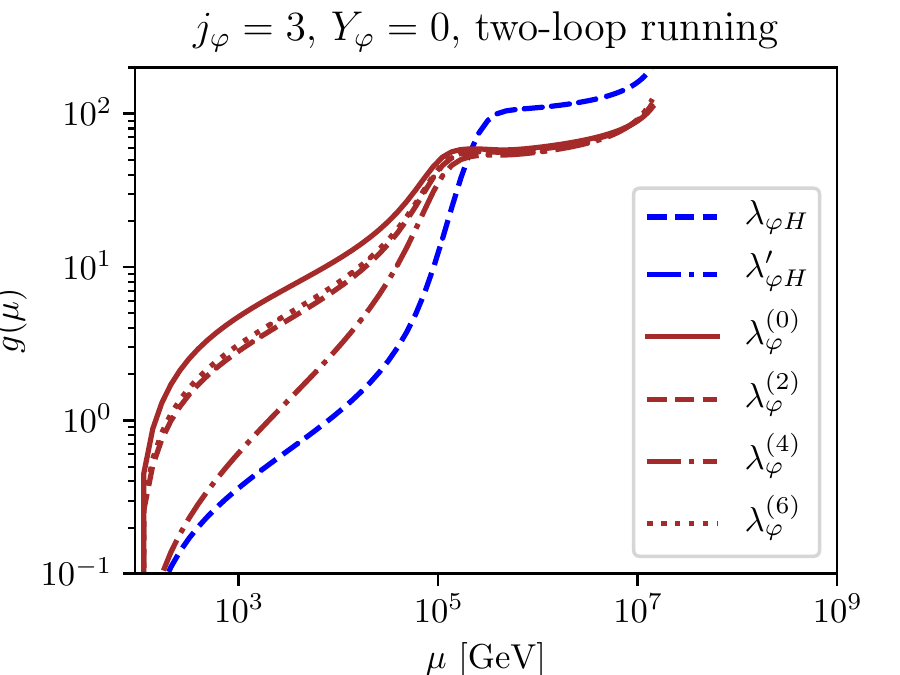}
\caption{Two-loop running of scalar quartic couplings for $\jp=3$ with
  $Y_\varphi=0$. The notation is the same as in
  Fig.~\ref{fig:1loop:scalar}.}
\label{fig:2loop:scalar}
\end{figure}

Next, we study the impact of the two-loop corrections to the RG
evolution of the scalar couplings in the same scenario, see
Fig.~\ref{fig:2loop:scalar}. Again, we display the results for the two
sets of initial conditions. Note that the Landau pole around
$\mu=10^5\,$GeV is shifted to the higher scale $\mu=10^7\,$GeV, with a
plateau-like behaviour in between. However, these features appear at
Fig.~\ref{fig:2loop:scalar}. Again, we display the results for the two
sets of initial conditions. Note that the Landau pole around
$\mu=10^5\,$GeV is shifted to the higher scale $\mu=10^7\,$GeV, with a
plateau-like behaviour in between. However, these features appear at
non-perturbative values for the coupling constants and should
therefore not be taken too literally. The only significant change is
that the ``triplet'' Higgs-portal coupling $\lambda_{\varphi H}'$ turns
out to be asymptotically free.

\begin{figure}
\centering
\includegraphics[width=0.49\textwidth]{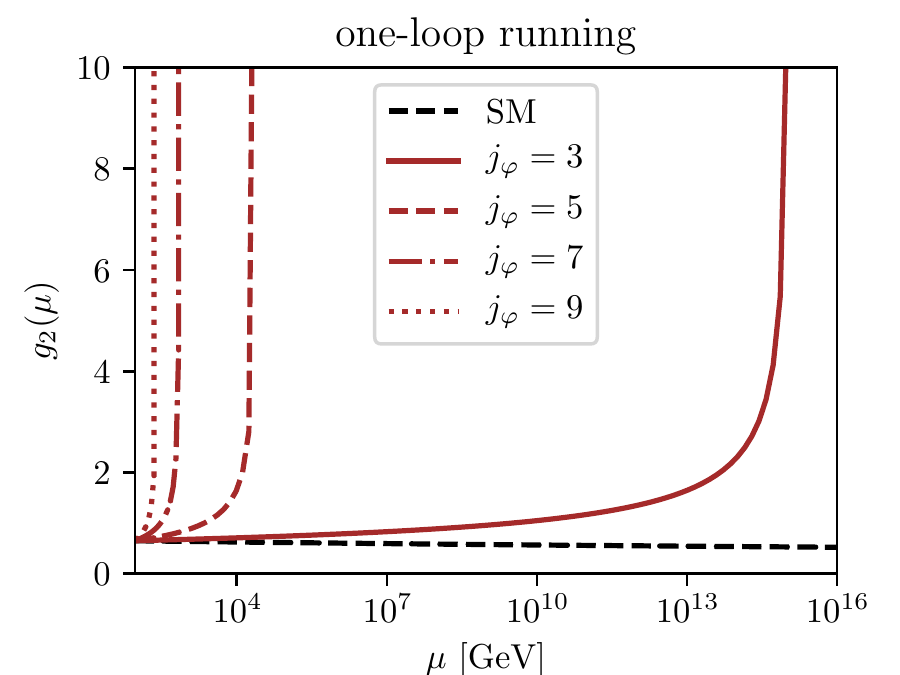}
\includegraphics[width=0.49\textwidth]{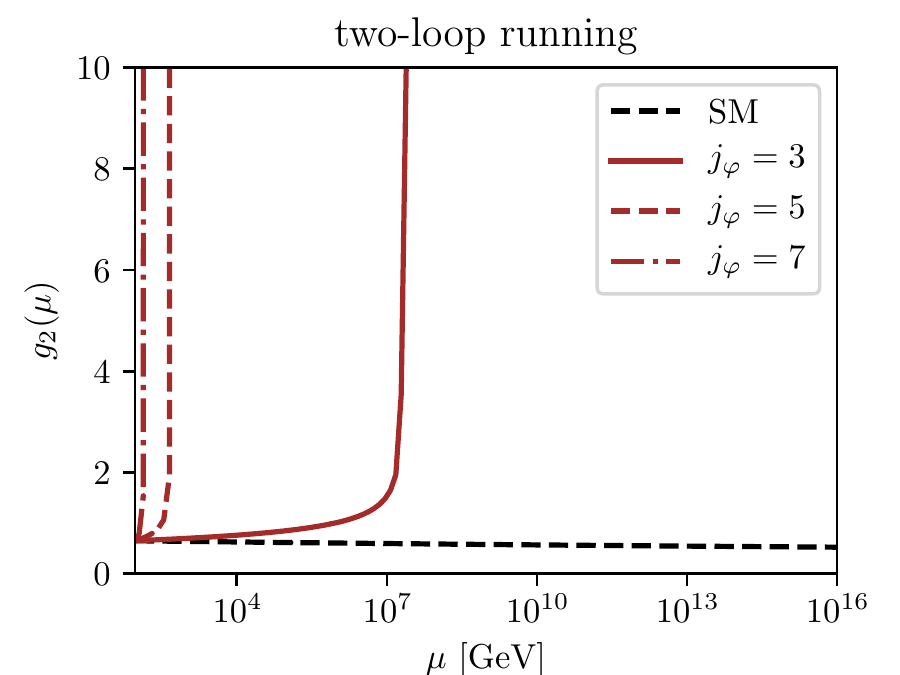}
\caption{Running of the $SU(2)$ gauge coupling $g_2$ at one-loop (left
  panel) and two-loop (right panel), for $Y_\varphi=0$. The black
  dashed line shows the SM result. The brown lines correspond to
  different representations of the complex scalar. Here, we assumed
  vanishing initial conditions for all non-SM scalar couplings at
  $\mu=M_Z$.}
\label{fig:g2}
\end{figure}

Finally, we examine the impact of the new scalar degrees of freedom on
the running of the SM couplings. We keep assuming vanishing
hypercharge for the new scalars, $Y_\varphi = 0$, and focus on the
evolution on the gauge coupling $g_2$ first. The running of $g_2$ is
displayed in Fig.~\ref{fig:g2}. In the left panel, we show the
one-loop evolution. We see that, at one-loop, the $SU(2)$ gauge
coupling exhibits a Landau pole at around $10^{15}\,$GeV for $\jp=3$
(MSDM), while for higher representations the Landau pole appears close
to or below the TeV scale. This behaviour has been qualitatively
described in, for instance, Ref.~\cite{Chao:2018xwz}. Looking at the
two-loop results in the right panel in Fig.~\ref{fig:g2}, we see that
the Landau pole for $\jp=3$ is significantly shifted down to
$10^7\,$GeV, while all other poles lie below the TeV
scale. Apparently, the SM extended by MSDM cannot be perturbative up
to the Planck scale.

\begin{figure}
\centering
\includegraphics[width=0.49\textwidth]{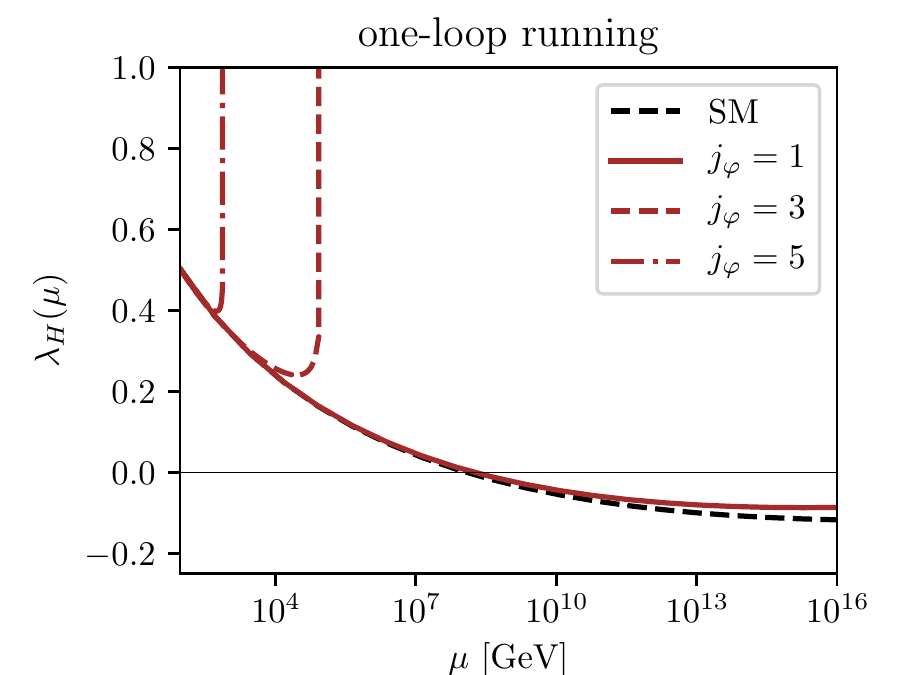}
\includegraphics[width=0.49\textwidth]{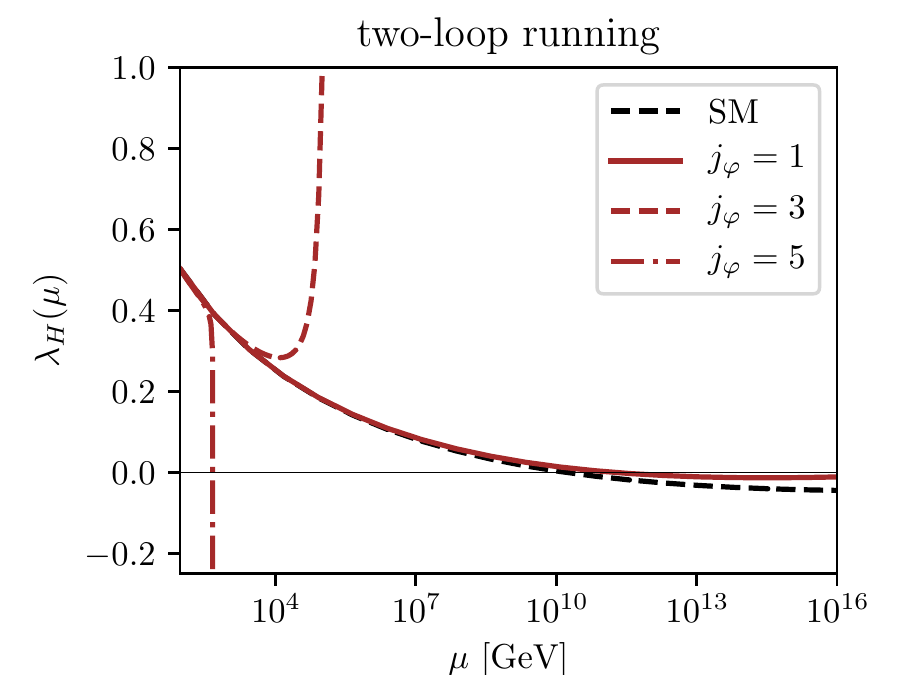}
\caption{Running of the quartic Higgs coupling $\lambda_H$ at one-loop
  (left panel) and two-loop (right panel), for $Y_\varphi=0$. The
  black dashed line shows the SM result. The brown lines correspond to
  different representations of the complex scalar. Here, we assumed
  vanishing initial conditions for all non-SM scalar couplings at
  $\mu=M_Z$.}
\label{fig:lamH}
\end{figure}

As our last example, we show the evolution of the quartic Higgs
coupling in Fig.~\ref{fig:lamH}. Again we display the one-loop results
in the left panel, and the two-loop results in the right panel. While
the SM evolution of $\lambda_H$ (black dashed line) is only marginally
affected by the presence of an additional scalar multiplet with
$j_\varphi=1$, higher representations lead to a drastic departure from
this picture. For $j_\varphi=3$ (MSDM), the Higgs quartic runs into a
Landau pole around $10^5\,$GeV, while the pole lies at the TeV scale
for $j_\varphi=5$. Interestingly, the two-loop results show that this
pole is in fact {\em negative}.

We relegate a more detailed discussion of the phenomenological
implications of these results to future work.

\section{Conclusions}
\label{sec:conclusions}

In this work, we constructed the form of the potential involving
four-point interaction of a complex scalar field furnishing a general
irreducible representation of the electroweak gauge group $SU(2)
\times U(1)$, in terms of Clebsch-Gordan coefficients. We presented
the beta functions determining the RG evolution of the scalar as well
as the SM couplings explicitly in terms of $SU(2)$ group invariants,
up to the two-loop level. As an important ingredient of our
calculation we proved a set of algebraic relations that we used to
express the results for the one- and two-loop Green's functions in
terms of our basis operators. For convenience, auxiliary files
containing the analytic results of the beta functions in the form of a
\texttt{python} module, as well as a \texttt{mathematica} package, are
available at
\begin{center}
\url{https://gitlab.com/complex-beta-function}\,.
\end{center}

Our results are completely general and might have applications in many
fields. As one example, we studied the RG flow of the self
interactions of scalar dark matter in minimal dark matter
models~\cite{Cirelli:2005uq}, and the impact of the scalar fields on
the RG evolution of the SM couplings. Moreover, the beta functions
will be a necessary ingredient in the RG analysis of scalar dark
matter interacting via higher dimension
operators~\cite{Bishara:2018vix, Brod:2017bsw}.

A generalization of our results in this direction would be to consider
the self interactions of fermionic dark matter. This case is more
complicated since the interactions start at mass dimension six, and
additional Fierz relations associated with the Dirac-matrix structure
restrict the form of all possible operators. This investigation is
relegated to future work.

\section*{Acknowledgments}
We thank Emmanuel Stamou and Jure Zupan for suggestions and comments
on the manuscript, and Jared Evans and Florian Hanisch for
discussions. JB acknowledges support in part by DOE grant
DE-SC0020047.

\appendix

\section{Analytic checks of our calculation}\label{app:checks}

As a check of our results we used a generalized $R_\xi$ gauge for the
$W$, $B$ and $G$ fields and verified that all beta functions are
gauge-parameter independent. For completeness, we provide here the
gauge-fixing and ghost terms in our Lagrangian:
\begin{equation}\begin{split}\label{eq:lagfull}
 \mathcal{L}_{\text{gf}} + \mathcal{L}_{\text{ghost}} =& 
	-\frac{1}{2\xi_W}\left(\partial_\mu W^{a\mu}\right)^2 -
 \frac{1}{2\xi_B}\left(\partial_\mu B^\mu\right)^2 
	- \frac{1}{2\xi_G}\left(\partial_\mu G^{A\,\mu}\right)^2\\
 & + \partial_\mu \bar{u}_W^a \partial^\mu u_W^a +
 g_2\epsilon^{abc}(\partial^\mu \bar{u}_W^a)W^b_\mu u_W^c \\
 & + \partial_\mu \bar{u}_G^A \partial^\mu u_G^A -
 g_sf^{ABC}(\partial^\mu \bar{u}_G^A)G^B_\mu u_G^C\,.
\end{split}\end{equation}

As a second consistency check of our calculation, we verified that all
two-loop counterterms are local, i.e. they do not contain any explicit
logarithms of the renormalization scale $\mu$. As a third check of our
calculation, we derive the explicit expressions of the beta function
in terms of the coupling counterterms (see below). The finiteness of
the beta function as $\epsilon \to 0$ yields consistency relations
that allow to calculate the quadratic pole of the two-loop coupling
renormalization constants in terms of the one-loop results. These
quadratic poles are in full agreement with the results of our
calculcation. For completeness, we provide the expressions for the
quadratic poles in App.~\ref{app:ren}.

In the remainder of this section, we derive the relation between the
beta function and the residua of the coupling renormalization
constants, as well as the relation between the linear one-loop poles
and the quadratic two-loop poles. As above, we denote the all
couplings generically by a coupling vector $g_i$. The bare couplings
$g_{i,0}$ are expressed in terms of the renormalized couplings $g_i$
as $g_{i,0} = \mu^{a_i\epsilon} Z_{g_i} g_i$, where $a_i = 1$ if $g_i$
is a gauge or Yukawa coupling, and $a_i=2$ if $g_i$ is a scalar
coupling (the coefficients $a_i$ are chosen such that all couplings
remain dimensionless in $d$ space-time dimensions). Here, $\mu$ is the
renormalization scale, and the $Z_{g_i}$ are the coupling
renormalization constants. We expand the $Z_{g_i}$ by order of pole
as
\begin{equation}
Z_{g_i}=1+\sum_{l=1}^\infty \frac{1}{\epsilon^l}Z_{g_i,l} \,,
\end{equation}
and use standard methods~\cite{Muta:2010xua} to express the beta
function in terms of the derivatives of the linear poles of the
coupling counterterms:
\begin{equation}\label{eq:beta:ep0}
\beta_{i}(g_j, \epsilon) = - a_i g_i \epsilon 
+ g_i \sum_k a_k g_k\frac{\partial Z_{g_i,1}}{\partial g_k}\,.
\end{equation}
The fact that the $1/\epsilon$ contributions to the beta function have
to cancel leads to the following consistency condition on the
counterterms:
\begin{equation}\label{eq:beta:ep1}
\sum_k a_k g_k\frac{\partial Z_{g_i,2}}{\partial g_k} =
\sum_k a_k g_k Z_{g_i,1}\frac{\partial Z_{g_i,1}}{\partial g_k} 
+ \sum_{km}a_k g_k g_m \frac{\partial Z_{g_m,1}}{\partial g_k}
  \frac{\partial Z_{g_i,1}}{\partial g_m} \,.
\end{equation}
Further conditions can be derived by requiring the cancelation of the
higher poles; however, they do not lead to additional constraints on
the two-loop counterterms. The relation~\eqref{eq:beta:ep1} is
made more explicit by expanding the counterterms by loop-order,
\begin{equation}
Z_{g_i} = 1+\sum_n \delta Z_{g_i}^{(n)} 
= 1+\sum_n\sum_l\frac{1}{\epsilon^l}\delta Z_{g_i,l}^{(n)} \,.
\end{equation}
Keeping only terms at two-loop order, and using the fact that the
counterterms are polynomials in the couplings, we arrive at
\begin{equation}
\sum_ka_k g_k \frac{\partial (\delta Z_{g_i,2}^{(2)})}{\partial g_k}
= 4\delta Z_{g_i,2}^{(2)} \,,
\qquad
\sum_ka_k g_k \frac{\partial (\delta Z_{g_i,2}^{(1)})}{\partial g_k}
= 2\delta Z_{g_i,2}^{(1)} \,.
\end{equation}
We then rewrite Eq.~\eqref{eq:beta:ep1} as
\begin{equation}
\delta Z_{g_i,2}^{(2)} = \frac{1}{2} \Big(\delta Z_{g_i,1}^{(1)}\Big)^2
 + \frac{1}{2} \sum_k g_k \delta Z_{g_k,1}^{(1)} 
\frac{\partial (\delta Z_{g_i,1}^{(1)})}{\partial g_k} \,.
\end{equation}
We checked explicitly that this relation is satisfied for all our
coupling counterterms.

\section{Renormalization constants}\label{app:ren}

In this appendix we collect all renormalization constants that were
needed in intermediate steps of the calculation, namely, all field and
artificial-mass counterterms. The $1/\epsilon$ pole parts of the
coupling counterterms give rise to the beta functions, as explained in
App.~\ref{app:checks}, and are not repeated here. For completeness,
however, we show the $1/\epsilon^2$ pole parts. The
$\overline{\mathrm{MS}}$ scheme is used throughout. For the one-loop
field renormalization constants we find
\begin{align}\label{eq:field1l}
\delta Z_{\varphi}^{(1)} & = 
	 \frac{g_1^2}{16\pi^2\epsilon}\frac{Y_\varphi^2}{4}(3-\xi_B)
        +\frac{g_2^2}{16\pi^2\epsilon}\mathcal{J}(\jp)(3-\xi_W) \,,\\
\delta Z_{H}^{(1)} & = \frac{g_1^2}{16\pi^2\epsilon} \frac{1}{4}\left(3-\xi_B\right)
	           + \frac{g_2^2}{16\pi^2\epsilon}\frac{3}{4}\left(3-\xi_W\right)
	           - \frac{1}{16\pi^2\epsilon}\big[
	             3\big(y_t^2+y_b^2+y_c^2 \big) + y_\tau^2\big]\,,\\
\delta Z_B^{(1)} & = 
	-\frac{g_1^2}{16\pi^2\epsilon} 
        \bigg(  \frac{Y_\varphi^2}{12}\mathcal{D}(\jp)
              + \frac{20}{9} n_g + \frac{1}{6} \bigg) \,,\\
\delta Z_W^{(1)} &=
	\frac{g_2^2}{16\pi^2\epsilon}\left(
		\frac{13}{3} - \frac{1}{6} - \xi_W
		- \frac{1}{9}\mathcal{J}(\jp)\mathcal{D}(\jp)
                - \frac{4}{3} n_g \right)\,,\\
\delta Z_{G}^{(1)} &=
        \frac{g_s^2}{16\pi^2\epsilon}\left(
                \frac{13}{2}
              - \frac{3}{2} \xi_G
              - n_g \right)\,,\\
\delta Z_{u_W}^{(1)} &=
	\frac{g_2^2}{16\pi^2\epsilon} \frac{1}{2} 
        \left(3 - \xi_W \right)\,,\\
\delta Z_{Q_{L,i}}^{(1)} &=
       -\frac{g_1^2}{16\pi^2\epsilon}\frac{\xi_B}{36}
       -\frac{g_2^2}{16\pi^2\epsilon}\frac{3\xi_W}{4}
       -\frac{g_s^2}{16\pi^2\epsilon} C_F \xi_G
       -\frac{1}{16\pi^2\epsilon}
        \left(\frac{y_{u_i}^2}{2}+\frac{y_{d_i}^2}{2}\right)\,,\\
\delta Z_{u_{R,i}}^{(1)} &=
       -\frac{g_1^2}{16\pi^2\epsilon}\frac{4\xi_B}{9}
       -\frac{g_s^2}{16\pi^2\epsilon} C_F \xi_G
       -\frac{y_{u_i}^2}{16\pi^2\epsilon}\,,\\
\delta Z_{d_{R,i}}^{(1)} &=
       -\frac{g_1^2}{16\pi^2\epsilon}\frac{\xi_B}{9}
       -\frac{g_s^2}{16\pi^2\epsilon} C_F \xi_G
       -\frac{y_{d_i}^2}{16\pi^2\epsilon}\,,\\
\delta Z_{L_{L,i}}^{(1)} &=
       -\frac{g_1^2}{16\pi^2\epsilon}\frac{\xi_B}{4}
       -\frac{g_2^2}{16\pi^2\epsilon}\frac{3\xi_W}{4}
       -\frac{1}{16\pi^2\epsilon}
        \frac{y_{\ell_i}^2}{2}\,,\\
\delta Z_{\ell_{R,i}}^{(1)} &=
       -\frac{g_1^2}{16\pi^2\epsilon} \xi_B
       -\frac{y_{\ell_i}^2}{16\pi^2\epsilon}\,,
\end{align}
where $C_F = 4/3$. At one-loop, the artificial-mass counterterms are
\begin{align}
  \delta Z_{M_\text{IRA}^2,\varphi}^{(1)} 
  & = \frac{1}{16\pi^2\epsilon} 
      {\sum_J}^\prime \lambda^{(J)} \frac{\mathcal{D}(J)}{\mathcal{D}(\jp)}
      + \frac{1}{16\pi^2\epsilon} \frac{\lambda_{\varphi H}}{2} \,, \\
  \delta Z_{M_\text{IRA}^2,H}^{(1)} 
  & =  \frac{\lambda_{\varphi H}}{16\pi^2\epsilon} \frac{\mathcal{D}(\jp)}{4}
     + \frac{\lambda_H}{16\pi^2\epsilon} \frac{3}{2}
     - \frac{3}{16\pi^2\epsilon}\big[
       3\big(y_t^2+y_b^2+y_c^2 \big) + y_\tau^2\big]\,,\\
  \delta Z_{M_\text{IRA}^2,B}^{(1)}
  & = \frac{g_1^2}{16\pi^2\epsilon} 
      \bigg(  \frac{Y_\varphi^2}{12}\mathcal{D}(\jp)
            - \frac{40}{9} n_g + \frac{1}{6} \bigg) \,,\\
  \delta Z_{M_\text{IRA}^2,W}^{(1)} 
  & = \frac{g_2^2}{16\pi^2\epsilon}
      \bigg(\frac{1}{9} \mathcal{J}(\jp)\mathcal{D}(\jp) 
      - \frac{1}{2}\xi_W - \frac{29}{6}
      - \frac{8}{3} n_g
      + \frac{1}{6} \bigg) \,,\\
  \delta Z_{M_\text{IRA}^2,G}^{(1)} &=
        \frac{g_s^2}{16\pi^2\epsilon}\left(
              - \frac{3}{4} 
              - \frac{9}{4} \xi_G
              - 3 n_g \right)\,,\\
  \delta Z_{M_\text{IRA}^2,u_W}^{(1)}
  & = \frac{g_2^2}{16\pi^2\epsilon} 
      \frac{1}{2} \left(\xi_W - 3\right) \,.
\end{align}
We find the following quadratic poles for the two-loop contributions
to the scalar and gauge coupling counterterms:
\begin{align}
  \delta Z_{g_1}^{(2)} &= 
    \quad \frac{g_1^4}{(16\pi^2)^2\epsilon^2}
            \bigg(   \frac{Y_\varphi^4}{384} \mathcal{D}(\jp)^2
                   + \frac{Y_\varphi^2}{96} \mathcal{D}(\jp)
                   + \frac{5}{36} Y_\varphi^2 \mathcal{D}(\jp) n_g
            \bigg)\,,\\
  \delta Z_{g_2}^{(2)} &= 
    \quad \frac{g_2^4}{(16\pi^2)^2\epsilon^2}
        \bigg(   \frac{1}{216} \mathcal{D}(\jp)^2 \mathcal{J}(\jp)^2
               + \frac{1}{72} \mathcal{D}(\jp) \mathcal{J}(\jp)
                 \big( 8 n_g - 43 \big) \bigg)\,,\\
  \delta Z_{y_t}^{(2)} &= 
    \quad - \frac{17}{576} \frac{g_1^4}{(16\pi^2)^2\epsilon^2}
            \mathcal{D}(\jp) Y_\varphi^2
          - \frac{1}{16} \frac{g_2^4}{(16\pi^2)^2\epsilon^2}
            \mathcal{J}(\jp) \mathcal{D}(\jp)\,,\\
  \delta Z_{y_b}^{(2)} &= 
    \quad - \frac{5}{576} \frac{g_1^4}{(16\pi^2)^2\epsilon^2}
            \mathcal{D}(\jp) Y_\varphi^2
          - \frac{1}{16} \frac{g_2^4}{(16\pi^2)^2\epsilon^2}
            \mathcal{J}(\jp) \mathcal{D}(\jp)\,,\\
  \delta Z_{y_c}^{(2)} &= 
    \quad - \frac{17}{576} \frac{g_1^4}{(16\pi^2)^2\epsilon^2}
            \mathcal{D}(\jp) Y_\varphi^2
          - \frac{1}{16} \frac{g_2^4}{(16\pi^2)^2\epsilon^2}
            \mathcal{J}(\jp) \mathcal{D}(\jp)\,,\\
  \delta Z_{y_\tau}^{(2)} &= 
    \quad - \frac{5}{64} \frac{g_1^4}{(16\pi^2)^2\epsilon^2}
            \mathcal{D}(\jp) Y_\varphi^2
          - \frac{1}{16} \frac{g_2^4}{(16\pi^2)^2\epsilon^2}
            \mathcal{J}(\jp) \mathcal{D}(\jp)\,,\\
  \begin{split}
  \lambda_i^{(J)} \delta Z_{\lambda_i^{(J)}}^{(2)} &= 
		\frac{1}{(16\pi^2)^2\epsilon^2}
                \bigg(   
			+ g_1^6 \Lambda^{(J)(2)}_{i,60}
			+ g_1^4g_2^2 \Lambda^{(J)(2)}_{i,42}
			+ g_2^2g_2^4 \Lambda^{(J)(2)}_{i,24}
			+ g_2^6 \Lambda^{(J)(2)}_{i,06}\\
       & \hspace{6.4em} + g_1^4\Lambda^{(J)(2)}_{i,40}
                        + g_1^2g_2^2 \Lambda^{(J)(2)}_{i,22}
			+ g_2^4\Lambda^{(J)(2)}_{i,04}\\
       & \hspace{6.4em} + g_1^2 \Lambda^{(J)(2)}_{i,20}
			+ g_2^2 \Lambda^{(J)(2)}_{i,02}
                        + \Lambda^{(J)(2)}_{i,00}
                \bigg)\,,
  \end{split}
\end{align}
with $\lambda_i = \lambda_{\varphi}^{(J)}, \lambda_{\varphi H},
\lambda_{\varphi H}^\prime, \lambda_{H}$, and coefficients
\begin{align}
  \Lambda^{(J)(2)}_{\varphi,60} &=   Y_\varphi^6 \bigg(   \frac{\mathcal{D}(\jp)}{16} - \frac{9}{16} \bigg)
                        + Y_\varphi^4 \bigg(   \frac{1}{8} + \frac{5}{3} n_g \bigg) \,,\\
  \begin{split}
  \Lambda^{(J)(2)}_{\varphi,42} &=
        Y_\varphi^4 \left[  \mathcal{J}(\jp) \left(\frac{9}{4}
                                - \frac{1}{4} \mathcal{D}(\jp)\right)
                        - \mathcal{J}(J) \left(\frac{9}{4}
                                - \frac{1}{8} \mathcal{D}(\jp)\right) \right] \\
       & \quad + Y_\varphi^2 \bigg(   \frac{1}{4} + \frac{10}{3} n_g \bigg) 
                         \Big( \mathcal{J}(J) - 2 \mathcal{J}(\jp)\Big) \,,
  \end{split}\\
  \begin{split}
  \Lambda^{(J)(2)}_{\varphi,24} &=
        Y_\varphi^2 \bigg[  \mathcal{J}(\jp) \left(  9 \mathcal{J}(\jp)
                                         - \frac{1}{3} \mathcal{D}(\jp) \mathcal{J}(\jp)
                                         + 26 - 4 n_g \right) \\
  & \hspace{3em}   - \mathcal{J}(J) \left( \frac{9}{4} \mathcal{J}(J)
                                         - \frac{1}{6} \mathcal{D}(\jp) \mathcal{J}(\jp)
                                         + 13 - 2 n_g \right) \bigg] \,,
  \end{split}\\
  \begin{split}
  \Lambda^{(J)(2)}_{\varphi,06} &= \left( \frac{\mathcal{J}(J)}{2} - \mathcal{J}(\jp) \right)
                              \left( \frac{\mathcal{J}(J)}{2} - \mathcal{J}(\jp) + \frac{1}{2} \right) \\
  & \quad \times              \left( \frac{4}{3} \mathcal{J}(\jp) \mathcal{D}(\jp)
                                     - 36 \mathcal{J}(\jp) - 86 + 16 n_g \right) \,,
  \end{split}\\
  \Lambda^{(J)(2)}_{\varphi,40} &=
        \frac{Y_\varphi^4}{4} \left( 9 \lambda_\varphi^{(J)}
                                   - \frac{\mathcal{D}(\jp)}{4} \lambda_\varphi^{(J)} 
                                   + 3 {\sum_{J_1}}^\prime \lambda_\varphi^{(J_1)}
                                     \frac{\mathcal{D}(J_1)}{\mathcal{D}(\jp)} \right)
              - Y_\varphi^2 \bigg(   \frac{1}{8} + \frac{5}{3} n_g \bigg) 
                                 + \frac{3\lambda_{\varphi H}}{8}Y_\varphi^2\,,\\
  \begin{split}
  \Lambda^{(J)(2)}_{\varphi,22} &= Y_{\varphi}^2 {\sum_{J_1}}^\prime \lambda_\varphi^{(J_1)}
                        \frac{\mathcal{D}(J_1)}{\mathcal{D}(\jp)}
        \bigg[ \frac{3}{2} \frac{\mathcal{J}(J_1)\mathcal{J}(J)}
                                {\mathcal{J}(\jp)}
               -3 \big( \mathcal{J}(J_1) + \mathcal{J}(J) \big)
               +6 \mathcal{J}(\jp) \bigg]  \\
                     & \quad + Y_{\varphi}^2 \lambda_\varphi^{(J)}
                       \bigg( 6 \mathcal{J}(\jp) + 3 \mathcal{J}(J)\bigg)
                             + \frac{3}{8} Y_{\varphi} \lambda_{\varphi H}'
                       \big( \mathcal{J}(J) - 2 \mathcal{J}(\jp) \big) \,,
  \end{split}\\
  \begin{split}
  \Lambda^{(J)(2)}_{\varphi,04} &= \lambda_\varphi^{(J)}
        \bigg[     \mathcal{J}(\jp) \left(  \frac{37}{2}
                                        + 36 \mathcal{J}(\jp)
                                        - 4 n_g
                                        - \frac{1}{3} \mathcal{D}(\jp) \mathcal{J}(\jp) \right) \\
  & \hspace{4em} + \mathcal{J}(J) \left(  3 
                                        + \frac{3}{2} \mathcal{J}(J)
                                        - 6 \mathcal{J}(\jp) \right) \bigg]\\
  & \quad + {\sum_{J_1,J_2}}^\prime K(J_1, J_2, J) \lambda_\varphi^{(J_2)} 
                               \Big[ 6 \mathcal{J}(J_1)^2 
                                      - 24 \mathcal{J}(\jp) \mathcal{J}(J_1) \Big]  
	  + \frac{3\lambda_{\varphi H}}{2}\mathcal{J}(\jp)\\
  & \quad + {\sum_{J_1}}^\prime \lambda_\varphi^{(J_1)} \frac{\mathcal{D}(J_1)}{\mathcal{D}(\jp)}
                      \bigg[ 6 \mathcal{J}(\jp)
                            +12 \mathcal{J}(\jp)^2
                            -3 \mathcal{J}(J)
                            -3 \mathcal{J}(J_1)
                            +\frac{3}{2} 
                             \frac{\mathcal{J}(J_1)\mathcal{J}(J)}
                                  {\mathcal{J}(\jp)} \bigg] \,,
  \end{split}\\
  \begin{split}
  \Lambda^{(J)(2)}_{\varphi,20} &=
        - \frac{9Y_\varphi^2}{8} \left(4 {\sum_{J_1,J_2}}^\prime K(J_1, J_2, J)
                                    \lambda_\varphi^{(J_1)}\lambda_\varphi^{(J_2)} 
                                  + \big(\lambda_\varphi^{(J)}\big)^2\right) \\
        & \quad - \frac{3}{128} \big( \lambda_{\varphi H}' \big)^2 
                       \Big( \mathcal{J}(J) - 2 \mathcal{J}(\jp) \Big)
                       \big( 1 + 2 Y_\varphi^2 \big) 
		       -\frac{3\lambda_{\varphi H}^2}{16}\big(1+2Y_\varphi^2\big)\,,
  \end{split}\\
  \begin{split}
  \Lambda^{(J)(2)}_{\varphi,02} &=
                - \mathcal{J}(\jp) \bigg[ 18 {\sum_{J_1,J_2}}^\prime K(J_1, J_2, J)
                                       \lambda_\varphi^{(J_1)}\lambda_\varphi^{(J_2)} 
                                     + \frac{9}{2} \big(\lambda_\varphi^{(J)}\big)^2 \bigg] \\
        & \quad - \frac{3}{16} \big( \lambda_{\varphi H}' \big)^2 
                \bigg( \mathcal{J}(\jp) + \frac{3}{8} \bigg)
                \Big( \mathcal{J}(J) - 2 \mathcal{J}(\jp) \Big)
		-\frac{3\lambda_{\varphi H}^2}{2}\bigg(
		\mathcal{J}(\jp)+\frac{3}{8}\bigg)\,,
  \end{split}\\
  \begin{split}
  \Lambda^{(J)(2)}_{\varphi,00} &= 
        \frac{\big(\lambda_\varphi^{(J)}\big)^3}{4} 
        + {\sum_{J_1,J_2}}^\prime K(J_1, J_2, J)
        \lambda_\varphi^{(J_1)}\lambda_\varphi^{(J_2)}
        \big( \lambda_\varphi^{(J_1)} + \lambda_\varphi^{(J)}\big)  \\
  & \quad + 4 {\sum_{J_1,J_2,J_3,J_4}}^\prime K(J_1, J_2, J_3) K(J_3, J_4, J) 
        \lambda_\varphi^{(J_1)} \lambda_\varphi^{(J_2)} \lambda_\varphi^{(J_4)} \\
  & \quad + \frac{\lambda_\varphi^{(J)} \big(\lambda_{\varphi H}' \big)^2}{32}
            \big( \mathcal{J}(J) - \mathcal{J}(\jp) \big) \\
  & \quad + \frac{\big(\lambda_{\varphi H}' \big)^2}{32 \mathcal{D}(\jp)}
            {\sum_{J_1}}^\prime \lambda_\varphi^{(J_1)} 
            \bigg(   \frac{\mathcal{J}(J)\mathcal{J}(J_1)\mathcal{D}(J_1)}{\mathcal{J}(\jp)} \\
  & \hspace{10em}   - 2 \big( \mathcal{J}(J_1) + \mathcal{J}(J) \big) \mathcal{D}(J_1)
                   + 4 \mathcal{D}(J_1) \mathcal{J}(\jp) \bigg) \\
  & \quad + \frac{\big(\lambda_{\varphi H}' \big)^2}{64}
            \big( \lambda_H + 6 y_t^2 + 6 y_t^2 + 6 y_t^2 + 2 y_\tau^2 \big)
            \big( \mathcal{J}(J) - 2 \mathcal{J}(\jp)\big)\\ 
  & \quad + \frac{\lambda_{\varphi H}\big(\lambda_{\varphi H}'\big)^2}{32}
	    \big(\mathcal{J}(J)-\mathcal{J}(\jp)\big)
	    + \frac{3}{8}\lambda_{\varphi H}\lambda_H
	    + \frac{\lambda_{\varphi H}^3}{8}\\
  & \quad + \frac{\lambda_{\varphi H}^2}{4}\bigg(\frac{3\lambda_{\varphi}^{(J)}}{2}
	  + 3y_t^2+3y_b^2+3y_c^2+y_\tau^2\bigg)
	  + \frac{\lambda_{\varphi H}^2}{2}
            {\sum_{J_1}}'\lambda_{\varphi}^{(J)}
            \frac{\mathcal{D}(J)}{\mathcal{D}(\jp)}\,.
  \end{split}
\end{align}
For the Higgs-portal couplings we find
\begin{align}
  \Lambda^{(2)}_{\varphi H,60} &= 
          Y_\varphi^4 \bigg(\frac{1}{8}\mathcal{D}(\jp) - \frac{9}{16} \bigg)
        - Y_\varphi^2 \bigg(   \frac{5}{16} - \frac{10}{3} n_g \bigg) \,,\\
  \Lambda^{(2)}_{\varphi H,42} &=
        - Y_\varphi^2 \bigg(\frac{9}{4}\mathcal{J}(\jp) + \frac{27}{16} \bigg) \,,\\
  \Lambda^{(2)}_{\varphi H,24} &= 
        - \frac{9}{4} \mathcal{J}(\jp) \big( 1 + Y_\varphi^2 \big) \,,\\
  \Lambda^{(2)}_{\varphi H,06} &= 
        - \mathcal{J}(\jp) \bigg(   \frac{199}{4} - 8 n_g
                                + 9 \mathcal{J}(\jp)
                                - \frac{2}{3} \mathcal{J}(\jp) \mathcal{D}(\jp) \bigg) \,,\\
  \begin{split}
  \Lambda^{(2)}_{\varphi H,40} &=
          Y_\varphi^2 \bigg(   \frac{9}{8} \lambda_H
                           + \frac{1}{4} \big( 9 y_t^2 + 9 y_b^2 + 9 y_c^2 + 3 y_\tau^2 \big)
			   + \frac{3}{4} {\sum_{J_1}}^\prime \lambda_{\varphi}^{(J_1)}
                             \frac{\mathcal{D}(J_1)}{\mathcal{D}(\jp)} \bigg)\\
  & \quad + Y_\varphi^2 \lambda_{\varphi H} 
            \bigg( \frac{5}{4} - \frac{\mathcal{D}(\jp)}{32} \bigg)
          + \frac{Y_\varphi^4}{32} \lambda_{\varphi H} \big( 15 + 5 \mathcal{D}(\jp) \big)
          + \lambda_{\varphi H} \bigg[
                   \frac{25}{32}
                 - \frac{5}{6} n_g \big( 1 + Y_\varphi^2 \big) \bigg] \,,
  \end{split}\\
  \Lambda^{(2)}_{\varphi H,22} &= 
            \frac{3\mathcal{J}(\jp)}{2} \lambda_{\varphi H}'Y_\varphi
	  + \lambda_{\varphi H}\left(\frac{45}{16}+\frac{27}{16}Y_\varphi^2
	  +\frac{9}{4}\mathcal{J}(\jp)+\frac{15}{4}\mathcal{J}(\jp)Y_\varphi^2\right)\,,\\
  \begin{split}
  \Lambda^{(2)}_{\varphi H,04} &= 
          \mathcal{J}(\jp) \bigg(   \frac{9}{2} \lambda_H
                                + 3 \big( 3 y_t^2 + 3 y_b^2 + 3 y_t^2 + y_\tau^2 \big)
				+ 3 {\sum_{J_1}}^\prime \lambda_{\varphi}^{(J_1)}
                                    \frac{\mathcal{D}(J_1)}{\mathcal{D}(\jp)} \bigg)\\
  & \quad + \lambda_{\varphi H}\bigg[\frac{393}{32} 
	  - \frac{3}{2} n_g + \mathcal{J}(\jp)\bigg(
	  19 - \frac{1}{2}N_{L_L} - \frac{3}{2}N_{Q_L}
	  - \frac{1}{8}\mathcal{D}(\jp)\bigg) \\
  & \hspace{5em} + \frac{15}{2}\mathcal{J}(\jp)^2 + \frac{5}{6}\mathcal{J}(\jp)^2\mathcal{D}(\jp)\bigg]\,,
  \end{split}\\
  \begin{split}
  \Lambda^{(2)}_{\varphi H,20} &= 
          - \frac{9\big( \lambda_{\varphi H}' \big)^2}{64}
            \mathcal{J}(\jp) \big( 1 + Y_\varphi^2 \big)
          - \frac{9\lambda_{\varphi H}^2}{16} \big( 1 + Y_\varphi^2 \big)\\
  & \quad - \frac{\lambda_{\varphi H}}{8} 
            \big( 35 y_t^2 + 23 y_b^2 + 35 y_c^2 + 21 y_\tau^2 \big)
          - \frac{3\lambda_{\varphi H}}{4} Y_\varphi^2
            \big( 3 y_t^2 + 3 y_b^2 + 3 y_c^2 + y_\tau^2 \big)\\
  & \quad - \frac{9}{8} \lambda_{\varphi H} \lambda_{H} \big( 2 + Y_\varphi^2 \big)
          - \frac{3}{4} \big( 1 + 2 Y_\varphi^2 \big)
            \lambda_{\varphi H} {\sum_{J_1}}^\prime  \lambda_\varphi^{(J_1)}
            \frac{\mathcal{D}(J_1)}{\mathcal{D}(\jp)} \,,
  \end{split}\\
  \begin{split}
  \Lambda^{(2)}_{\varphi H,02} &= 
          - \frac{\big( \lambda_{\varphi H}' \big)^2}{64}
            \mathcal{J}(\jp) \big( 27 + 36 \mathcal{J}(\jp) \big)
          - \frac{\lambda_{\varphi H}^2}{16} \big( 27 + 36 \mathcal{J}(\jp) \big)\\
  & \quad - \lambda_{\varphi H}
            \big( 9 y_t^2 + 9 y_b^2 + 9 y_c^2 + 3 y_\tau^2 \big)
            \bigg( \frac{9}{8} + \mathcal{J}(\jp) \bigg)\\
  & \quad - \frac{1}{4} \lambda_{\varphi H} \lambda_{H} \big( 27 + 18 \mathcal{J}(\jp) \big)
          - \bigg( \frac{9}{4} + 6 \mathcal{J}(\jp) \bigg)
            \lambda_{\varphi H} {\sum_{J_1}}^\prime  \lambda_\varphi^{(J_1)}
            \frac{\mathcal{D}(J_1)}{\mathcal{D}(\jp)} \,,
  \end{split}\\
  \begin{split}
  \Lambda^{(2)}_{\varphi H,00} &= 
            \bigg(   \frac{3\big(\lambda_{\varphi H}'\big)^2}{16} \mathcal{J}(\jp) 
                   + 3\lambda_{\varphi H} \lambda_{H}
                   + \frac{3\lambda_{\varphi H}^2}{4}
                   + \lambda_{\varphi H} {\sum_{J_1}}^\prime \lambda_\varphi^{(J_1)}
                                       \frac{\mathcal{D}(J_1)}{\mathcal{D}(\jp)}
            \bigg)\\
  & \quad   \times \big( 3 y_t^2 + 3 y_b^2 + 3 y_c^2 + y_\tau^2 \big)\\
  & \quad - 12 \lambda_{\varphi H} g_s^2 \big( y_t^2 + y_b^2 + y_c^2 \big)
          + \frac{\lambda_{\varphi H}}{4}
            \big( 9 y_t^4 + 9 y_b^4 + 9 y_c^4 - 5 y_\tau^4 \big)\\
  & \quad + 6 \lambda_{\varphi H} y_\tau^2 \big( y_t^2 + y_b^2 + y_c^2 \big)
          + \frac{\lambda_{\varphi H}}{2} 
            \big( 27 y_t^2 y_b^2 + 36 y_t^2 y_c^2 + 36 y_b^2 y_c^2 \big)\\
  & \quad + \frac{5\mathcal{J}(\jp)}{32} \big(\lambda_{\varphi H}'\big)^2 \lambda_{H}
          + \frac{27}{8} \lambda_{\varphi H} \lambda_{H}^2
          + \lambda_{\varphi H} \big(\lambda_{\varphi H}'\big)^2
            \bigg( \frac{13}{64} + \frac{\mathcal{D}(\jp)}{128} \bigg) \mathcal{J}(\jp)\\
  & \quad + \frac{9}{8} \lambda_{\varphi H}^2 \lambda_{H}
          + 5 \lambda_{\varphi H}^3
            \bigg( \frac{1}{16} + \frac{\mathcal{D}(\jp)}{32} \bigg)\\
  & \quad + \lambda_{\varphi H} \bigg(   \frac{3}{4} {\sum_{J_1}}^\prime \big(\lambda_\varphi^{(J_1)}\big)^2
                                      \frac{\mathcal{D}(J_1)}{\mathcal{D}(\jp)}
                                    + {\sum_{J_1,J_2}}^\prime
                                      \lambda_\varphi^{(J_1)} \lambda_\varphi^{(J_2)}
                                      \frac{\mathcal{D}(J_1)\mathcal{D}(J_2)}{\mathcal{D}(\jp)^2}
                             \bigg)\\
  & \quad + \frac{\big(\lambda_{\varphi H}'\big)^2}{16} {\sum_{J_1}}^\prime \lambda_\varphi^{(J_1)}                  
               \frac{\big(\mathcal{J}(J_1) - \mathcal{J}(\jp)\big)\mathcal{D}(J_1)}
                    {\mathcal{D}(\jp)}\\
  & \quad + \bigg(   \frac{3}{2} \lambda_{\varphi H} \lambda_{H}
                   + \frac{3}{4} \lambda_{\varphi H}^2 \bigg)
            {\sum_{J_1}}^\prime \lambda_\varphi^{(J_1)}
            \frac{\mathcal{D}(J_1)}{\mathcal{D}(\jp)}\,;
  \end{split}
\end{align}
\begin{align}
  \Lambda^{\prime\,(2)}_{\varphi H,60} &= 0 \,,\\
  \Lambda^{\prime\,(2)}_{\varphi H,42} &=
          Y_\varphi^3 \bigg(\frac{1}{2}\mathcal{D}(\jp) - \frac{9}{2} \bigg)
        - Y_\varphi   \bigg(   \frac{7}{2} - \frac{40}{3} n_g \bigg) \,,\\
  \Lambda^{\prime\,(2)}_{\varphi H,24} &=
        - Y_\varphi   \bigg(   \frac{113}{2} - 8 n_g
                           + 18 \mathcal{J}(\jp)
                           - \frac{2}{3} \mathcal{D}(\jp) \mathcal{J}(\jp) \bigg) \,,\\
  \Lambda^{\prime\,(2)}_{\varphi H,60} &= 0 \,,\\
  \Lambda^{\prime\,(2)}_{\varphi H,40} &=
          - \frac{1}{32} Y_\varphi^2 \big( 1 + Y_\varphi^2 \big) \lambda_{\varphi H}' \mathcal{D}(\jp) 
          + \frac{15}{32} Y_\varphi^4
          + \lambda_{\varphi H}' \bigg[
                   \frac{13}{32} + \frac{5}{4} Y_\varphi^2
                 - \frac{5}{6} n_g \big( 1 + Y_\varphi^2 \big) \bigg] \,,\\
  \begin{split}
  \Lambda^{\prime\,(2)}_{\varphi H,22} &=
            \lambda_{\varphi H}'
            \bigg[   \frac{33}{16} + \frac{9}{4} \mathcal{J}(\jp)
                   + Y_\varphi^2 \bigg(   \frac{1}{2} \mathcal{J}(\jp) \mathcal{D}(\jp) 
                                      + \frac{15}{4} \mathcal{J}(\jp)
                                      + \frac{3}{16} \bigg) \bigg]
	+ 3Y_\varphi\left(\lambda_H + 2\lambda_{\varphi H}\right)\\
  & \quad + Y_\varphi \big( 18 y_t^2 + 18 y_b^2 + 18 y_c^2 + 6 y_\tau^2 \big) 
          + 3 Y_\varphi {\sum_{J_1}}^\prime \lambda_\varphi^{(J_1)} 
                      \frac{\big(\mathcal{J}(J_1)-2\mathcal{J}(\jp)\big)\mathcal{D}(J_1)}
                           {\mathcal{J}(\jp)\mathcal{D}(\jp)} \,,
  \end{split}\\
  \begin{split}
  \Lambda^{\prime\,(2)}_{\varphi H,04} &= \lambda_{\varphi H}'
            \bigg[   \frac{501}{32} - \frac{3}{2} n_g \\
  & \qquad \qquad + \mathcal{J}(\jp) 
             \bigg(   13 - \frac{1}{8} \mathcal{D}(\jp)
                    - \frac{1}{2} \big( N_{L_L} + 3 N_{Q_L} \big) 
                    + \mathcal{J}(\jp)
                      \bigg( \frac{15}{2} - \frac{1}{6} \mathcal{D}(\jp) \bigg) \bigg) \bigg]\,,
  \end{split}\\
  \begin{split}
  \Lambda^{\prime\,(2)}_{\varphi H,20} &= 
          - \frac{\lambda_{\varphi H}'}{8} 
            \big( 35 y_t^2 + 23 y_b^2 + 35 y_c^2 + 21 y_\tau^2 \big)
          - \frac{3\lambda_{\varphi H}'}{4} Y_\varphi^2
            \big( 3 y_t^2 + 3 y_b^2 + 3 y_c^2 + y_\tau^2 \big)\\
  & \quad - \frac{3}{8} \lambda_{\varphi H}' \lambda_{H} \big( 2 + Y_\varphi^2 \big)
          - \frac{9}{8} \lambda_{\varphi H}' \lambda_{\varphi H} \big( 1 + Y_\varphi^2 \big)\\
  & \quad - \frac{3}{8} \big( 1 + 2 Y_\varphi^2 \big)
            \lambda_{\varphi H}' {\sum_{J_1}}^\prime \lambda_\varphi^{(J_1)}
            \frac{\big(\mathcal{J}(J_1) - 2 \mathcal{J}(\jp)\big)
                  \mathcal{D}(J_1)}{\mathcal{J}(\jp)\mathcal{D}(\jp)} \,,
  \end{split}\\
  \begin{split}
  \Lambda^{\prime\,(2)}_{\varphi H,02} &= 
          - \frac{\lambda_{\varphi H}' \lambda_{\varphi H}}{8}
            \mathcal{J}(\jp) \big( 27 + 36 \mathcal{J}(\jp) \big)
          - \frac{\lambda_{\varphi H}' \lambda_H}{4} \big( 9 + 6 \mathcal{J}(\jp) \big)\\
  & \quad - \lambda_{\varphi H}'
            \big( 9 y_t^2 + 9 y_b^2 + 9 y_c^2 + 3 y_\tau^2 \big)
            \bigg( \frac{9}{8} + \mathcal{J}(\jp) \bigg)\\
  & \quad - \lambda_{\varphi H}'\bigg(\frac{9}{8} + 3\mathcal{J}(\jp)\bigg)
	    {\sum_{J_1}}'\lambda_{\varphi}^{(J_1)}
            \frac{\left(\mathcal{J}(J_1)-2\mathcal{J}(\jp)\right)\mathcal{D}(J_1)}
                 {\mathcal{J}(\jp)\mathcal{D}(\jp)}\,,
  \end{split}\\
  \begin{split}
  \Lambda^{\prime\,(2)}_{\varphi H,00} &= 
            \bigg[   \lambda_{\varphi H}' \lambda_{H}
                   + \frac{3\lambda_{\varphi H} \lambda_{\varphi H}'}{2}
                   + \frac{1}{2} {\sum_{J_1}}^\prime \lambda_\varphi^{(J_1)}
                     \bigg(     \frac{\mathcal{J}(J_1)\mathcal{D}(J_1)}
                                     {\mathcal{J}(\jp)\mathcal{D}(\jp)}
                            - 2 \frac{\mathcal{D}(J_1)}
                                     {\mathcal{D}(\jp)}
                     \bigg)
            \bigg]\\
  & \quad   \times \big( 3 y_t^2 + 3 y_b^2 + 3 y_c^2 + y_\tau^2 \big)\\
  & \quad - 12 \lambda_{\varphi H}' g_s^2 \big( y_t^2 + y_b^2 + y_c^2 \big)
          + \frac{3\lambda_{\varphi H}'}{4}
            \big( 11 y_t^4 + 11 y_b^4 + 11 y_c^4 + y_\tau^4 \big)\\
  & \quad + 6 \lambda_{\varphi H}' y_\tau^2 \big( y_t^2 + y_b^2 + y_c^2 \big)
          + \frac{\lambda_{\varphi H}'}{2} 
            \big( 27 y_t^2 y_b^2 + 36 y_t^2 y_c^2 + 36 y_b^2 y_c^2 \big)\\
  & \quad + \frac{5}{4} \lambda_{\varphi H} \lambda_{\varphi H}' \lambda_{H}
          + \frac{7}{8} \lambda_{\varphi H}' \lambda_{H}^2
          + \lambda_{\varphi H}' \lambda_{\varphi H}^2
            \bigg( \frac{13}{16} + \frac{\mathcal{D}(\jp)}{32} \bigg)\\
  & \quad + \big( \lambda_{\varphi H} \big)^3
            \bigg(   \frac{\mathcal{J}(\jp)\mathcal{D}(\jp)}{128}
                   + \frac{5\mathcal{J}(\jp)}{64}
                   - \frac{1}{64} \bigg)\\
  & \quad + \frac{\lambda_{\varphi H}'}{4}
            {\sum_{J_1}}^\prime \big(\lambda_\varphi^{(J_1)}\big)^2
            \bigg(   \frac{\mathcal{J}(J_1)\mathcal{D}(J_1)}
                          {\mathcal{J}(\jp)\mathcal{D}(\jp)}
                   - \frac{\mathcal{D}(J_1)}
                          {\mathcal{D}(\jp)}
            \bigg)\\
  & \quad + \frac{\lambda_{\varphi H}'}{4\mathcal{D}(\jp)^2}
            {\sum_{J_1,J_2}}^\prime \lambda_\varphi^{(J_1)} \lambda_\varphi^{(J_2)}
            \bigg(   \frac{\mathcal{J}(J_1)\mathcal{D}(J_1)\mathcal{J}(J_2)\mathcal{D}(J_2)}
                          {\mathcal{J}(\jp)^2}\\
  & \hspace{6em}   -2\frac{\big(\mathcal{J}(J_1)+\mathcal{J}(J_2)\big)\mathcal{D}(J_1)\mathcal{D}(J_2)}
                          {\mathcal{J}(\jp)}
                   +4 \mathcal{D}(J_1)\mathcal{D}(J_2)
            \bigg)\\
  & \quad + \frac{\lambda_{\varphi H}' \lambda_H}{4}
            {\sum_{J_1}}^\prime \lambda_\varphi^{(J_1)}
            \bigg(   \frac{\mathcal{J}(J_1)\mathcal{D}(J_1)}
                          {\mathcal{J}(\jp)\mathcal{D}(\jp)}
                   -2\frac{\mathcal{D}(J_1)}
                          {\mathcal{D}(\jp)}
            \bigg)\\
  & \quad + \frac{\lambda_{\varphi H} \lambda_{\varphi H}'}{2}
            {\sum_{J_1}}^\prime \lambda_\varphi^{(J_1)}
            \bigg(   \frac{\mathcal{J}(J_1)\mathcal{D}(J_1)}
                          {\mathcal{J}(\jp)\mathcal{D}(\jp)}
                   - \frac{\mathcal{D}(J_1)}
                          {\mathcal{D}(\jp)}
            \bigg)\,.
  \end{split}
\end{align}
Our results for the quartic Higgs self coupling are
\begin{align}
  \Lambda^{(2)}_{H,60} &= 
          \frac{1}{16} Y_\varphi^2 \mathcal{D}(\jp)
        - \frac{7}{16} + \frac{5}{3} n_g \,,\\
  \Lambda^{(2)}_{H,42} &= 
          \frac{1}{16} Y_\varphi^2 \mathcal{D}(\jp)
        - \frac{43}{16} + \frac{5}{3} n_g \,,\\
  \Lambda^{(2)}_{H,24} &= 
          \frac{1}{12} \mathcal{J}(\jp) \mathcal{D}(\jp)
	- \frac{167}{16} + n_g \,,\\
  \Lambda^{(2)}_{H,06} &= 
          \frac{1}{4} \mathcal{J}(\jp) \mathcal{D}(\jp)
        - \frac{339}{16} + 3 n_g \,,\\
  \begin{split}
  \Lambda^{(2)}_{H,40} &= 
            \frac{3}{16} \lambda_{\varphi H} Y_\varphi^2 \mathcal{D}(\jp)
	  + \frac{13}{4} \lambda_{H}
          - \frac{1}{16} \lambda_H Y_\varphi^2 \mathcal{D}(\jp)
          - \lambda_H \frac{5}{3} n_g \\
  & \quad + \frac{3}{4} \big( 3 y_t^2 + 3 y_b^2 + 3 y_c^2 + y_\tau^2 \big)\,,
  \end{split}\\
  \Lambda^{(2)}_{H,22} &= 
            \frac{45}{4} \lambda_{H}
          + \frac{\lambda_{\varphi H}'}{8} Y_\varphi \mathcal{J}(\jp) \mathcal{D}(\jp)
          + \frac{3}{2} \big( 3 y_t^2 + 3 y_b^2 + 3 y_c^2 + y_\tau^2 \big)\,,\\
  \begin{split}
  \Lambda^{(2)}_{H,04} &= 
            \frac{3}{4} \lambda_{\varphi H} \mathcal{J}(\jp) \mathcal{D}(\jp)
          + \frac{9}{4} \big( 3 y_t^2 + 3 y_b^2 + 3 y_c^2 + y_\tau^2 \big)\\
  & \quad - \lambda_H \bigg(   3 n_g
                             - 33
                             + \frac{1}{4} \mathcal{J}(\jp) \mathcal{D}(\jp)\bigg)\,,
  \end{split}\\
  \begin{split}
  \Lambda^{(2)}_{H,20} &= 
          - \frac{27}{4} \lambda_{H}^2
          - \frac{3}{32} \lambda_{\varphi H}^2 \big( 2 + Y_\varphi^2 \big) \mathcal{D}(\jp)
          - \frac{1}{128} \big(\lambda_{\varphi H}'\big)^2 
                         \big( 2 + Y_\varphi^2 \big) \mathcal{J}(\jp) \mathcal{D}(\jp)\\
  & \quad - \frac{\lambda_{H}}{4} \big( 53 y_t^2 + 41 y_b^2 + 53 y_c^2 + 27 y_\tau^2 \big)
          + 26 y_t^4 + 14 y_b^4 + 26 y_c^4 + 18 y_\tau^4 \,,
  \end{split}\\
  \begin{split}
  \Lambda^{(2)}_{H,02} &= 
          - \frac{81}{4} \lambda_{H}^2
          - \frac{1}{16} \lambda_{\varphi H}^2 \big( 9 + 6 \mathcal{J}(\jp) \big) \mathcal{D}(\jp)
          - \frac{1}{64} \big(\lambda_{\varphi H}'\big)^2 
                         \big( 3 + 2 \mathcal{J}(\jp) \big) \mathcal{J}(\jp) \mathcal{D}(\jp)\\
  & \quad - \frac{45}{4} \lambda_{H} \big( 3 y_t^2 + 3 y_b^2 + 3 y_c^2 + y_\tau^2 \big)
          + 54 y_t^4 + 54 y_b^4 + 54 y_c^4 + 18 y_\tau^4 \,,
  \end{split}\\
  \begin{split}
  \Lambda^{(2)}_{H,00} &= 
            9 \lambda_{H}^3
          + \frac{\mathcal{D}(\jp)}{16} \lambda_{\varphi H}^3
          + \frac{9}{16} \lambda_{\varphi H}^2 \lambda_{H} \mathcal{D}(\jp)
          + \frac{7}{192} \big(\lambda_{\varphi H}'\big)^2 \lambda_{H} 
            \mathcal{J}(\jp) \mathcal{D}(\jp)\\
  & \quad + \frac{5}{192} \big(\lambda_{\varphi H}'\big)^2 \lambda_{\varphi H} 
            \mathcal{J}(\jp) \mathcal{D}(\jp)
          + \frac{\lambda_{\varphi H}^2}{8}
            {\sum_{J_1}}^\prime \lambda_\varphi^{(J_1)} \mathcal{D}(J_1)\\
  & \quad + \frac{\big(\lambda_{\varphi H}'\big)^2}{192}
            {\sum_{J_1}}^\prime \lambda_\varphi^{(J_1)}
            \mathcal{D}(J_1) \big( \mathcal{J}(J_1) - 2 \mathcal{J}(\jp)\big)\\
  & \quad + \bigg(   9 \lambda_{H}^2 
                   + \frac{\mathcal{D}(\jp)}{4} \lambda_{\varphi H}^2
                   + \frac{\mathcal{J}(\jp)\mathcal{D}(\jp)}{48}
                     \big(\lambda_{\varphi H}'\big)^2 \bigg)
            \big( 3 y_t^2 + 3 y_b^2 + 3 y_c^2 + y_\tau^2 \big)\\
  & \quad - \frac{\lambda_{H}}{2} 
            \big( 9 y_t^4 + 9 y_b^4 + 9 y_c^4 + 15 y_\tau^4 
                  - 90 y_t^2 y_b^2 - 108 y_t^2 y_c^2 - 108 y_b^2 y_c^2 \big)\\
  & \quad + 18 \lambda_{H} \big( y_t^2 + y_b^2 + y_c^2 \big) y_\tau^2 
          - 24 \lambda_{H} g_s^2 \big( y_t^2 + y_b^2 + y_c^2 \big)\\
  & \quad - 90 \big( y_t^6 + y_b^6 + y_c^6 \big) - 14 y_\tau^6 
          - 24 \big( y_t^2 + y_b^2 + y_c^2 \big) y_\tau^4
          - 24 \big( y_t^4 + y_b^4 + y_c^4 \big) y_\tau^2\\
  & \quad + 96 g_s^2 \big( y_t^4 + y_b^4 + y_c^4 \big)
          - 54 \big( y_t^4 y_b^2 + y_t^2 y_b^4 \big) 
          - 72 \big( y_t^4 y_c^2 + y_t^2 y_c^4 \big)\\
  & \quad - 72 \big( y_b^4 y_c^2 + y_b^2 y_c^4 \big) \,.
  \end{split}
\end{align}
The full two-loop contributions to the fermion, scalar, and Higgs field
renormalization constants are
\begin{align}
\begin{split}
  \delta Z_{t_R}^{(2)} &=
	\frac{g_1^4}{(16\pi^2)^2\epsilon}
	\bigg(\frac{11}{54} + \frac{20}{27}n_g 
		+ \frac{1}{36}\mathcal{D}(\jp)Y_\varphi^2\bigg)
	+\frac{8}{81}\frac{g_1^4}{(16\pi^2)^2\epsilon^2}\xi_B^2\\
  & \quad - \frac{g_s^4}{(16\pi^2)^2\epsilon}
	\bigg(\frac{67}{6}+4\xi_G + \frac{1}{2}\xi_G^2-\frac{4}{3}n_g\bigg)
	+ \frac{g_s^4}{(16\pi^2)^2\epsilon^2}
	\bigg(3\xi_G + \frac{17}{9}\xi_G^2\bigg)\\
  & \quad + \frac{8}{9}\frac{g_1^2g_s^2}{(16\pi^2)^2\epsilon}
	+ \frac{16}{27}\frac{g_1^2 g_s^2}{(16\pi^2)^2\epsilon^2}\xi_B\xi_G
	- \frac{y_t^2}{(16\pi^2)^2\epsilon}\bigg(\frac{49}{144}g_1^2
		+\frac{51}{16}g_2^2 - \frac{8}{3}g_s^2\bigg) \\
  & \quad + \frac{y_t^2}{(16\pi^2)^2\epsilon^2}\bigg[\bigg(\frac{17}{24}
		+ \frac{4}{9}\xi_B\bigg)g_1^2 + \frac{9}{8}g_2^2
		+ \bigg(4+\frac{4}{3}\xi_G\bigg)g_s^2\bigg] \\
  & \quad + \frac{y_t^2}{(16\pi^2)^2\epsilon}\bigg(
	\frac{19}{8}y_t^2 + \frac{19}{8}y_b^2 + \frac{9}{4}y_c^2
	+ \frac{3}{4}y_\tau^2\bigg)\\
  & \quad - \frac{y_t^2}{(16\pi^2)^2\epsilon^2}\bigg(\frac{7}{4}y_t^2
	+ \frac{3}{4}y_b^2 + \frac{3}{2}y_c^2 + \frac{1}{2}y_\tau^2\bigg)
\end{split}\\
\begin{split}
  \delta Z_{b_R}^{(2)} &=
	\frac{g_1^4}{(16\pi^2)^2\epsilon}
	\bigg(\frac{5}{216} + \frac{5}{27}n_g 
		+ \frac{1}{144}\mathcal{D}(\jp)Y_\varphi^2\bigg)
	+\frac{1}{162}\frac{g_1^4}{(16\pi^2)^2\epsilon^2}\xi_B^2\\
  & \quad - \frac{g_s^4}{(16\pi^2)^2\epsilon}
	\bigg(\frac{67}{6} + 4\xi_G + \frac{1}{2}\xi_G^2-\frac{4}{3}n_g\bigg)
	+ \frac{g_s^4}{(16\pi^2)^2\epsilon^2}
	\bigg(3\xi_G + \frac{17}{9}\xi_G^2\bigg)\\
  & \quad + \frac{2}{9}\frac{g_1^2g_s^2}{(16\pi^2)^2\epsilon}
	+ \frac{4}{27}\frac{g_1^2 g_s^2}{(16\pi^2)^2\epsilon^2}\xi_B\xi_G
	- \frac{y_b^2}{(16\pi^2)^2\epsilon}\bigg(\frac{133}{144}g_1^2
		+\frac{51}{16}g_2^2 - \frac{8}{3}g_s^2\bigg) \\
  & \quad + \frac{y_b^2}{(16\pi^2)^2\epsilon^2}\bigg[\bigg(\frac{5}{24}
		+ \frac{1}{9}\xi_B\bigg)g_1^2 + \frac{9}{8}g_2^2
		+ \bigg(4+\frac{4}{3}\xi_G\bigg)g_s^2\bigg] \\
  & \quad + \frac{y_b^2}{(16\pi^2)^2\epsilon}\bigg(
	\frac{19}{8}y_t^2 + \frac{19}{8}y_b^2 + \frac{9}{4}y_c^2
	+ \frac{3}{4}y_\tau^2\bigg)\\
  & \quad - \frac{y_t^2}{(16\pi^2)^2\epsilon^2}\bigg(\frac{3}{4}y_t^2
	+ \frac{7}{4}y_b^2 + \frac{3}{2}y_c^2 + \frac{1}{2}y_\tau^2\bigg)
\end{split}\\
\begin{split}
  \delta Z_{c_R}^{(2)} &=
	\frac{g_1^4}{(16\pi^2)^2\epsilon}
	\bigg(\frac{11}{54} + \frac{20}{27}n_g 
		+ \frac{1}{36}\mathcal{D}(\jp)Y_\varphi^2\bigg)
	+\frac{8}{81}\frac{g_1^4}{(16\pi^2)^2\epsilon^2}\xi_B^2\\
  & \quad - \frac{g_s^4}{(16\pi^2)^2\epsilon}
	\bigg(\frac{67}{6}+4\xi_G + \frac{1}{2}\xi_G^2-\frac{4}{3}n_g\bigg)
	+ \frac{g_s^4}{(16\pi^2)^2\epsilon^2}
	\bigg(3\xi_G + \frac{17}{9}\xi_G^2\bigg)\\
  & \quad + \frac{8}{9}\frac{g_1^2g_s^2}{(16\pi^2)^2\epsilon}
	+ \frac{16}{27}\frac{g_1^2 g_s^2}{(16\pi^2)^2\epsilon^2}\xi_B\xi_G
	- \frac{y_c^2}{(16\pi^2)^2\epsilon}\bigg(\frac{49}{144}g_1^2
		+\frac{51}{16}g_2^2 - \frac{8}{3}g_s^2\bigg) \\
  & \quad + \frac{y_c^2}{(16\pi^2)^2\epsilon^2}\bigg[\bigg(\frac{17}{24}
		+ \frac{4}{9}\xi_B\bigg)g_1^2 + \frac{9}{8}g_2^2
		+ \bigg(4+\frac{4}{3}\xi_G\bigg)g_s^2\bigg] \\
  & \quad + \frac{y_c^2}{(16\pi^2)^2\epsilon}\bigg(
	\frac{9}{4}y_t^2 + \frac{9}{4}y_b^2 + \frac{19}{8}y_c^2
	+ \frac{3}{4}y_\tau^2\bigg)\\
  & \quad - \frac{y_c^2}{(16\pi^2)^2\epsilon^2}\bigg(\frac{3}{2}y_t^2
	+ \frac{3}{2}y_b^2 + \frac{7}{4}y_c^2 + \frac{1}{2}y_\tau^2\bigg)
\end{split}\\
\begin{split}
  \delta Z_{Q_{t,L}}^{(2)} &=
	\frac{g_1^4}{(16\pi^2)^2\epsilon}
	\bigg(\frac{7}{1728} + \frac{5}{108}n_g 
		+ \frac{1}{576}\mathcal{D}(\jp)Y_\varphi^2\bigg)
	+\frac{1}{2592}\frac{g_1^4}{(16\pi^2)^2\epsilon^2}\xi_B^2\\
  & \quad - \frac{g_2^4}{(16\pi^2)^2\epsilon}\bigg(\frac{267}{64}
	+ \frac{3}{2}\xi_W + \frac{3}{16}\xi_W^2 - \frac{3}{4}n_g
	- \frac{1}{16}\mathcal{J}(\jp)\mathcal{D}(\jp)\bigg)\\
  & \quad + \frac{g_2^4}{(16\pi^2)^2\epsilon^2}\bigg(\frac{9}{8}\xi_W
	+ \frac{21}{32}\xi_W^2\bigg)
	- \frac{g_s^4}{(16\pi^2)^2\epsilon}
	\bigg(\frac{67}{6}+4\xi_G + \frac{1}{2}\xi_G^2-\frac{4}{3}n_g\bigg)\\
  & \quad + \frac{g_s^4}{(16\pi^2)^2\epsilon^2}
	\bigg(3\xi_G + \frac{17}{9}\xi_G^2\bigg)
 	+ \frac{1}{32}\frac{g_1^2g_2^2}{(16\pi^2)^2\epsilon}
	+ \frac{1}{48}\frac{g_1^2g_2^2}{(16\pi^2)^2\epsilon^2}\xi_B\xi_W\\
  & \quad + \frac{1}{18}\frac{g_1^2g_s^2}{(16\pi^2)^2\epsilon}
	+ \frac{1}{27}\frac{g_1^2 g_s^2}{(16\pi^2)^2\epsilon^2}\xi_B\xi_G
	+ \frac{3}{2}\frac{g_2^2 g_s^2}{(16\pi^2)^2\epsilon}
	+ \frac{g_2^2 g_s^2}{(16\pi^2)^2\epsilon^2}\xi_W\xi_G\\
  & \quad - \frac{1}{(16\pi^2)^2\epsilon}
	\bigg[\bigg(\frac{139}{288}y_t^2 + \frac{151}{288}y_b^2\bigg)g_1^2
		+ \frac{33}{32}\big(y_t^2 + y_b^2\big)g_2^2 
		- \frac{4}{3}\big(y_t^2 + y_b^2\big)g_s^2\bigg] \\
  & \quad + \frac{1}{(16\pi^2)^2\epsilon^2}
	\bigg[\bigg(\frac{17}{48}y_t^2 + \frac{5}{48}y_b^2
		+ \frac{1}{72}\big(y_t^2+y_b^2\big)\xi_B\bigg)g_1^2 
		\\
 & \qquad \qquad \qquad + \bigg(\frac{9}{16}+\frac{3}{8}\xi_W\bigg)
		\big(y_t^2+y_b^2\big)g_2^2
		+ \bigg(2+\frac{2}{3}\xi_G\bigg)
		\big(y_t^2+y_b^2\big)g_s^2\bigg] \\
  & \quad + \frac{y_t^2}{(16\pi^2)^2\epsilon}\bigg(
	\frac{5}{4}y_t^2 + \frac{9}{8}y_b^2 + \frac{9}{8}y_c^2
	+ \frac{3}{8}y_\tau^2\bigg)
	- \frac{y_t^2}{(16\pi^2)^2\epsilon^2}\bigg(y_t^2
	+ \frac{1}{4}y_t^2 + \frac{3}{4}y_c^2 + \frac{1}{4}y_\tau^2\bigg)\\
  & \quad + \frac{y_b^2}{(16\pi^2)^2\epsilon}\bigg(
	\frac{9}{8}y_b^2 + \frac{5}{4}y_b^2 + \frac{9}{8}y_c^2
	+ \frac{3}{8}y_\tau^2\bigg) - \frac{y_b^2}{(16\pi^2)^2\epsilon^2}
	\bigg(\frac{1}{4}y_t^2
	+ y_b^2 + \frac{3}{4}y_c^2 + \frac{1}{4}y_\tau^2\bigg)
\end{split}\\
\begin{split}
  \delta Z_{Q_{c,L}}^{(2)} &=
	\frac{g_1^4}{(16\pi^2)^2\epsilon}
	\bigg(\frac{7}{1728} + \frac{5}{108}n_g 
		+ \frac{1}{576}\mathcal{D}(\jp)Y_\varphi^2\bigg)
	+\frac{1}{2592}\frac{g_1^4}{(16\pi^2)^2\epsilon^2}\xi_B^2\\
  & \quad - \frac{g_2^4}{(16\pi^2)^2\epsilon}\bigg(\frac{267}{64}
	+ \frac{3}{2}\xi_W + \frac{3}{16}\xi_W^2 - \frac{3}{4}n_g
	- \frac{1}{16}\mathcal{J}(\jp)\mathcal{D}(\jp)\bigg)\\
  & \quad + \frac{g_2^4}{(16\pi^2)^2\epsilon^2}\bigg(\frac{9}{8}\xi_W
	+ \frac{21}{32}\xi_W^2\bigg)
	- \frac{g_s^4}{(16\pi^2)^2\epsilon}
	\bigg(\frac{67}{6}+4\xi_G + \frac{1}{2}\xi_G^2-\frac{4}{3}n_g\bigg)\\
  & \quad + \frac{g_s^4}{(16\pi^2)^2\epsilon^2}
	\bigg(3\xi_G + \frac{17}{9}\xi_G^2\bigg)
 	+ \frac{1}{32}\frac{g_1^2g_2^2}{(16\pi^2)^2\epsilon}
	+ \frac{1}{48}\frac{g_1^2g_2^2}{(16\pi^2)^2\epsilon^2}\xi_B\xi_W\\
  & \quad + \frac{1}{18}\frac{g_1^2g_s^2}{(16\pi^2)^2\epsilon}
	+ \frac{1}{27}\frac{g_1^2 g_s^2}{(16\pi^2)^2\epsilon^2}\xi_B\xi_G
	+ \frac{3}{2}\frac{g_2^2 g_s^2}{(16\pi^2)^2\epsilon}
	+ \frac{g_2^2 g_s^2}{(16\pi^2)^2\epsilon^2}\xi_W\xi_G\\
  & \quad - \frac{y_c^2}{(16\pi^2)^2\epsilon}
	\bigg(\frac{139}{288}g_1^2
		+ \frac{33}{32}g_2^2 
		- \frac{4}{3}g_s^2\bigg) \\
  & \quad + \frac{y_c^2}{(16\pi^2)^2\epsilon^2}
	\bigg[\bigg(\frac{17}{48}
		+ \frac{1}{72}\xi_B\bigg)g_1^2 
		+ \bigg(\frac{9}{16}+\frac{3}{8}\xi_W\bigg)g_2^2
		+ \bigg(2+\frac{2}{3}\xi_G\bigg)g_s^2\bigg] \\
  & \quad + \frac{y_c^2}{(16\pi^2)^2\epsilon}\bigg(
	\frac{9}{8}y_t^2 + \frac{9}{8}y_b^2 + \frac{5}{4}y_c^2
	+ \frac{3}{8}y_\tau^2\bigg)
	- \frac{y_c^2}{(16\pi^2)^2\epsilon^2}\bigg(\frac{3}{4}y_t^2
	+ \frac{3}{4}y_b^2 + y_c^2 + \frac{1}{4}y_\tau^2\bigg)
\end{split}\\
\begin{split}
  \delta Z_{\tau_R}^{(2)} &=
	\frac{g_1^4}{(16\pi^2)^2\epsilon}
	\bigg(\frac{7}{8} + \frac{5}{3}n_g 
		+ \frac{1}{16}\mathcal{D}(\jp)Y_\varphi^2\bigg)
	+\frac{1}{2}\frac{g_1^4}{(16\pi^2)^2\epsilon^2}\xi_B^2\\
  & \quad + \frac{11}{16}\frac{y_\tau^2g_1^2}{(16\pi^2)^2\epsilon}
	+ \frac{y_\tau^2g_1^2}{(16\pi^2)^2\epsilon^2}\bigg(\frac{15}{8}
		+ \xi_B\bigg)
	- \frac{51}{16}\frac{y_\tau^2g_2^2}{(16\pi^2)^2\epsilon}
	+ \frac{9}{8}\frac{y_\tau^2g_1^2}{(16\pi^2)^2\epsilon^2}\\
  & \quad + \frac{y_\tau^2}{(16\pi^2)^2\epsilon}\bigg(
	\frac{9}{4}y_t^2 + \frac{9}{4}y_b^2 + \frac{9}{4}y_c^2
	+ \frac{7}{8}y_\tau^2\bigg)\\
  & \quad - \frac{y_\tau^2}{(16\pi^2)^2\epsilon^2}\bigg(\frac{3}{2}y_t^2
	+ \frac{3}{2}y_b^2 + \frac{3}{2}y_c^2 + \frac{3}{4}y_\tau^2\bigg)
\end{split}\\
\begin{split}
  \delta Z_{L_{\tau,L}}^{(2)} &=
	\frac{g_1^4}{(16\pi^2)^2\epsilon}
	\bigg(\frac{5}{64} + \frac{5}{12}n_g 
		+ \frac{1}{64}\mathcal{D}(\jp)Y_\varphi^2\bigg)
	+\frac{1}{32}\frac{g_1^4}{(16\pi^2)^2\epsilon^2}\xi_B^2\\
  & \quad - \frac{g_2^4}{(16\pi^2)^2\epsilon}\bigg(\frac{267}{64}
	+ \frac{3}{2}\xi_W + \frac{3}{16}\xi_W^2 - \frac{3}{4}n_g
	- \frac{1}{16}\mathcal{J}(\jp)\mathcal{D}(\jp)\bigg)\\
  & \quad + \frac{g_2^4}{(16\pi^2)^2\epsilon^2}\bigg(\frac{9}{8}\xi_W
	+ \frac{21}{32}\xi_W^2\bigg)
 	+ \frac{9}{32}\frac{g_1^2g_2^2}{(16\pi^2)^2\epsilon}
	+ \frac{3}{16}\frac{g_1^2g_2^2}{(16\pi^2)^2\epsilon^2}\xi_B\xi_W\\
  & \quad - \frac{y_\tau^2}{(16\pi^2)^2\epsilon}
	\bigg(\frac{7}{32}g_1^2
		+ \frac{33}{32}g_2^2\bigg) \\
  & \quad + \frac{y_\tau^2}{(16\pi^2)^2\epsilon^2}
	\bigg[\bigg(\frac{15}{16}
		+ \frac{1}{8}\xi_B\bigg)g_1^2 
		+ \bigg(\frac{9}{16}+\frac{3}{8}\xi_W\bigg)g_2^2\bigg] \\
  & \quad + \frac{y_\tau^2}{(16\pi^2)^2\epsilon}\bigg(
	\frac{9}{8}y_t^2 + \frac{9}{8}y_b^2 + \frac{9}{8}y_c^2
	+ \frac{1}{2}y_\tau^2\bigg)\\
  & \quad - \frac{y_\tau^2}{(16\pi^2)^2\epsilon^2}\bigg(\frac{3}{4}y_t^2
	+ \frac{3}{4}y_b^2 + \frac{3}{4}y_c^2 + \frac{1}{2}y_\tau^2\bigg)
\end{split}\\
\begin{split}
  \delta Z_\varphi^{(2)} &= 
          - \frac{g_1^4}{(16\pi^2)^2\epsilon}
        \bigg[\frac{Y_\varphi^4}{64} 
              \bigg( 3 + \frac{11}{3} \mathcal{D}(\jp) \bigg)  \\
  & \hspace{7em} + Y_\varphi^2 \bigg( \frac{11}{96} + \frac{5N_{L_L}}{48} + \frac{5N_\ell}{24}
                 + \frac{5N_{Q_L}}{144} + \frac{5N_u}{18} + \frac{5N_d}{72} \bigg) \bigg] \\
  & \quad + \frac{g_1^4}{(16\pi^2)^2\epsilon^2}
        \bigg[\frac{Y_\varphi^4}{32} 
              \big( 9 - 6 \xi_B + \xi_B^2 + \mathcal{D}(\jp) \big) \\
  & \hspace{7em} + Y_\varphi^2 \bigg( \frac{1}{16} + \frac{N_{L_L}}{8} + \frac{N_\ell}{4}
                 + \frac{N_{Q_L}}{24} + \frac{N_u}{3} + \frac{N_d}{12} \bigg) \bigg] \\
  & \quad + \frac{g_2^4}{(16\pi^2)^2\epsilon} \mathcal{J}(\jp)
        \bigg[   \frac{269}{24} - 2 \xi_W - \frac{\xi_W^2}{4} 
               - \frac{5N_{L_L}}{12} - \frac{5N_{Q_L}}{4} \\
  & \hspace{9em} - \mathcal{J}(\jp) \bigg( \frac{3}{4} + \frac{11}{36} \mathcal{D}(\jp) \bigg) \bigg] \\
  & \quad + \frac{g_2^4}{(16\pi^2)^2\epsilon^2} \mathcal{J}(\jp)
        \bigg[   \frac{\xi_W^2}{2} + \frac{3\xi_W}{2} - \frac{43}{4}
               + \frac{N_{L_L}}{2} + \frac{3N_{Q_L}}{2} \\
  & \hspace{9em} + \mathcal{J}(\jp) \bigg( \frac{\mathcal{D}(\jp)}{6} 
                 + \frac{\xi_W^2}{2} - 3\xi_W + \frac{9}{2} \bigg) \bigg] \\
  & \quad - \frac{3}{8} \frac{g_1^2 g_2^2}{(16\pi^2)^2\epsilon}
            Y_\varphi^2 \mathcal{J}(\jp)
          + \frac{g_1^2 g_2^2}{(16\pi^2)^2\epsilon^2}
            \left[\frac{Y_\varphi^2}{4} \mathcal{J}(\jp) 
                  \big( 9 - 3(\xi_W + \xi_B) + \xi_W \xi_B \big) \right] \\
  & \quad - \frac{1}{8} {\sum_J}^\prime \frac{\big(\lambda_\varphi^{(J)}\big)^2}{(16\pi^2)^2\epsilon}
           \frac{\mathcal{D}(J)}{\mathcal{D}(\jp)}
          - \frac{1}{32} \frac{\lambda_{\varphi H}^2}{(16\pi^2)^2\epsilon}
          - \frac{\mathcal{J}(\jp)}{128}
            \frac{\big(\lambda_{\varphi H}'\big)^2}{(16\pi^2)^2\epsilon} \,,
\end{split}\\
\begin{split}
  \delta Z_H^{(2)} &=
  \quad \frac{g_1^4}{(16\pi^2)^2\epsilon}
        \bigg[- \frac{1}{192}
              \big( 31 + 11 \mathcal{D}(\jp) Y_\varphi^2 \big) \\
  & \hspace{7em} - \frac{5N_{L_L}}{48} - \frac{5N_\ell}{24}
                 - \frac{5N_{Q_L}}{144} - \frac{5N_u}{18} - \frac{5N_d}{72} \bigg] \\
  & \quad + \frac{g_1^4}{(16\pi^2)^2\epsilon^2}
        \bigg[\frac{1}{32}
              \Big( 11 - 6 \xi_B + \xi_B^2 + \mathcal{D}(\jp) Y_\varphi^2 \Big) \\
  & \hspace{7em} + \frac{N_{L_L}}{8} + \frac{N_\ell}{4}
                 + \frac{N_{Q_L}}{24} + \frac{N_u}{3} + \frac{N_d}{12} \bigg] \\
  & \quad + \frac{g_2^4}{(16\pi^2)^2\epsilon}
        \left[   \frac{511}{64} - \frac{3}{2} \xi_W - \frac{3}{16}\xi_W^2
               - \frac{5N_{L_L}}{16} - \frac{15N_{Q_L}}{16}
               - \frac{11}{48} \mathcal{J}(\jp) \mathcal{D}(\jp) \right] \\
  & \quad + \frac{g_2^4}{(16\pi^2)^2\epsilon^2}
        \left[ -\frac{177}{32} +\frac{21}{32}\xi_W^2 -\frac{9}{16}\xi_W 
               + \frac{3N_{L_L}}{8} + \frac{9N_{Q_L}}{8}
               + \frac{\mathcal{J}(\jp)\mathcal{D}(\jp)}{8} \right] \\
  & \quad - \frac{9}{32} \frac{g_1^2 g_2^2}{(16\pi^2)^2\epsilon}
          + \frac{1}{16} \frac{g_1^2 g_2^2}{(16\pi^2)^2\epsilon^2}
            \big( 27 - 9(\xi_W + \xi_B) + 3\xi_W \xi_B \big) \\
  & \quad - \frac{3}{16} \frac{\lambda_H^2}{(16\pi^2)^2\epsilon}
          - \frac{\mathcal{D}(\jp)}{64} \frac{\lambda_{\varphi H}^2}{(16\pi^2)^2\epsilon}
          - \frac{\mathcal{D}(\jp) \mathcal{J}(\jp)}{256}
            \frac{\big(\lambda_{\varphi H}'\big)^2}{(16\pi^2)^2\epsilon} \\
  & \quad + \frac{27y_t^4 + 27y_b^4 + 27y_c^4 - 6 y_t^2 y_b^2 + 9y_\tau^4}{8(16\pi^2)^2\epsilon} 
          - \frac{9y_t^4 + 9y_b^4 + 9y_c^4 - 18 y_t^2 y_b^2 + 3y_\tau^4}{4(16\pi^2)^2\epsilon^2} \\
  & \quad - \frac{10g_s^2}{(16\pi^2)^2\epsilon} \big( y_t^2 + y_b^2 + y_c^2 \big)
          + \frac{12g_s^2}{(16\pi^2)^2\epsilon^2} \big( y_t^2 + y_b^2 + y_c^2 \big) \\
  & \quad - \frac{15}{16} \frac{g_2^2}{(16\pi^2)^2\epsilon}
            \big( 3y_t^2 + 3y_b^2 + 3y_c^2 + y_\tau^2 \big) \\
  & \quad - \frac{g_2^2}{8(16\pi^2)^2\epsilon^2}
            \big( 3y_t^2 + 3y_b^2 + 3y_c^2 + y_\tau^2 \big) \big( 9 - 6 \xi_W \big) \\
  & \quad - \frac{g_1^2}{48(16\pi^2)^2\epsilon}
            \big( 85y_t^2 + 25y_b^2 + 85y_c^2 + 75y_\tau^2 \big) \\
  & \quad + \frac{g_1^2}{8(16\pi^2)^2\epsilon^2}
            \Big[ \big( 6y_t^2 + 6y_b^2 + 6y_c^2 + 2y_\tau^2 \big) \xi_B
                  - \big( y_t^2 + 13y_b^2 + y_c^2 - 9y_\tau^2 \big) \Big] \,.
\end{split}
\end{align}
For the two-loop contributions to the electroweak gauge-boson field
renormalization constants we consider only contributions of the scalar
multiplet. We find
\begin{align}
  \delta Z_B^{(2)} &= - \frac{1}{8} \frac{g_1^4}{(16\pi^2)^2\epsilon}
                       Y_\varphi^4 \mathcal{D}(\jp)
                     - \frac{1}{2} \frac{g_1^2 g_2^2}{(16\pi^2)^2\epsilon} 
                       Y_\varphi^2 \mathcal{J}(\jp)\mathcal{D}(\jp) \,,\\
\begin{split}
  \delta Z_W^{(2)} &= 
    \quad   \frac{g_2^4}{(16\pi^2)^2\epsilon}
            \left(\frac{1}{12}\mathcal{J}(\jp) \mathcal{D}(\jp)
          - \frac{2}{3}\mathcal{J}(\jp)^2 \mathcal{D}(\jp) \right) \\
  & \quad + \frac{g_2^4}{(16\pi^2)^2\epsilon^2}
            \frac{1}{18} \mathcal{J}(\jp) \mathcal{D}(\jp) ( 3 + 2\xi_W)
          - \frac{g_1^2 g_2^2}{(16\pi^2)^2\epsilon}
            \frac{1}{6} Y_\varphi^2 \mathcal{J}(\jp) \mathcal{D}(\jp) \,.
\end{split}
\end{align}

\addcontentsline{toc}{section}{References}
\bibliographystyle{JHEP}
\bibliography{references}

\providecommand{\href}[2]{#2}\begingroup\raggedright\begin{thebibliography}{10}

\bibitem{Cirelli:2005uq}
M.~Cirelli, N.~Fornengo and A.~Strumia, \emph{{Minimal dark matter}},
  \href{https://doi.org/10.1016/j.nuclphysb.2006.07.012}{\emph{Nucl. Phys. B}
  {\bfseries 753} (2006) 178--194},
  [\href{https://arxiv.org/abs/hep-ph/0512090}{{\ttfamily hep-ph/0512090}}].

\bibitem{Collins:1984xc}
J.~C. Collins, \emph{{Renormalization}: {An Introduction to Renormalization,
  The Renormalization Group, and the Operator Product Expansion}}, vol.~26 of
  \emph{Cambridge Monographs on Mathematical Physics}.
\newblock Cambridge University Press, Cambridge, 1986,
  \href{https://doi.org/10.1017/CBO9780511622656}{10.1017/CBO9780511622656}.

\bibitem{Bishara:2018vix}
F.~Bishara, J.~Brod, B.~Grinstein and J.~Zupan, \emph{{Renormalization Group
  Effects in Dark Matter Interactions}},
  \href{https://doi.org/10.1007/JHEP03(2020)089}{\emph{JHEP} {\bfseries 03}
  (2020) 089}, [\href{https://arxiv.org/abs/1809.03506}{{\ttfamily
  1809.03506}}].

\bibitem{Brod:2017bsw}
J.~Brod, A.~Gootjes-Dreesbach, M.~Tammaro and J.~Zupan, \emph{{Effective Field
  Theory for Dark Matter Direct Detection up to Dimension Seven}},
  \href{https://doi.org/10.1007/JHEP10(2018)065}{\emph{JHEP} {\bfseries 10}
  (2018) 065}, [\href{https://arxiv.org/abs/1710.10218}{{\ttfamily
  1710.10218}}].

\bibitem{Silveira:1985rk}
V.~Silveira and A.~Zee, \emph{{SCALAR PHANTOMS}},
  \href{https://doi.org/10.1016/0370-2693(85)90624-0}{\emph{Phys. Lett. B}
  {\bfseries 161} (1985) 136--140}.

\bibitem{McDonald:1993ex}
J.~McDonald, \emph{{Gauge singlet scalars as cold dark matter}},
  \href{https://doi.org/10.1103/PhysRevD.50.3637}{\emph{Phys. Rev. D}
  {\bfseries 50} (1994) 3637--3649},
  [\href{https://arxiv.org/abs/hep-ph/0702143}{{\ttfamily hep-ph/0702143}}].

\bibitem{Burgess:2000yq}
C.~Burgess, M.~Pospelov and T.~ter Veldhuis, \emph{{The Minimal model of
  nonbaryonic dark matter: A Singlet scalar}},
  \href{https://doi.org/10.1016/S0550-3213(01)00513-2}{\emph{Nucl. Phys. B}
  {\bfseries 619} (2001) 709--728},
  [\href{https://arxiv.org/abs/hep-ph/0011335}{{\ttfamily hep-ph/0011335}}].

\bibitem{Chao:2018xwz}
W.~Chao, G.-J. Ding, X.-G. He and M.~Ramsey-Musolf, \emph{{Scalar Electroweak
  Multiplet Dark Matter}},
  \href{https://doi.org/10.1007/JHEP08(2019)058}{\emph{JHEP} {\bfseries 08}
  (2019) 058}, [\href{https://arxiv.org/abs/1812.07829}{{\ttfamily
  1812.07829}}].

\bibitem{Machacek:1984zw}
M.~E. Machacek and M.~T. Vaughn, \emph{{Two Loop Renormalization Group
  Equations in a General Quantum Field Theory. 3. Scalar Quartic Couplings}},
  \href{https://doi.org/10.1016/0550-3213(85)90040-9}{\emph{Nucl. Phys.}
  {\bfseries B249} (1985) 70--92}.

\bibitem{Jack:1984vj}
I.~Jack and H.~Osborn, \emph{{General Background Field Calculations With
  Fermion Fields}},
  \href{https://doi.org/10.1016/0550-3213(85)90088-4}{\emph{Nucl. Phys. B}
  {\bfseries 249} (1985) 472--506}.

\bibitem{Arason:1991ic}
H.~Arason, D.~Castano, B.~Keszthelyi, S.~Mikaelian, E.~Piard, P.~Ramond et~al.,
  \emph{{Renormalization group study of the standard model and its extensions.
  1. The Standard model}},
  \href{https://doi.org/10.1103/PhysRevD.46.3945}{\emph{Phys. Rev. D}
  {\bfseries 46} (1992) 3945--3965}.

\bibitem{Luo:2002ti}
M.-x. Luo, H.-w. Wang and Y.~Xiao, \emph{{Two loop renormalization group
  equations in general gauge field theories}},
  \href{https://doi.org/10.1103/PhysRevD.67.065019}{\emph{Phys. Rev.}
  {\bfseries D67} (2003) 065019},
  [\href{https://arxiv.org/abs/hep-ph/0211440}{{\ttfamily hep-ph/0211440}}].

\bibitem{Luo:2003}
M.-x. Luo and Y.~Xiao, \emph{{Two loop renormalization group equations in the
  Standard Model}},
  \href{https://doi.org/10.1103/PhysRevLett.90.011601}{\emph{Phys. Rev. Lett.}
  {\bfseries 90} (2003) 011601},
  [\href{https://arxiv.org/abs/hep-ph/0207271}{{\ttfamily hep-ph/0207271}}].

\bibitem{Chetyrkin:2012rz}
K.~Chetyrkin and M.~Zoller, \emph{{Three-loop $\beta$-functions for top-Yukawa
  and the Higgs self-interaction in the Standard Model}},
  \href{https://doi.org/10.1007/JHEP06(2012)033}{\emph{JHEP} {\bfseries 06}
  (2012) 033}, [\href{https://arxiv.org/abs/1205.2892}{{\ttfamily 1205.2892}}].

\bibitem{Mihaila:2012}
L.~Mihaila, J.~Salomon and M.~Steinhauser, \emph{{Renormalization constants and
  beta functions for the gauge couplings of the Standard Model to three-loop
  order}}, \href{https://doi.org/10.1103/PhysRevD.86.096008}{\emph{Phys. Rev.
  D} {\bfseries 86} (2012) 096008},
  [\href{https://arxiv.org/abs/1208.3357}{{\ttfamily 1208.3357}}].

\bibitem{Bednyakov:2013cpa}
A.~Bednyakov, A.~Pikelner and V.~Velizhanin, \emph{{Three-loop Higgs
  self-coupling beta-function in the Standard Model with complex Yukawa
  matrices}},
  \href{https://doi.org/10.1016/j.nuclphysb.2013.12.012}{\emph{Nucl. Phys. B}
  {\bfseries 879} (2014) 256--267},
  [\href{https://arxiv.org/abs/1310.3806}{{\ttfamily 1310.3806}}].

\bibitem{Bednyakov:2014}
A.~Bednyakov, A.~Pikelner and V.~Velizhanin, \emph{{Three-loop SM
  beta-functions for matrix Yukawa couplings}},
  \href{https://doi.org/10.1016/j.physletb.2014.08.049}{\emph{Phys. Lett. B}
  {\bfseries 737} (2014) 129--134},
  [\href{https://arxiv.org/abs/1406.7171}{{\ttfamily 1406.7171}}].

\bibitem{Davies:2019onf}
J.~Davies, F.~Herren, C.~Poole, M.~Steinhauser and A.~E. Thomsen, \emph{{Gauge
  Coupling $\beta$ Functions to Four-Loop Order in the Standard Model}},
  \href{https://doi.org/10.1103/PhysRevLett.124.071803}{\emph{Phys. Rev. Lett.}
  {\bfseries 124} (2020) 071803},
  [\href{https://arxiv.org/abs/1912.07624}{{\ttfamily 1912.07624}}].

\bibitem{weinberg_2015}
S.~Weinberg, \emph{Lectures on Quantum Mechanics}.
\newblock Cambridge University Press, 2~ed., 2015,
  \href{https://doi.org/10.1017/CBO9781316276105}{10.1017/CBO9781316276105}.

\bibitem{Chetyrkin:1997fm}
K.~G. Chetyrkin, M.~Misiak and M.~Munz, \emph{{Beta functions and anomalous
  dimensions up to three loops}},
  \href{https://doi.org/10.1016/S0550-3213(98)00122-9}{\emph{Nucl. Phys. B}
  {\bfseries 518} (1998) 473--494},
  [\href{https://arxiv.org/abs/hep-ph/9711266}{{\ttfamily hep-ph/9711266}}].

\bibitem{Vermaseren:2000nd}
J.~A.~M. Vermaseren, \emph{{New features of FORM}},
  \href{https://arxiv.org/abs/math-ph/0010025}{{\ttfamily math-ph/0010025}}.

\bibitem{Davydychev:1992mt}
A.~I. Davydychev and J.~B. Tausk, \emph{{Two loop selfenergy diagrams with
  different masses and the momentum expansion}},
  \href{https://doi.org/10.1016/0550-3213(93)90338-P}{\emph{Nucl. Phys.}
  {\bfseries B397} (1993) 123--142}.

\bibitem{Nogueira:1991ex}
P.~Nogueira, \emph{{Automatic Feynman graph generation}},
  \href{https://doi.org/10.1006/jcph.1993.1074}{\emph{J. Comput. Phys.}
  {\bfseries 105} (1993) 279--289}.

\bibitem{Messiah:1962}
A.~Messiah, \emph{9j symbols},  in \emph{Quantum Mechanics Vol. II},
  ch.~Appendix C III, pp.~1066--1068.
\newblock North-Holland, Amsterdam, Netherlands, 1962.

\bibitem{Trueman:1995ca}
T.~Trueman, \emph{{Spurious anomalies in dimensional renormalization}},
  \href{https://doi.org/10.1007/BF02907437}{\emph{Z. Phys. C} {\bfseries 69}
  (1996) 525--536}, [\href{https://arxiv.org/abs/hep-ph/9504315}{{\ttfamily
  hep-ph/9504315}}].

\bibitem{Sartore:2020gou}
L.~Sartore and I.~Schienbein, \emph{{PyR@TE 3}},
  \href{https://arxiv.org/abs/2007.12700}{{\ttfamily 2007.12700}}.

\bibitem{PDG2020}
{\scshape Particle Data Group} collaboration, P.~A. Zyla et~al., \emph{{Review
  of Particle Physics}}, {\emph{Prog. Theor. Exp. Phys} {\bfseries 083C01}
  (2020) }.

\bibitem{Degrassi:2012ry}
G.~Degrassi, S.~Di~Vita, J.~Elias-Miro, J.~R. Espinosa, G.~F. Giudice,
  G.~Isidori et~al., \emph{{Higgs mass and vacuum stability in the Standard
  Model at NNLO}}, \href{https://doi.org/10.1007/JHEP08(2012)098}{\emph{JHEP}
  {\bfseries 08} (2012) 098},
  [\href{https://arxiv.org/abs/1205.6497}{{\ttfamily 1205.6497}}].

\bibitem{Johansson:2015cca}
H.~Johansson and C.~Forssén, \emph{{Fast and accurate evaluation of Wigner 3j,
  6j, and 9j symbols using prime factorisation and multi-word integer
  arithmetic}}, \href{https://doi.org/10.1137/15M1021908}{\emph{SIAM J. Sci.
  Statist. Comput.} {\bfseries 38} (2016) A376--A384},
  [\href{https://arxiv.org/abs/1504.08329}{{\ttfamily 1504.08329}}].

\bibitem{Wodtke:1999}
A.~Wodtke and J.~Halpern, ``{Mathematica notebook to calculate Wigner 9-j
  Symbols}.'' \url{https://library.wolfram.com/infocenter/MathSource/481/},
  December, 1999.

\bibitem{Cirelli:2009uv}
M.~Cirelli and A.~Strumia, \emph{{Minimal Dark Matter: Model and results}},
  \href{https://doi.org/10.1088/1367-2630/11/10/105005}{\emph{New J. Phys.}
  {\bfseries 11} (2009) 105005},
  [\href{https://arxiv.org/abs/0903.3381}{{\ttfamily 0903.3381}}].

\bibitem{Muta:2010xua}
T.~Muta, \emph{{Foundations of Quantum Chromodynamics: An Introduction to
  Perturbative Methods in Gauge Theories, (3rd ed.)}}, vol.~78 of \emph{World
  scientific Lecture Notes in Physics}.
\newblock World Scientific, Hackensack, N.J., 3rd~ed., 2010.

\end{thebibliography}\endgroup

\end{document}